\documentclass[]{emulateapj} 
\usepackage{amsmath}

\newcommand{\emm}[1]{\ensuremath{#1}}
\newcommand{\emr}[1]{\emm{\mathrm{#1}}}
\newcommand{\unit}[1]{\emr{\,#1}}

\newcommand{\pc}{\unit{pc}}
 
\newcommand{\D}{\unit{Debye}}

\newcommand{\pscm}{\unit{cm^{-2}}}

\newcommand{\kms}{\unit{km\,s^{-1}}}
 
\newcommand{\K}{\unit{K}}
\newcommand{\mK}{\unit{mK}} 

\newcommand{\kHz}{\unit{kHz}} 
\newcommand{\MHz}{\unit{MHz}}
\newcommand{\GHz}{\unit{GHz}}

\newcommand{\Av}{\emm{A_{v}}}

\newcommand{\Tex}{\emm{T_\emr{ex}}}

\newcommand{\chem}[1]{\ensuremath{\mathrm{#1}}}
\newcommand{\Ht}{\chem{H_{2}}} 
\newcommand{\NHt}{N(\Ht)}

\newcommand{\dV}{\emm{\Delta V}}

\newcommand{\sciexp}[2]{\emm{#1\times10^{#2}}}
\newcommand{\datapoint}[3]{\emm{#1_{-#2}^{+#3}}}
\renewcommand{\datapoint}[3]{\emm{#1(-#2,+#3)}}

\newcommand{\ie} {{i.e.}} 
\newcommand{\eg} {{e.g.}}

\newcommand{\dix}[1]{10^{#1}}

\newcommand{\TMC}{TMC-1}

\newcommand{\FigPlotAb}{ 
\begin{figure*}
	\centering
	\includegraphics[width=0.8\hsize{}]{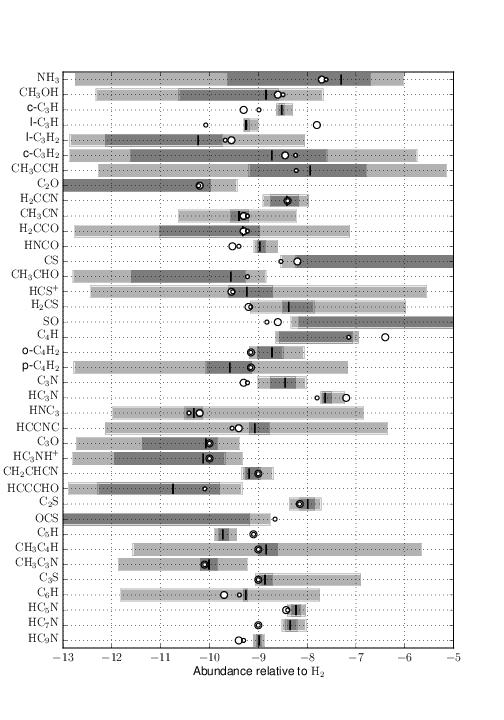} 
	
	\caption{Graphical representation of the computed abundances of
	molecules in \TMC. The thick black vertical lines are the medians of the
	abundance distributions, while the dark (resp. light) gray rectangles
	represent the $68\%$ and $95\%$ confidence intervals, respectively.
	Large open circles are the values presented in \citet{Ohishi.1998},
	while smaller open circles are the values compiled in
	\citet{Agundez.2013}} 
	\label{fig:plot_ab} 
\end{figure*}
}

\newcommand{\FigPlotAbIso}{ 
\begin{figure*}
	\centering
	\includegraphics[width=\hsize{}]{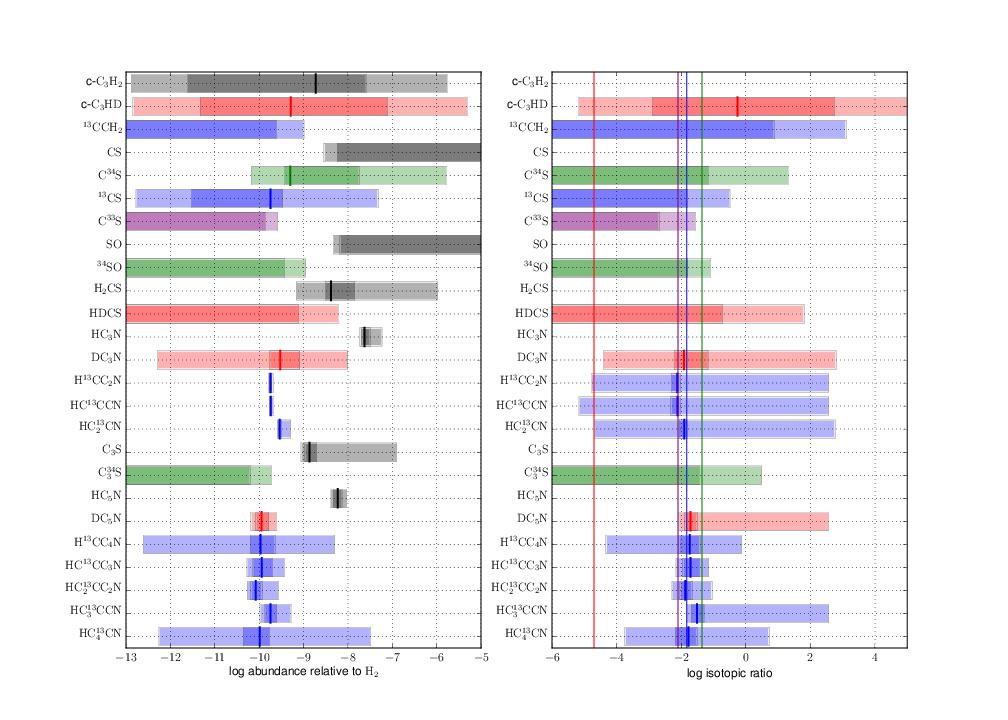}
	
	\caption{Graphical representation of the computed abundances of
	isotopologues in \TMC. The left plot shows the computed abundances with
	the same definition of the shading as for Fig.~\ref{fig:plot_ab}. The
	right plot shows the ratios of the isotopologue to the molecule with the
	main isotope. Species are color coded, with main isotopologues in black,
	isotopologues of deuterium in red, isotopologues of \chem{^{13}C} in
	blue, isotopologues of \chem{^{34}S} in green, and isotopologues of
	\chem{^{33}S} in purple. The solid vertical lines spanning the full
	height of the plot correspond to constant values of the isotopic ratio
	of $\chem{D/H} = \sciexp{2}{-5}$, $\chem{^{13}C/^{12}C} = 1/68$,
	$\chem{^{34}S/^{32}S} = \sciexp{4.4}{-2}$, $\chem{^{33}S/^{32}S} =
	\sciexp{7.9}{-3}$.
	\label{fig:plot_ab_iso}} 
\end{figure*}
}

\newcommand{\TabOutliers}{ 
\begin{table*}
	\caption{
	\label{tab.outliers}List of Lines that Have an Outlier Probability
	Larger than 0.75.} 
	\begin{center}
		{\small 
		\begin{tabular}
			{lr@{ =
			}lrrrr}
			
			\hline Species & \multicolumn{2}{c}{Transition} & Observed & Outlier &
			Modeled & Observed \\
			& \multicolumn{2}{c}{ } & Frequency & Probability
			& Intensity & Intensity \\
			
			& \multicolumn{2}{c}{} & (\MHz) & & (K \kms) & (K \kms) \\
			
			\hline \chem{HC_3N} & $J'-J'', F'-F'' $& 1-0, 2-1 & $9098.329$ & 0.90 &
			$1.3\pm0.1$ & $2.638\pm0.020$ \\
			\chem{HC_3N} & $J'-J'', F'-F'' $& 5-4,
			6-5 \& 5-4 \& 4-3 & $45490.332$ & 0.93 & $7.2\pm0.4$ & $5.239\pm0.020$
			\\
			\chem{HC_7N} & $J'-J'' $& 11-10 & $12407.995$ & 0.999 & $0.56\pm0.25$
			& $0.228\pm0.009$ \\
			\chem{HC_7N} & $J'-J'' $& 12-11 & $13535.999$ &
			0.999 & $0.58\pm0.24$ & $0.212\pm0.011$ \\
			\chem{C_4H} & $N'-N'' J'-J''
			F'-F'' $& 4-3, 9/2-7/2, 4-3 & $38049.629$ & 0.94 & $0.58\pm0.24$ &
			$0.212\pm0.011$ \\
			
			\hline \hline 
		\end{tabular}
		} 
	\end{center}
\end{table*}
}

\newcommand{\TabPrior}{ 
\begin{table*}
	\caption{ 
	\label{tab:prior}Prior
	parameter distributions} 
	\begin{center}
		\begin{tabular}
			{llr@{ = }llc}
			\hline Parameter & Distribution & \multicolumn{2}{c}{Expression} & Lower
			bound & Upper bound \\
			\hline $\log N$ & Uniform & $p(\log N) $ & $
			[(\log N)_{max}-(\log N)_{min}]^{-1}$ & 9 & 17\\
			$\Tex$ & Gaussian &
			$p(\Tex)$ & $\mathcal{N}(5,2) $ & 2.73 & 20\\
			\dV & Gaussian & $p(\dV) $
			& $\mathcal{N}(0.5,0.2) $ & 0 & 3\\
			$\log \sigma$ & Uniform & $p(\log
			\sigma) $ & $[(\log \sigma)_{max}-(\log \sigma)_{min}]^{-1}$ & -3 & 1\\
			$\log \sigma$ & Gaussian$^{\chem{(a)}}$ & $p(\log \sigma) $ &
			$\mathcal{N}(-1.2,0.4) $ & -3 & 1\\
			$Y_o$ & Uniform & $p(Y_o)$ &
			$[(Y_o)_{max}-(Y_o)_{min}]^{-1}$ & 0 & 10\\
			$\log \sigma_o$ & Uniform &
			$p(\log \sigma_o) $ & $ [(\log \sigma_o)_{max}-(\log
			\sigma_o)_{min}]^{-1}$ & -3 & 2\\
			$\{q_i\}$ & Binomial & $p(\{q_i\}) $ &
			$ {\prod_{i=1}^N \left[1-P_o\right]^{q_i} P_o^{\left[1-q_i\right]}}$ & &
			\\
			$P_o$ & Uniform & $p(P_o)$ & $1$ & 0 & 1\\
			\hline 
		\end{tabular}
	\end{center}
	(a): The gaussian prior is used when only one line of a
	given molecule is detected.
\end{table*}

}

\slugcomment{}

\begin{document}

\title{A new reference chemical composition for \TMC}

\author{P. Gratier} \affil{Laboratoire d'astrophysique de Bordeaux,
Univ. Bordeaux, CNRS, B18N, allée Geoffroy Saint-Hilaire, 33615 Pessac,
France} 
\and 
\author{L. Majumdar} \affil{Laboratoire d'astrophysique de
Bordeaux, Univ. Bordeaux, CNRS, B18N, allée Geoffroy Saint-Hilaire,
33615 Pessac, France} \affil{ Indian Centre for Space Physics,
Chalantika 43, Garia Station Road, Kolkata 700084, India} 
\and
\author{M.
Ohishi} \affil{Department of Astronomical Science, The Graduate
University for Advanced Studies (SOKENDAI), Osawa 2-21-1, Mitaka, Tokyo
181-8588, Japan\\
National Astronomical Observatory of Japan, Osawa
2-21-1, Mitaka, Tokyo 181-8588, Japan } 
\and 
\author{E. Roueff}
\affil{LERMA, Observatoire de Paris, PSL Research University, CNRS,
UMR8112, Place Janssen, F-92190, Meudon Cedex, France\\
Sorbonne
Universit\'{e}s, UPMC Univ. Paris 6, 4 Place Jussieu, F-75005, Paris,
France} 
\and 
\author{J. C. Loison} \affil{Universit\'{e} de Bordeaux,
Institut des Sciences Mol\'{e}culaires, UMR 5255, F-33400 Talence,
France} 
\and 
\author{K. M. Hickson} \affil{Universit\'{e} de Bordeaux,
Institut des Sciences Mol\'{e}culaires, UMR 5255, F-33400 Talence,
France} 
\and 
\author{V. Wakelam} \affil{Laboratoire d'astrophysique de
Bordeaux, Univ. Bordeaux, CNRS, B18N, allée Geoffroy Saint-Hilaire,
33615 Pessac, France} 

\begin{abstract}
	Recent detections of complex organic molecules in dark
	clouds have rekindled interest in the astrochemical modeling of these
	environments. Because of its relative closeness and rich molecular
	complexity, \TMC\ has been extensively observed to study the chemical
	processes taking place in dark clouds. We use local thermodynamical
	equilibrium radiative transfer modeling coupled with a Bayesian
	statistical method which takes into account outliers to analyze the data
	from the Nobeyama spectral survey of \TMC\ between 8 and 50\GHz. We
	compute the abundance relative to molecular hydrogen of 57 molecules,
	including 19 isotopologues in \TMC\ along with their associated
	uncertainty. The new results are in general agreement with previous
	abundance determination from \citeauthor{Ohishi.1998} and the values
	reported in the review from \citeauthor{Agundez.2013}. However, in some
	cases, large opacity and low signal to noise effects allow only upper or
	lower limits to be derived, respectively.
\end{abstract}


\section{Introduction} 
\label{sec:introduction}

Complex organic molecules \citep[COMs; \eg, \chem{CH_3OH},
\chem{CH_3CHO}, \chem{HCOOCH_3}, \chem{CH_3OCH_3}, see][]{Herbst.2009}
were detected in various sources, \eg\ hot cores or hot corinos or warm
star-forming regions such as SgB2 \citep{Cummins.1986}, OMC-1
\citep{Blake.1987}, or IRAS 16293-2422 \citep{Bottinelli.2004}.
\citet{Matthews.1985} and \citet{Friberg.1988} also detected some of
these COMs (\chem{CH_3OH} and \chem{CH_3CHO}) in the \TMC\ and L134N
dark clouds. Recently, \citet{Cernicharo.2012} detected the methoxy
radical (\chem{CH_3O}) in the B1 cold dense core. \citet{Bacmann.2012}
also reported the detections of a variety of COMs (\chem{CH_3OH},
\chem{CH_3OCH_3}, \chem{CH_3OCHO}, \chem{CH_2CO}) in the cold prestellar
core L1689b along with few more recent detections of additional COMs
(\chem{C_3O}, t-\chem{HCOOH}) in the L1544 prestellar core reported by
\citet{Vastel.2014}. These detections have triggered a renewed interest
in the study of the astrochemistry of these environments
\citep{Garrod.2006, Herbst.2006, Vasyunin.2013, Ruaud.2015}.
Among the different COMs, \chem{CH_3OH} is the only one for which the
production is well understood. Its formation goes through the the
successive hydrogenation of carbon monoxide accreted from the gas phase
on ice mantles of interstellar dust grains \citep{Watanabe.2002}. For
several other COMs, studies by \citet{Vasyunin.2013} and
\citet{Ruaud.2015} have shown that non-thermal desorption processes can
bring COMs to the gas phase from the grain surfaces in regions where the
grain temperatures remain low ($T_{\emr{dust}}<15\K$) and the UV
radiation is also very low ($\Av>10$) for a standard interstellar
radiation field.

Astrochemical models are based on microphysics processes that can be
studied in laboratories or computed by quantum chemistry theories.
Nevertheless, these models must be benchmarked on template sources that
are extensively studied. \TMC\ is very close \citep[140\pc;
][]{Elias.1978, Kenyon.1994} with a very rich molecular composition and
has thus been extensively observed to understand dark cloud chemistry.
In order to improve astrochemical models, their outputs must be compared
to the widest set of observations as possible. Spectral surveys enable
the unbiased detection of a large number (several tens) of species,
often with many detected lines per species, while at the same time
minimizing relative calibration errors. With the advent of new wideband
receivers and spectrometers, spectral surveys are \emph{de facto} the
default observation mode at millimeter facilities. To interpret this
wealth of data, new statistical tools have to be devised.

In this article, we propose to use a Bayesian approach to compute
abundance values along with uncertainty estimates associated with all
the detected molecules in the Nobeyama spectral survey of \TMC\ from
\citet{Kaifu.2004}.

The paper is organized as follows. We present the observational and
spectroscopic data used in Sect~\ref{sec:data}, the methods to determine
abundances in Sect.~\ref{sec:ab_methods}, and the derived values of
abundances in Sect.~\ref{sec:results}. Finally, these results are
discussed and compared with previous abundance values from the
literature in Sect.~\ref{sec:ab_discussion}. The Appendix presents, in
detail, the Bayesian statistical method used and the individual results
for all molecules.

\section{Data} 
\label{sec:data}

\subsection{Observational data}

\label{sec:observational_data} 

\citet{Kaifu.2004} used the Nobeyama 45m dish between 1984 and 1996 to
observe a full spectral survey of the Cyanopolyyne peak position of
\TMC\ (for the rest of the article, we will use \TMC\ as a shorthand for
\TMC(CP)) between 8.8 and 50\GHz\ with a spectral resolution of 37\kHz\
(this corresponds to a velocity resolution ranging from 1.26\kms\ to
0.22\kms) with a typical sensitivity of 10\mK\ per channel.

The observational data from \citet{Kaifu.2004} consists of integrated
intensities associated with a single (or several in the case of
unresolved hyperfine structure) line of a given molecule. These
integrated intensities are computed by fitting Gaussian line profiles to
the spectral data. No uncertainty is given for these integrated
intensities but the authors give a measured line width $\dV$ and a local
noise level per channel $\sigma$. From these values, we compute the
noise on the integrated intensity, assuming a gaussian profile, using
the following formula:
\begin{equation}
	\sigma_I = \sqrt{\frac{2\pi}{8\ln2}} \, \sigma \dV
	\simeq 1.064 \, \sigma \dV 
\end{equation}

,where the numerical term comes from the integral of a gaussian function
of $FWHM=\dV$

The data from \citet{Kaifu.2004} take into account the coupling with the
source (they assume \TMC\ is a gaussian source of FWHM $160''$). Thus,
the column densities and abundances we compute correspond to this
spatial scale.

\subsection{Spectroscopic data} 

\label{sec:spectroscopic_data} 

Under the local thermal equilibrium hypothesis, integrated intensities
are dependent on a set of microphysic parameters: the Einstein
coefficients of spontaneous de-excitation between levels $i$ and $j$:
$Aij$, the degeneracy of the upper level $i$, $g_i$, the energy of the
upper level $i$, $E_i$, and the partition function $Q(T)$ measuring the
number of occupied states at temperature $T$. These values are available
from online databases, mainly the CDMS \citep{Muller.2005a} and the JPL
Spec databases \citep{Pickett.1998}. The online databases often only
give values of the partition functions down to a temperature of 9.75~\K\
but the measured excitation temperatures in dark clouds are often lower
than that. Using the energy levels $E_i$ and the corresponding level
degeneracies $g_i$, we compute when necessary the values of the
partition function at 2.375 and 5~\K\ using the following formula
\begin{equation}
	\label{eq.partfunc} Q(T) = \sum_i g_i
	\exp^{-\frac{E_i}{kT}}.
\end{equation}

We systematically check that the value obtained by this sum for 9.75~\K\
is similar to the tabulated value of the databases.

Between these tabulated values, the logarithm of the partition function
is interpolated linearly in temperature. The last two columns of
Table~\ref{tab:ab} give the source of the spectroscopic parameters and
the lowest temperature for which the partition function values are
tabulated in the databases.

\section{Methods} 
\label{sec:ab_methods}

\subsection{Local Thermal Equilibrium}
\label{sec:local_thermal_equilibrium}

We determine the column densities and abundances in the local
thermodynamical equilibrium (LTE) hypothesis. For a given molecule, we
compute the integrated intensities for each line, using the hypothesis
of gaussian opacity profiles as a function of frequency. The modeled
opacity is a function of the species column density, excitation
temperature, and line width. The integrated intensity is derived from
the opacity assuming a temperature of the cosmic microwave background of
2.73~\K.

Some lines detected in \citet{Kaifu.2004} are attributed to several
blended transitions of a given molecule. In this case, we sum the
computed intensity for the identified transitions. Following
\citet{Ohishi.1998}, the column density of a species $X$, noted $[X]$
are converted into abundance assuming a constant molecular hydrogen
column density of $\NHt=\dix{22}\pscm$:
\begin{equation}
	[X] = N(\chem{X})/\NHt 
\end{equation}

\subsection{Bayesian approach: outlier detection}
\label{sec:outlier_detection}

We devise a Bayesian approach which enables to retrieve the LTE model
parameter distributions. The advantages of a Bayesian approach over a
more traditional optimization approach are threefold.
\begin{enumerate}
	
	\item The full distribution of parameters can be recovered. This means
	that upper abundance limits in the case of marginal detections and lower
	abundance limits in the case of highly optically thick lines are readily
	obtained.
	
	\item Prior knowledge can be included for some parameters. For example,
	with a single detected line, the abundance can only be constrained by
	assuming a prior value of the excitation temperature. A Bayesian
	approach enables us to include a prior \emph{distribution} of excitation
	temperatures. The derived abundance then takes into account the prior
	uncertainty on the knowledge of the excitation temperature.
	
	\item Additional parameters can be added to the model, for example, to
	identify data points that are outlier to the physical model. These
	additional parameters are marginalized out in the end to obtain a
	posterior distribution of the parameters of interest; in our case, the
	molecular abundances.
\end{enumerate}

For a few of the molecules, some of the integrated intensities were
clearly outliers and could not be modeled consistently with the rest of
the data set.
This can arise when a line has either: (1) erroneous spectroscopic
parameters, (2) been misattributed to the modeled species, or (3) when a
line from another species is blended with one from the modeled species.
The method we have developed automatically takes into account the
probability of each data point of being an outlier. This enables us to
recover the parameters for the modeled species such as the abundance and
excitation temperature in a robust way. Although the method is applied
to radiative transfer modeling of molecular lines, it is quite general
and can be readily applied to a wide range of problems where outlier
detection is needed.

The model parameters are divided into two groups: the first concerning
the LTE modeling, and the second concerning the modeling of outliers.
The LTE model parameters are the the column density $N$, the excitation
temperature $T_{\emr{ex}}$, the line width $\dV$, and a scatter term
$\sigma$ that enables us to describe both the possible underestimation
of the observational uncertainty and the fact that the LTE model is
simpler than would be needed to perfectly describe the observations. The
outlier model parameters are the two parameters (mean integrated
intensity $Y_o$ and dispersion around this mean $\sigma_o$) describing
the gaussian distribution from which the outliers are supposed to be
originating.
Additionally, each data point is attributed a binary parameter $q_i$
which can take a value of either $0$ if the point is an outlier or $1$
if the point is not. There are thus as many $q_i$ as there are data
points. A final parameter is $P_o$ the probability that a point is an
outlier considering the binomial distribution of the $q_i$ parameters.
It possible to analytically marginalize the $q_i$ parameters while at
the same time recovering the posterior probability of a point being an
outlier. The full model thus has 10 free parameters and does not become
more complicated to solve as the number of data points increase.

The mathematical details of this approach, in particular the expression
of the likelihood used, are presented in Appendix~\ref{app.Bayes}.

When only a single line is detected for a given species, the
identification of outliers makes no sense. In this case, the number of
parameters is limited to four: $N$, $T_{\emr{ex}}$ , $\dV$, and $\log
\sigma$.
Since with only one line the additional scatter parameter is completely
unconstrained, an informative prior consisting of a gaussian function of
mean $-1.2$ and standard deviation $0.4$ was used. This is the median
and sample standard deviation of the values of $\log \sigma$ determined
from the 26 species with more than 2 detected lines.

The Bayesian approach requires us to choose prior distributions for all
model parameters. Some of them can be rather uninformative. For example
the prior distribution of the logarithm of the column density is chosen
to be uniform between 9 and 17 (with the column density in units of
\pscm). Similarly, the prior probability of the outliers $P_o$ is chosen
as uniform between 0 and 1. Conversely, we use informative priors on two
specific parameters.
Following previous studies \citep{Ohishi.1998} that found excitation
temperatures around 5~\K\ in \TMC, we use a gaussian prior on $\Tex$ of
mean 5~\K\ and standard deviation 2~\K\ along with a lower temperature
limit of 2.73~\K\ corresponding to the background temperature.
Similarly, we choose a gaussian distribution of mean 0.5~\kms\ and
standard deviation 0.2~\kms\ for the prior distribution of line widths.
Table~\ref{tab:prior} summarizes these choices.

\TabPrior{}

\section{Results} 

\label{sec:results} 

Table~\ref{tab:ab} summarizes the derived abundances for observed
molecules toward \TMC\ by means of the Bayesian approach. In this table,
column (1) is the species name, column (2) is the median value of the
abundance along with the associated 68\% ($1\sigma$) confidence
interval, column (3) is similar to column (2) for the excitation
temperature, column (4) is the peak signal-to-noise ratio, the ratio
between the intensity of the brightest line of each molecule to the
noise computed for the fit of this molecule, column (5) is the range of
opacities spanned by the lines of a given species, column (6) is the
source of the spectroscopic data, and column (7) is the minimum
temperature for which the partition functions are tabulated in the
spectroscopic databases; below these temperatures, the partition
functions are extrapolated according to equation~\ref{eq.partfunc}.
Individual fits to the observed data are shown in
Appendix~\ref{app.fits}

\subsection{Main isotopes} 

\label{sub:main_isotopes}

Figure~\ref{fig:plot_ab} gives a graphical representation of the
computed abundances for the main isotope-bearing molecules.

\FigPlotAb{}  
\begin{table*}
	\caption{ 
	\label{tab:ab}Best-fit
	Parameters by the Bayesian Approach for Main Isotope-Bearing Species}
	{\scriptsize 
	\begin{center}
		\begin{tabular}
			{lrrrrrrr} \hline Species &
			$\log N^{a}$ & \Tex & PSNR & $\tau$ range & Number of & Min Partition &
			Appendix \\
			& ($cm^{-2}$) & (K) & & & Transitions & Function Temp &
			Figure \\
			\hline \chem{NH_3} & \datapoint{14.70}{2.33}{0.61} & \datapoint{6.16}{1.92}{1.76} & $26.8$ & $0.00-0.73$ &   6 & 9.375\dag & \ref{NH3} \\ 
\chem{CH_3OH} & \datapoint{13.16}{1.79}{0.25} & \datapoint{5.22}{1.43}{1.95} & $2.0$ & $0.00-0.31$ &   1 & 9.375\dag & \ref{CH3OH} \\ 
\mbox{c-}\chem{C_3H} & \datapoint{13.48}{0.05}{0.07} & \datapoint{5.73}{1.29}{1.58} & $12.7$ & $0.02-0.12$ &   6 & 9.375\dag & \ref{c-C3H} \\ 
\mbox{l-}\chem{C_3H} & \datapoint{12.75}{0.03}{0.07} & \datapoint{5.90}{1.40}{1.56} & $14.5$ & $0.03-0.14$ &   6 & 9.375 & \ref{C3H} \\ 
\mbox{l-}\chem{C_3H_2} & \datapoint{11.77}{1.90}{0.50} & \datapoint{6.02}{1.97}{1.77} & $1.6$ & $0.01-0.04$ &   4 & 9.375 & \ref{H2C3} \\ 
\mbox{c-}\chem{C_3H_2} & \datapoint{13.27}{2.90}{1.13} & \datapoint{5.45}{1.53}{1.86} & $6.6$ & $0.07-0.59$ &   4 & 2.725 & \ref{c-C3H2} \\ 
\chem{CH_3CCH} & \datapoint{14.06}{1.26}{1.16} & \datapoint{5.47}{1.89}{2.03} & $2.6$ & $0.08-0.39$ &   3 & 9.375 & \ref{CH3CCH} \\ 
\chem{C_2O} & $<12.57$ & \datapoint{5.50}{1.57}{1.71} & $0.4$ & $<0.02$ &   1 & 9.375\dag & \ref{C2O} \\ 
\chem{H_2CCN} & \datapoint{13.58}{0.34}{0.24} & \datapoint{3.46}{0.35}{1.66} & $5.3$ & $0.01-1.74$ &  38 & 9.375 & \ref{CH2CN} \\ 
\chem{CH_3CN} & \datapoint{12.61}{0.18}{0.19} & \datapoint{5.04}{1.27}{1.85} & $3.2$ & $0.04-0.20$ &   5 & 2.725 & \ref{CH3CN} \\ 
\chem{H_2CCO} & \datapoint{12.68}{1.71}{0.35} & \datapoint{9.50}{4.68}{3.46} & $1.6$ & $0.01-0.01$ &   3 & 2.725 & \ref{H2CCO} \\ 
\chem{HNCO} & \datapoint{13.03}{0.05}{0.10} & \datapoint{5.26}{1.09}{1.58} & $10.5$ & $0.03-0.24$ &   5 & 9.375\dag & \ref{HNCO} \\ 
\chem{CS} & $>13.46$ & \datapoint{4.45}{0.52}{1.60} & $24.2$ & $>3.96$ &   1  & 2.725 & \ref{CS} \\ 
\chem{CH_3CHO} & \datapoint{12.43}{2.03}{0.31} & \datapoint{5.24}{1.49}{1.80} & $1.7$ & $0.04-0.04$ &   2 & 9.375\dag & \ref{CH3CHO} \\ 
\chem{HCS^+} & \datapoint{12.76}{0.38}{0.53} & \datapoint{5.33}{1.74}{1.62} & $5.5$ & $0.10-3.37$ &   1 & 9.375 & \ref{HCSp} \\ 
\chem{H_2CS} & \datapoint{13.62}{0.13}{0.53} & \datapoint{5.16}{1.55}{1.55} & $7.7$ & $0.26-6.74$ &   1 & 2.725 & \ref{H2CS} \\ 
\chem{SO} & $>13.67$ & \datapoint{4.44}{1.21}{2.06} & $11.8$ & $>0.50$ &   1  & 2.725 & \ref{SO} \\ 
\chem{C_4H} & $[13.43,14.94]^{b}$ & \datapoint{4.79}{0.39}{0.62} & $26.4$ & $0.03-2.78$ &  27  & 9.375 & \ref{C4H} \\ 
\mbox{o-}\chem{C_4H_2} & \datapoint{13.28}{0.30}{0.23} & \datapoint{5.04}{0.77}{1.52} & $6.5$ & $0.08-0.25$ &   8 & 9.375 & \ref{o-H2C4} \\ 
\mbox{p-}\chem{C_4H_2} & \datapoint{12.42}{0.49}{0.45} & \datapoint{4.63}{1.18}{1.78} & $3.6$ & $0.10-0.22$ &   4 & 9.375 & \ref{p-H2C4} \\ 
\chem{C_3N} & \datapoint{13.55}{0.30}{0.21} & \datapoint{3.31}{0.22}{0.64} & $3.6$ & $0.13-0.94$ &  12 & 9.375 & \ref{C3N} \\ 
\chem{HC_3N} & \datapoint{14.37}{0.06}{0.13} & \datapoint{5.22}{0.58}{0.80} & $33.0$ & $0.04-19.50$ &  20 & 2.725 & \ref{HC3N} \\ 
\chem{HNC_3} & \datapoint{11.68}{0.18}{0.19} & \datapoint{5.69}{1.67}{1.97} & $3.5$ & $0.05-0.12$ &   4 & 9.375\dag & \ref{HNCCC} \\ 
\chem{HCCNC} & \datapoint{12.93}{0.11}{0.31} & \datapoint{5.41}{1.45}{1.80} & $5.0$ & $0.04-0.27$ &   5 & 9.375\dag & \ref{HCCNC} \\ 
\chem{C_3O} & \datapoint{11.92}{1.29}{0.25} & \datapoint{5.03}{1.23}{1.77} & $1.7$ & $0.04-0.04$ &   2 & 9.375 & \ref{C3O} \\ 
\chem{HC_3NH^+} & \datapoint{11.87}{1.81}{0.45} & \datapoint{5.55}{1.49}{1.85} & $1.4$ & $0.01-0.01$ &   2 & 9.375 & \ref{HC3NHp} \\ 
\chem{CH_2CHCN} & \datapoint{12.81}{0.08}{0.22} & \datapoint{5.00}{1.36}{2.07} & $3.9$ & $0.04-0.21$ &  12 & 9.375 & \ref{CH2CHCN} \\ 
\chem{HCCCHO} & \datapoint{11.26}{1.54}{0.95} & \datapoint{5.29}{1.42}{1.39} & $0.7$ & $0.00-0.08$ &   1 & 9.375 & \ref{HCCCHO} \\ 
\chem{C_2S} & \datapoint{14.01}{0.16}{0.14} & \datapoint{4.19}{0.34}{0.60} & $9.9$ & $0.28-5.08$ &   8 & 9.375 & \ref{C2S} \\ 
\chem{OCS} & $<13.26$ & \datapoint{5.27}{1.39}{1.88} & $0.7$ & $<0.06$ &   1 & 9.375 & \ref{OCS} \\ 
\chem{C_5H} & \datapoint{12.27}{0.10}{0.12} & \datapoint{4.80}{0.72}{0.94} & $4.7$ & $0.03-0.05$ &  16 & 9.375 & \ref{C5H} \\ 
\chem{CH_3C_4H} & \datapoint{13.17}{0.17}{0.23} & \datapoint{6.85}{2.01}{1.68} & $2.7$ & $0.01-0.04$ &   8 & 2.725 & \ref{CH3C4H} \\ 
\chem{CH_3C_3N} & \datapoint{11.99}{0.17}{0.18} & \datapoint{6.26}{1.71}{1.63} & $2.8$ & $0.01-0.05$ &   8 & 2.725 & \ref{CH3C3N} \\ 
\chem{C_3S} & \datapoint{13.14}{0.11}{0.16} & \datapoint{5.19}{0.70}{0.90} & $7.4$ & $0.22-0.83$ &   7 & 2.725 & \ref{C3S} \\ 
\chem{C_6H} & \datapoint{12.74}{0.05}{0.04} & \datapoint{5.95}{0.45}{0.35} & $6.1$ & $0.04-0.09$ &  42 & 9.375 & \ref{C6H} \\ 
\chem{HC_5N} & \datapoint{13.77}{0.09}{0.10} & \datapoint{6.39}{0.51}{0.54} & $9.5$ & $0.20-2.33$ &  16 & 9.375 & \ref{HC5N} \\ 
\chem{HC_7N} & \datapoint{13.66}{0.10}{0.14} & \datapoint{6.14}{0.41}{0.32} & $14.3$ & $0.04-3.52$ &  34 & 9.375 & \ref{HC7N} \\ 
\chem{HC_9N} & \datapoint{13.02}{0.06}{0.06} & \datapoint{8.51}{1.15}{1.07} & $4.6$ & $0.00-0.03$ &  16 & 9.375 & \ref{HC9N} \\ 
 \hline 
		\end{tabular}
	\end{center}
	} Spectroscopic data from the CDMS database except for
	species marked by a $\dag$ which are from form the JPL database.
	(a):
	The abundance $[X]$ can be derived using $N(X)/N(\Ht)$ with $N(\Ht) =
	10^{22}\pscm$\\
	(b): Values in bracket are the 1$\sigma$ confidence
	interval. See the text for details concerning these specific
	molecules.\\
\end{table*}
{}

Three main classes of outcomes can be identified. For some molecules,
all observed lines have high opacities, and in these cases only lower
limits can be placed on their abundances. This is the case for \chem{CS}
and \chem{SO}.
For other molecules, the signal-to-noise ratio is not sufficient enough
to obtain more than an upper limit on the abundances. This is the case
for \chem{C_2O} and \chem{OCS}. For all other molecules, it is possible
to determine an abundance along with an associated uncertainty.

\subsection{Isotopologues} 

\label{sub:isotopologues}

Twenty isotopologues are also detected in the Nobeyama survey of \TMC.
We have computed their abundances relative to \Ht\ and the ratio of the
minor isotopologues to the main isotopologues for each molecule. The
results are summarized in Fig.~\ref{fig:plot_ab_iso} and
Table~\ref{tab:ab_iso}.

\FigPlotAbIso{}  
\begin{table*}
	\caption{ 
	\label{tab:ab_iso}Best-fit Parameters by the Bayesian Approach
	for Rarer Isotope-Bearing Species} {\scriptsize 
	\begin{center}
		\begin{tabular}
			{lrrrrrrr} \hline Species & $\log N^{a}$ & \Tex & PSNR &
			$\tau$ range & Number of & Min Partition & Appendix \\
			& ($cm^{-2}$) &
			(K) & & & Transitions & Function Temp & Figure \\
			\hline
			\chem{^{13}CCH_2} & $<13.01$ & \datapoint{5.31}{1.49}{1.69} & $0.9$ & $<0.04$ &   2 & 9.375 & \ref{C13CCH2} \\ 
\mbox{c-}\chem{C_3HD} & \datapoint{12.71}{2.03}{2.19} & \datapoint{5.23}{1.91}{2.13} & $2.0$ & $0.11-0.28$ &   4 & 9.375 & \ref{C3HD} \\ 
\chem{^{13}CS} & \datapoint{12.26}{1.78}{0.27} & \datapoint{5.30}{1.66}{1.99} & $2.3$ & $0.00-0.51$ &   1 & 9.375 & \ref{thCS} \\ 
\chem{C^{33}S} & $<12.42$ & \datapoint{5.22}{1.40}{1.52} & $1.2$ & $<0.02$ &   1 & 9.375 & \ref{C33S} \\ 
\chem{C^{34}S} & \datapoint{12.70}{0.14}{1.55} & \datapoint{4.94}{1.61}{1.91} & $6.9$ & $0.40-546.68$ &   1 & 9.375 & \ref{C34S} \\ 
\chem{HDCS} & $<13.80$ & \datapoint{5.21}{1.57}{2.00} & $1.5$ & $<0.41$ &   1 & 9.375 & \ref{HDCS} \\ 
\chem{^{34}SO} & $<13.04$ & \datapoint{5.35}{1.55}{1.93} & $1.1$ & $<0.08$ &   1 & 9.375 & \ref{34SO} \\ 
\chem{DC_3N} & \datapoint{12.47}{0.24}{0.44} & \datapoint{5.53}{1.66}{1.90} & $4.3$ & $0.01-0.28$ &   5 & 9.375 & \ref{DC3N} \\ 
\chem{H^{13}CC_2N} & \datapoint{12.25}{0.03}{0.03} & \datapoint{6.54}{1.07}{1.32} & $14.6$ & $0.01-0.15$ &   6 & 9.375 & \ref{H13CCCN} \\ 
\chem{HC^{13}CCN} & \datapoint{12.26}{0.02}{0.02} & \datapoint{5.38}{0.64}{0.57} & $20.6$ & $0.02-0.20$ &   6 & 9.375 & \ref{HC13CCN} \\ 
\chem{HC_2^{13}CN} & \datapoint{12.47}{0.03}{0.03} & \datapoint{5.90}{0.93}{1.29} & $11.5$ & $0.03-0.21$ &   6 & 9.375 & \ref{HCC13CN} \\ 
\chem{DC_5N} & \datapoint{12.05}{0.13}{0.16} & \datapoint{5.80}{0.85}{0.96} & $6.5$ & $0.02-0.04$ &   6 & 9.375 & \ref{DC5N} \\ 
\chem{H^{13}CC_4N} & \datapoint{12.03}{0.23}{0.34} & \datapoint{6.31}{1.55}{1.66} & $3.8$ & $0.02-0.03$ &   3 & 9.375 & \ref{H13CCCCCN} \\ 
\chem{HC^{13}CC_3N} & \datapoint{12.06}{0.20}{0.25} & \datapoint{4.03}{0.58}{0.99} & $6.9$ & $0.03-0.08$ &   4 & 9.375 & \ref{HC13CCCCN} \\ 
\chem{HC_2^{13}CC_2N} & \datapoint{11.92}{0.11}{0.15} & \datapoint{5.49}{1.05}{1.09} & $4.8$ & $0.02-0.04$ &   6 & 9.375 & \ref{HCC13CCCN} \\ 
\chem{HC_3^{13}CCN} & \datapoint{12.25}{0.12}{0.14} & \datapoint{4.38}{0.45}{0.49} & $6.7$ & $0.03-0.12$ &   6 & 9.375 & \ref{HCCC13CCN} \\ 
\chem{HC_4^{13}CN} & \datapoint{12.01}{0.36}{0.23} & \datapoint{5.61}{1.33}{1.56} & $2.7$ & $0.03-0.04$ &   5 & 9.375 & \ref{HCCCC13CN} \\ 
 \hline 
		\end{tabular}
	\end{center}
	} (a): the
	abundance [X] can be derived using $N(X)/N(\Ht)$ with $N(\Ht) =
	10^{22}\pscm$\\
	Spectroscopic data from the CDMS database.
\end{table*}
{}

\section{Discussion} 

\label{sec:ab_discussion}

\subsection{Outlier lines} 
\label{sec:outlier_lines}

Using the Bayesian approach we have demonstrated that it is possible to
identify outliers in the data set: observed line intensities that are
not explainable by the global model best fitting the rest of the lines
for a given species.
Table~\ref{tab.outliers} summarizes the properties of the outlier lines.

\begin{figure*}
\includegraphics[width=16cm]{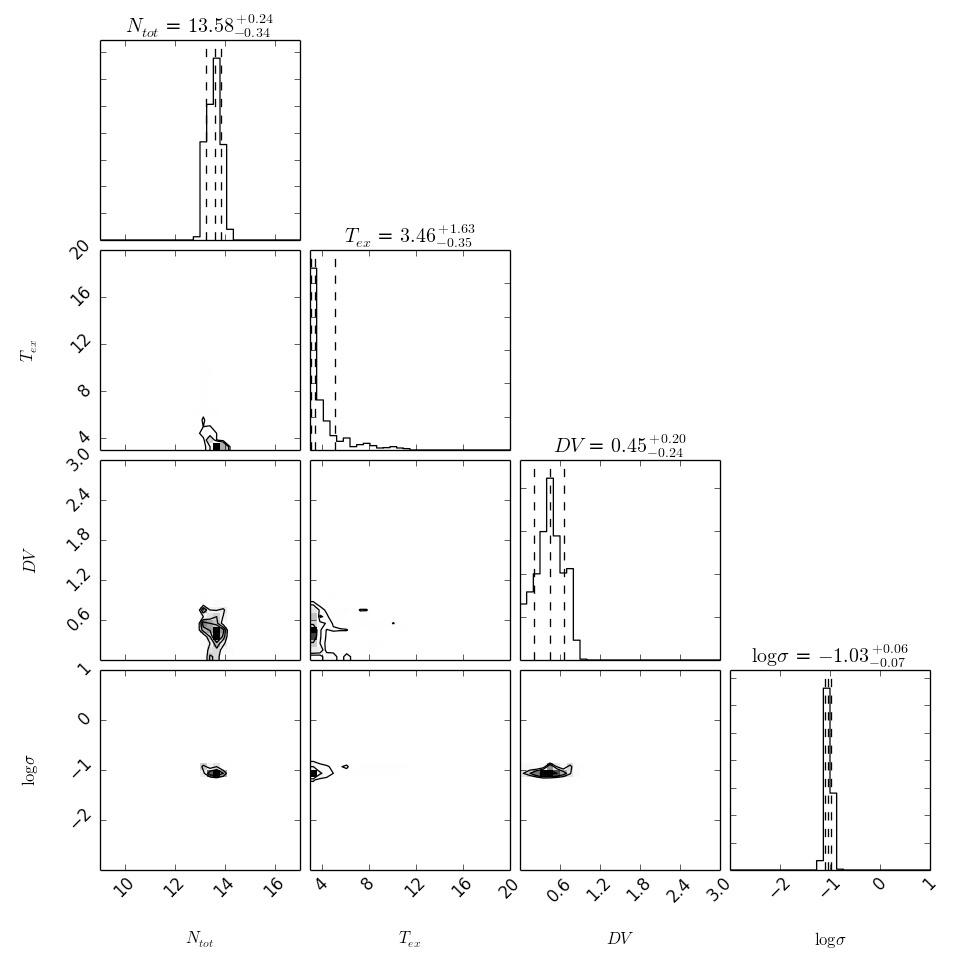}
\includegraphics[width=8cm]{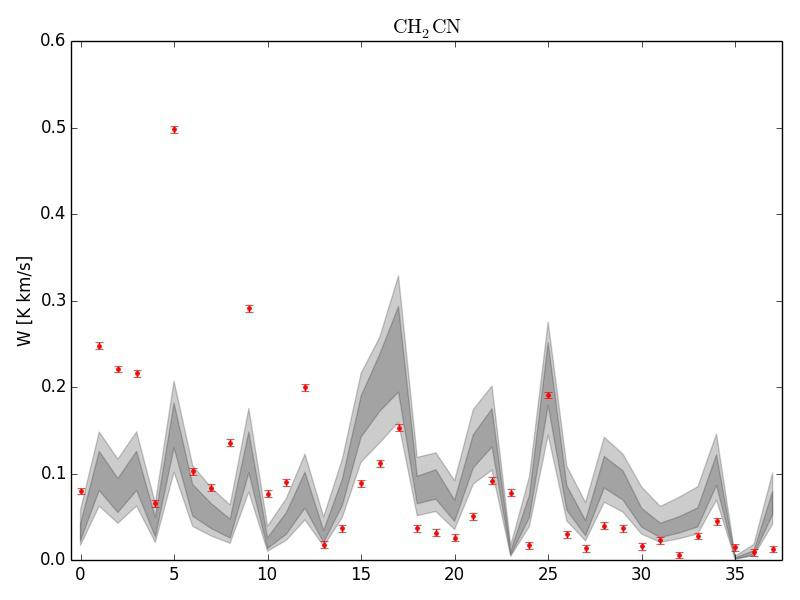}
\includegraphics[width=8cm]{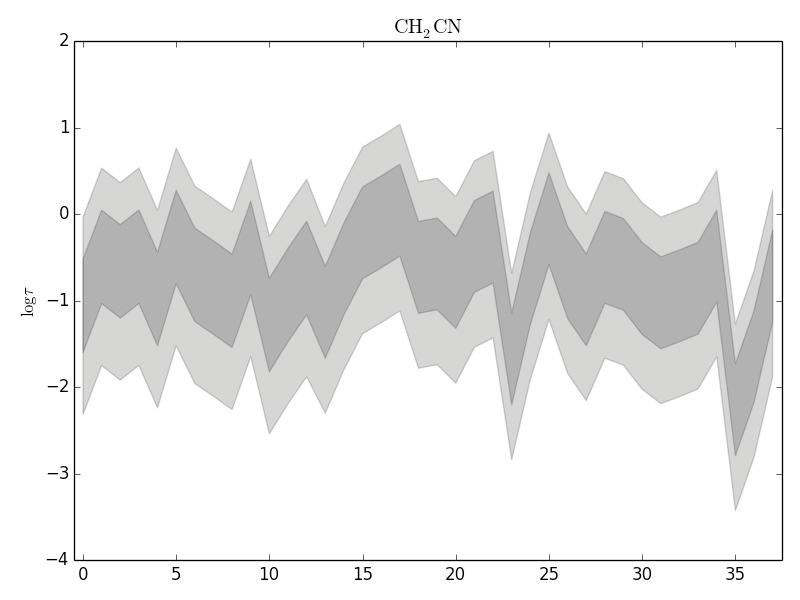}
\caption{\chem{H_2CCN}}
\label{CH2CN}
\end{figure*}

\begin{figure*}
\includegraphics[width=16cm]{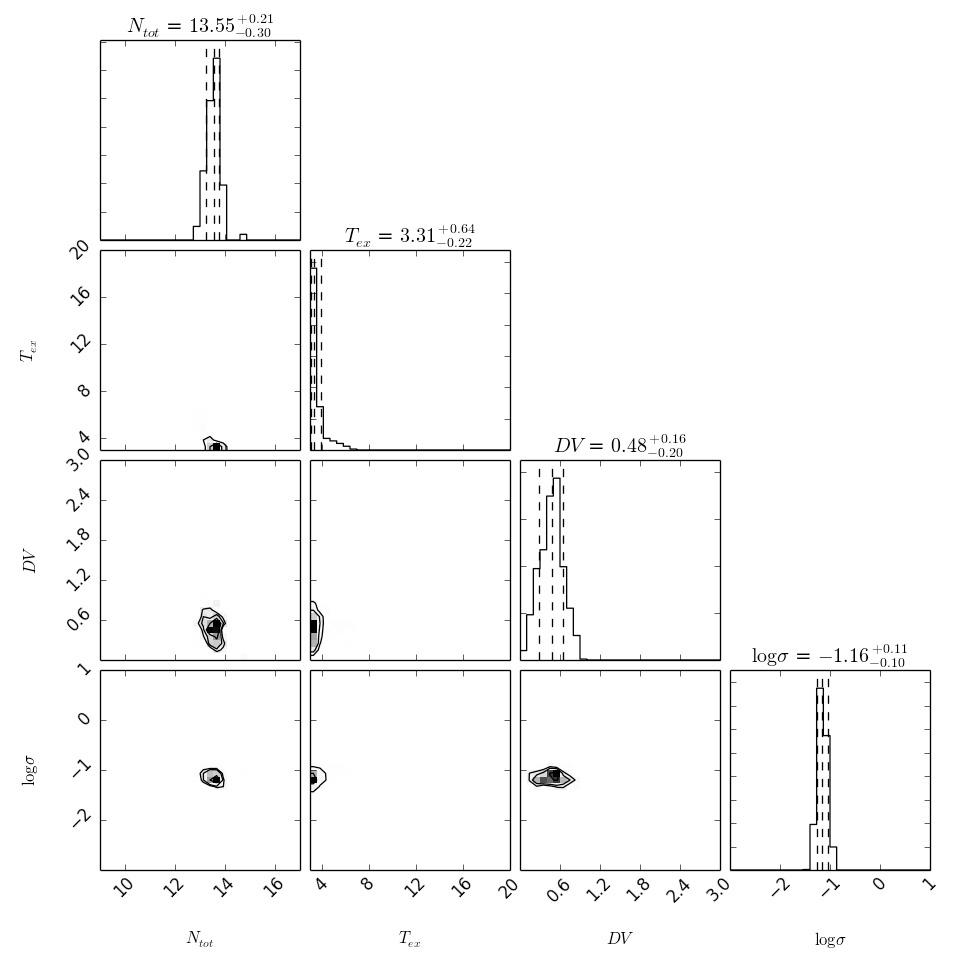}
\includegraphics[width=8cm]{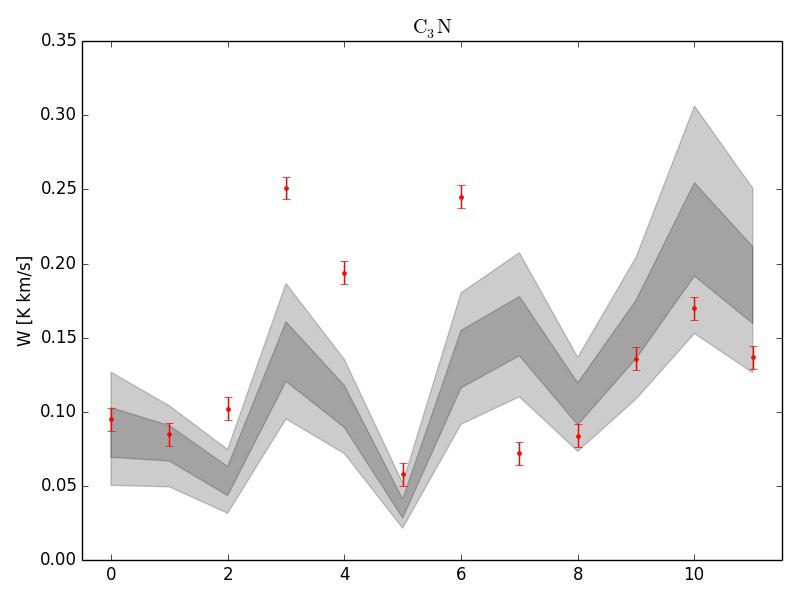}
\includegraphics[width=8cm]{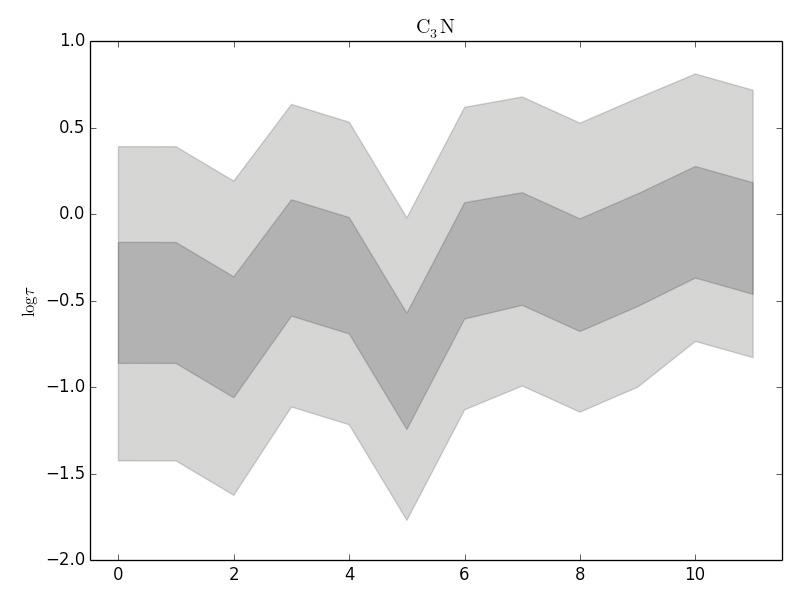}
\caption{\chem{C_3N}}
\label{C3N}
\end{figure*}

Two molecules stand out in the outlier detection scheme, \chem{H_2CCN}
and \chem{C_3N}. In both cases, about half of the observed intensities
are not compatible with the best-fit model (see
Figures~\ref{CH2CN} and \ref{C3N} respectively). With so many outlier
lines, the problem cannot arise solely from observational effects
(possible blended or misattributed lines). The origin of the issue is
possibly the values of the spectroscopic catalogs such as erroneous
normalized intensities or level degeneracies (the spectroscopy data for
\chem{H_2CCN} is only available from CDMS, while the same dataset is
present and identical both in CDMS and JPL catalogs for \chem{C_3N}).

In these two cases, the outlier detection scheme is not used. The
simpler model with just the three LTE parameters ($N$, $T_\emr{ex}$ and
$\Delta V$) and a scatter term is used as described in
Sect.~\ref{sec:outlier_detection} for the case when only a single line
of a molecule is detected. The sampling method adjusts by increasing the
scatter error added in quadrature to the observational uncertainty to
encompass all observed points at the expense of a larger uncertainty on
the other parameters, in particular the column density and thus the
abundance.

\TabOutliers{}

\subsection{Comparison with previous abundance determination} 

\label{sec:abundance_comparison_with_previous_values}

The derived column densities and abundances for these molecules were
published in \citet{Ohishi.1992} and \citet{Ohishi.1998} which have been
used since as references for abundances in \TMC\ and by extension for
dark clouds in general. In \citet{Ohishi.1998}, the derived abundances
have no associated uncertainty estimates and the authors, while
discussing the advantage of having detected lines for some species
spanning a wide range of opacities, do not discuss the possible effect
of high opacity or low signal-to-noise detection on the molecular
abundances, especially for species where the number of detected lines is
low (some species have only one detected line).

\citet{Ohishi.1998} have used the data later published in
\citet{Kaifu.2004} to compute the abundances of observed species. Since
the local thermal equilibrium approach is used in both our study and
theirs, the main difference should arise from the spectrometric data
used and from the fact that our Bayesian approach enables us to compute
uncertainties on the abundances and even upper or lower limits when the
signal to noise ratio is too low or the opacity is too large for all
detected lines of a given species. In Figure~\ref{fig:plot_ab}, the
values from \citet{Ohishi.1998} are shown as large white circles.
\citet{Agundez.2013} made a review of the literature to identify
molecular abundances in \TMC. The results they compiled appear as small
circles in Fig.~\ref{fig:plot_ab}. Most of the previously determined
abundances fall within the $95\%$ ($2\sigma$) confidence interval of our
Bayesian method. We now discuss notable differences, \ie{} species where
the old values fall outside our 95\% confidence interval.
Most of the abundance determination in \citet{Ohishi.1998} were carried
out without taking into account the hyperfine splitting of the
observations. This can lead to biased results when opacities vary across
the hyperfine components.

\subsubsection{Polyynes} 
\label{ssub:cnh} 

\chem{C_6H} has a similar abundance as the one derived by
\citet{Ohishi.1998}. \chem{C_5H} has a new abundance that is lower by
about an order of magnitude from previous values. The value we compute
for linear \chem{C_3H} falls between the values of \citet{Ohishi.1998}
and \citet{Agundez.2013}. The value reported for this species in
\citet{Agundez.2013} comes from IRAM 30m millimeter observations
\citep{Fosse.2001}. The difference of a factor of two in beam size
between the observations could explain the abundance difference. Cyclic
\chem{C_3H} is found to be about three times more abundant.

The case of C4H is particular because of the electronic
structure of this radical. Previous studies \citep{Woon.1995,
Mazzotti.2011} have established that \chem{C_4H} presents two low-lying
electronic states of ${^2}\Sigma$$^{+}$ and ${^2}\Pi$ symmetries. The
experimental spectra of \chem{C_4H} are consistent with the ground state
${^2}\Sigma$$^{+}$ \citep{Dismuke.1975,Gottlieb.1983,Shen.1990}.
Recently, \citet{Senent.2010} confirmed that \chem{C_4H} radical
presents two electronic states ${^2}\Sigma$$^{+}$ and ${^2}\Pi$ which
are near degenerate. Their computation confirms that the ground
electronic state of \chem{C_4H} radical is ${^2}\Sigma$$^{+}$, which
agrees with the findings of electron paramagnetic resonance (EPR) by
\citet{Dismuke.1975}. Their computation also confirms the short gap
between the two low-lying electronic states. They have found that they
are around 9 cm$^{-1}$ and 2800 cm$^{-1}$ at the (R)CCSD(T) and CASSCF
levels of quantum chemical theory. The dipole moment is specific to each
electronic state but since the two electronic states are very close to
each other and the molecule is not static, the motion will cause
electronic states to mix. However, since the concept of an average
dipole moment for a mixed state is not well defined, this issue needs
more studies from both the theoretical and experimental points of view.

The CDMS database gives a dipole value of 0.870~Debye, which corresponds
to the electronic ground state of the molecule. The computed dipole
value for the ${^2}\Pi$ electronic state is 4.3\D\ \citep{Senent.2010}.
Nevertheless, in Tab.~\ref{tab:ab} and Figure~\ref{fig:plot_ab} we show
the full range of abundances permitted by any value of the dipole moment
between the values computed for the $^2\Sigma^+$ and $^2\Pi$ states.

In the optically thin regime, the column density scales as the square of
the dipole. Thus, once the correct dipole moment of \chem{C_4H} has been
determined, an updated value of the abundance can be approximated using
the following expression:
\begin{equation}
	N(\chem{C_4H})_D = \left(\frac{D}{4.3}\right)^2
	N(\chem{C_4H})_{4.3} 
\end{equation}

where $ N(\chem{C_4H})_D$ is the column density considering a dipole
moment of value $D$ and $N(\chem{C_4H})_{4.3} = \sciexp{3}{13} \pscm$
the column density considering a dipole moment of value 4.3 Debye.

Of the detected molecules which posses distinct ortho and para states,
only \chem{C_4H_2} has a number of detected lines (12) that is large
enough that the abundances of the ortho and para forms can be determined
independently.
The median ortho-to-para abundance ratio is found to be around 6, but
this ratio is highly uncertain with a one sigma confidence interval
ranging from 1.7 to 70. The two ortho and para abundances bracket the
abundance previously determined by \citet{Ohishi.1998}.

\subsubsection{Cyanopolyynes} 

\label{ssub:hcnh} 

Cyanopolyynes are abundant in \TMC\ with the largest cyanopolyyne
present in the interstellar medium, \chem{HC_{11}N} first detected there
by \citet{Bell.1997}. Our new results are comparable with previous
abundance determinations in \citet{Ohishi.1998} except for \chem{HC_7N}
and \chem{HC_9N} which we find to be about three times more abundant
than was determined by \citet{Ohishi.1998}. \chem{C_3N} is found to be
six times more abundant but our derived abundance remains very uncertain
(see Sect.~\ref{sec:outlier_lines})

\subsubsection{Other molecules}

\chem{HNCO} is more abundant by a factor of 3.

\subsection{Isotopic ratios} 

\label{sub:isotopic_ratios}

\subsubsection{Sulfur} 

\label{ssub:sulfur}

For the two minor isotopes of sulfur \chem{^{33}S} and \chem{^{34}S},
only upper limits to the ratio can be computed. This arises because, for
SO and CS, the main isotopologues have only lower limits to their
abundances because of the high opacity of their lines. In the case of
\chem{C_3S}, it is because only an upper limit on the abundance of the
\chem{^{34}S} isotopologue itself can be computed because of the low
signal-to-noise ratio of the observed line. The derived upper limits are
in agreement with the cosmic abundance ratios of $\chem{^{34}S/^{32}S} =
\sciexp{4.4}{-2}$, $\chem{^{33}S/^{32}S} = \sciexp{7.9}{-3}$
\citep{Berglund.2011}.

\subsubsection{Carbon} 

\label{ssub:carbon}

The overall median value of \chem{^{12}C}/\chem{^{13}C} considering all
the $^{13}C$ bearing molecules, is 66. For the three $^{13}C$
isotopologues of \chem{HC_3N}, the median ratio is 130 with values
raging from 80 to 133.
For the five $^{13}C$ isotopologues of \chem{HC_5N}, the median ratio is
54 with values ranging from 32 to 75. Recently, \citet{Taniguchi.2016}
have observed $^{13}C$ isotopes of \chem{HC_5N} towards \TMC. Our
computed column densities tend to be somewhat larger than the ones they
derive although within the $1\sigma$ interval we estimate.

The upper limits on the carbon isotopic ratios for the two molecules
\chem{^{13}CS} and \chem{^{13}CCH_2} are compatible with the typical
value of 68 assumed in the ISM \citep{Milam.2005}.

\subsubsection{Hydrogen} 

\label{ssub:hydrogen}

In the observed sample, only three molecules have measured H/D ratios:
\chem{DC_3N} (H/D = 81(-68,+82)), \chem{DC_5N} (H/D = 52(-20,+26)) and
c-\chem{C_3HD}, with an additional upper limit on \chem{HDCS}. Excluding
c-\chem{C_3HD} which is very uncertain, the average H/D ratio is 64. Due
to differences in zero level energy, molecules tend to be enriched in
deuterium in cold ISM conditions \citep{Albertsson.2013}. The value we
derive are typical of those found in the cold dense ISM.

\section{Summary} 
\label{sec:conclusions}

In this work, we have developed a statistical Bayesian method to
determine the abundance of molecules in the interstellar medium in the
presence of possible outlier points. This method has been used on
previously published data from a spectral survey of \TMC\ by the
Nobeyama Observatory. We derive a new reference set of abundances for
\TMC\ which will be used to benchmark the chemistry of dark clouds.

\acknowledgments

Based on observations carried out with the Nobeyama Radio Observatory.
The Nobeyama Radio Observatory is an open-use facility for mm-wave
astronomy, and is being operated under the National Astronomical
Observatory of Japan (NAOJ).
P.G., L.M., and V.W. thanks ERC starting grant (3DICE, grant agreement
336474) for funding during this work. P.G.'s current postdoctoral
position is funded by the INSU/CNRS. V.W and J.-C.L. acknowledge the
French PCMI for funding of their research.

{\it Facilities:} \facility{National Astronomical Observatory of Japan
(NAOJ) 45m Radio Telescope at Nobeyama Radio Observatory}.

\appendix

\section{Bayesian method: detecting outliers} 
\label{app.Bayes}

In the Bayesian framework, the posterior $p(\theta|D)$ distribution is
obtained using Bayes' equation by multiplying the prior distribution
$p(\theta)$ by the likelihood $p(D|\theta)$.
\begin{equation}
	p(\theta|D) \propto p(\theta) p(D|\theta)
\end{equation}

Here $\theta$ is a vector of the parameters describing the model and $D$
is the set of observed data points.

We have devised a method that enables the automatic mitigation of
outlier points. This method was inspired by the discussions in
\citet{Hogg.2010} and \citet{Foreman-Mackey.2014} concerning the
detection of outliers. The likelihood is constructed by considering two
generative models, one for the data points that follow the model and the
second one for the outlier points.

The likelihood for the non-outliers points
$p_m(\{y_i\}_{i=1}^N|N,\dV,\Tex,\sigma)$ is constructed assuming that
the observed uncertainties for point $i$ are generated from a gaussian
of standard deviation $\sigma_{yi}$. To explain the additional scatter
around the modeled data, an additional noise term $\sigma$ is added in
quadrature to the observed uncertainties. We make the hypothesis that
the outlier points come from a gaussian distribution of mean $Y_o$ and
dispersion $\sigma_o$. The corresponding likelihood is written as
$p_o(\{y_i\}_{i=1}^N|Y_o,\sigma_o,I)$ in the following equations. Each
data point is associated with a number $q_i$ which can be either 0 if
the point is an outlier and 1 if it is not. We add an additional
parameter $P_o$ representing the overall probability of the data points
of being outliers.

Making the hypothesis of independent observations, the full likelihood
is obtained by multiplying the individual points' likelihood values.
Numerically, we consider the logarithm of the likelihood function.
\begin{equation}
	\label{eq.like} 
	\begin{split}
		\mathcal{L} = {} &
		p(D|\theta)\\
		\mathcal{L} = {} &
		p(\{y_i\}_{i=1}^N|N,\dV,\Tex,\sigma,\{q_i\}_{i=1}^N,Y_o,\sigma_o,I)\\
		\mathcal{L} = {} & \prod_{i=1}^N
		\left[p_m(\{y_i\}_{i=1}^N|N,\dV,\Tex,\sigma)\right]^{q_i} \times
		\left[p_o(\{y_i\}_{i=1}^N|Y_o,\sigma_o,I)\right]^{[1-q_i]} \\
		\mathcal{L} = {} & \prod_{i=1}^N \left[\frac{1}{\sqrt{2 \pi
		\left[\sigma_{yi}^2+\sigma^2\right]}}
		\exp\left(-\frac{\left[y_i-W_i(N,\dV,\Tex)\right]^2}{2\left[\sigma_{yi}^
		2+\sigma^2\right]} \right) \right]^{q_i} \\
		& \times
		\left[\frac{1}{\sqrt{2 \pi
		\left[\sigma_{yi}^2+\sigma^2+\sigma_o\right]}}\exp\left(-\frac{\left[y_i
		-Y_o\right]^2}{2\left[\sigma_{yi}^2+\sigma^2+\sigma_o\right]} \right)
		\right]^{\left[1-q_i\right]} 
	\end{split}
\end{equation}

The total number of parameters is thus $K+7$ where $K$ is the number of
observed lines. $N$,
$\dV$,$\Tex$,$\sigma$,$\{q_i\}_{i=1}^K$,$P_o$,$Y_o$,$\sigma_o$

By specifying a prior for the ${q_i}$ parameters, it is possible to
marginalize them out analytically. The marginalization can be written:
\begin{equation}
	p({y_i}|\theta, {\sigma_{y_i}}) =
	\sum_{{q_i}}\prod_{i=1}^{N} p(q_i)p(y_i|\theta, \sigma_{y_i}, q_i)
\end{equation}

where the sum is over all the possible permutations of the $q_i$ flags.

Assuming the following prior for $q_i$:

\[p(q_i) = \left \{ 
\begin{array}{ll}
	P_o & \mathrm{if}\,q_i=0 \\
	1-P_o
	& \mathrm{if}\,q_i=1 
\end{array}
\right.\]

where $P_o$ is the prior probability that a point is draw from the
outlier distribution. It is then possible to marginalize out the ${q_i}$
and obtain the following likelihood:
\begin{equation}
	\mathcal{L} = \prod_{i=1}^N \left[(1-P_o)
	p_m(\{y_i\}_{i=1}^N|N,\dV,\Tex,\sigma) + P_o
	p_o(\{y_i\}_{i=1}^N|Y_o,\sigma_o,I)\right] \\
\end{equation}

It is also possible to obtain the individual posterior distributions of
each $p(q_i|y)$ which is the probability that a point is either an
outlier or not.
For each data point $i$, this probability is written 
\begin{equation}
	p(q_i|y) = \int p(q_i,\theta|y) d\theta = \int p(q_i|\theta,y)
	p(\theta|y) d\theta 
\end{equation}
where $y$ has been simplified to
signify $\{y_i\}_{i=1}^N$.

Since establishing an analytical formulation of the posterior
distribution is intractable, the posterior distribution is obtained by
numerically sampling the posterior distribution using an MCMC algorithm.
We use the affine invariant ensemble sampling method developed by
\citet{Goodman.2010} in its Python implementation {\tt EMCEE}
\citep{Foreman-Mackey.2013}. This sampling method uses an ensemble of
walkers which sample simultaneously the posterior distribution. The
number of walker has been kept fixed at 100.

For each molecule, the sampling is run during a burnin time of about
10,000 steps at the end of which the convergence of the ensemble of
chains is checked by plotting the running mean of each parameter.

A final run of 100 steps yields 100,000 points from which the histograms
of the marginalized distribution and the two parameter correlations can
be plotted.

\section{Figures} 

\label{app.fits}

The following set of figures synthesize the results of the Bayesian
method for each species. The top triangle plot shows the marginalized
posterior distributions of each model parameters along with the median
and 16th and 84th percentiles. The set of two-dimension histogram in the
lower triangle of the matrix of plot are useful to identify possible
correlations between parameters.

In the bottom left, a comparison to the observations is plotted, shown
with red error bars, and the modeled integrated intensities for each
lines of a given species. The indices in abscissa correspond to the line
identifications from \citet{Kaifu.2004} in order of increasing
frequency.
The modeled intensity distribution is represented by the gray intervals
corresponding to 68\% and 95\% of the samples. The outlier probability
for each point is represented by the ratio of the white to black surface
of each data point. When at least one point has a probability of being
an outlier larger than 50\%, the 1 sigma interval of the gaussian
function that describe the outlier distribution is shown in light blue.

On the bottom right is plotted the distribution of the base 10 logarithm
of opacities for each line of a given species is plotted. Again, the
gray intervals represent 68\% and 95\% of the samples.

\begin{figure*}
\includegraphics[width=16cm]{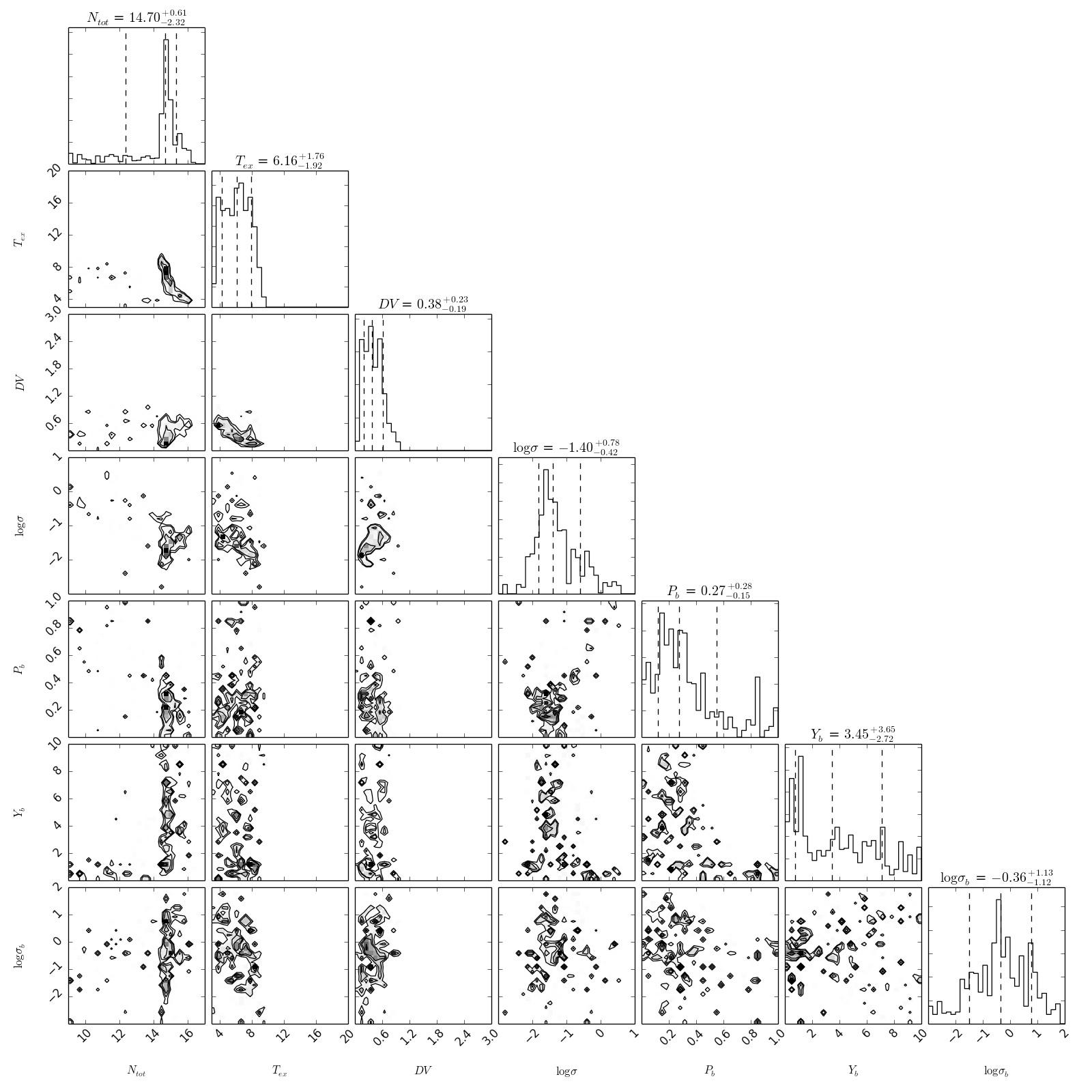}
\includegraphics[width=8cm]{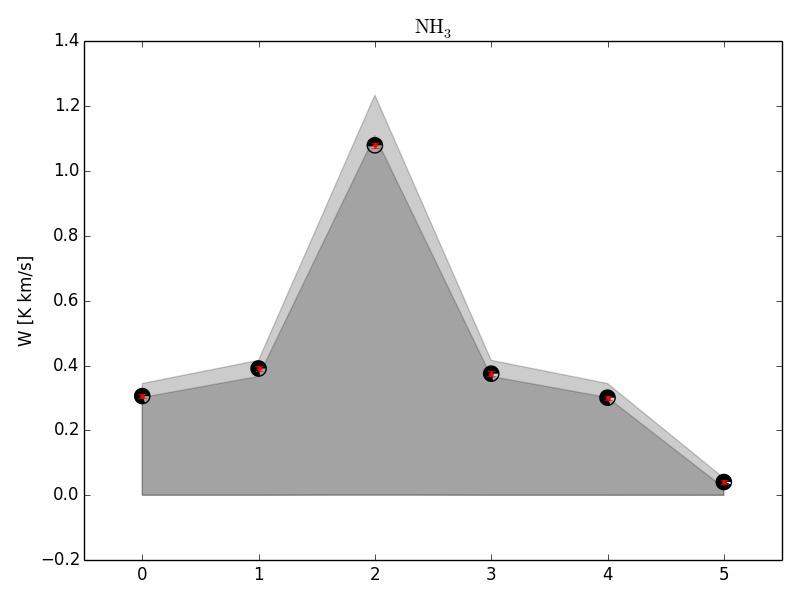}
\includegraphics[width=8cm]{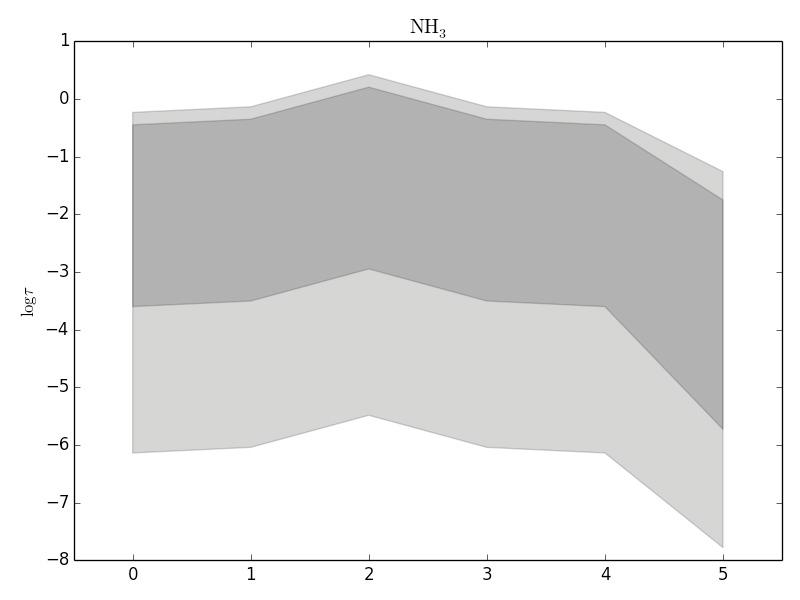}
\caption{\chem{NH_3}}
\label{NH3}
\end{figure*}
\clearpage
\begin{figure*}
\includegraphics[width=16cm]{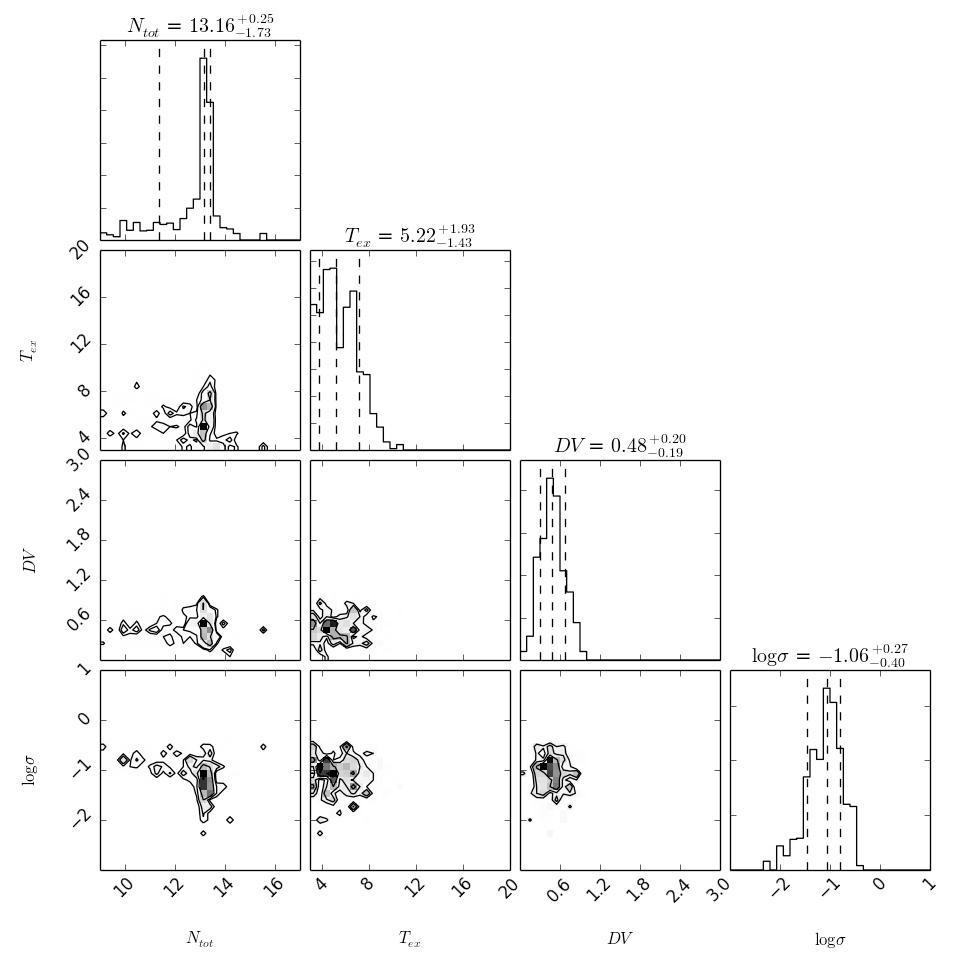}
\includegraphics[width=8cm]{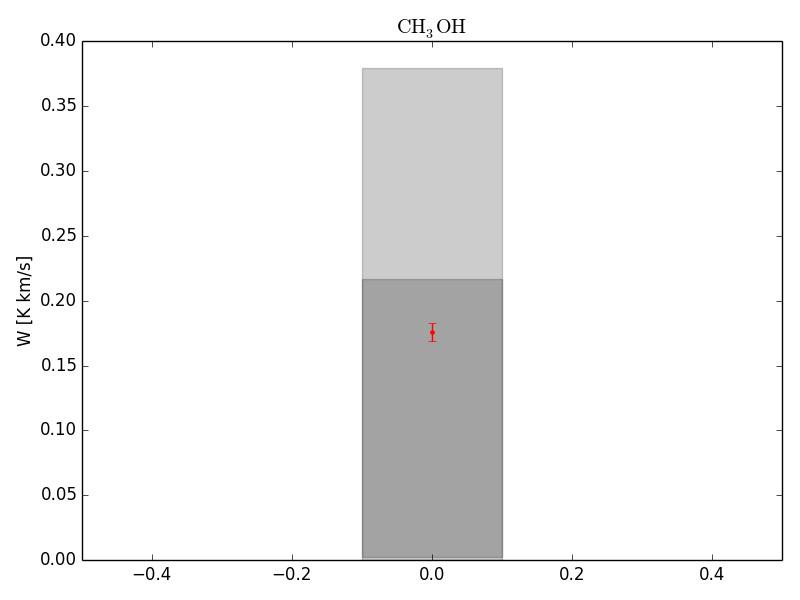}
\includegraphics[width=8cm]{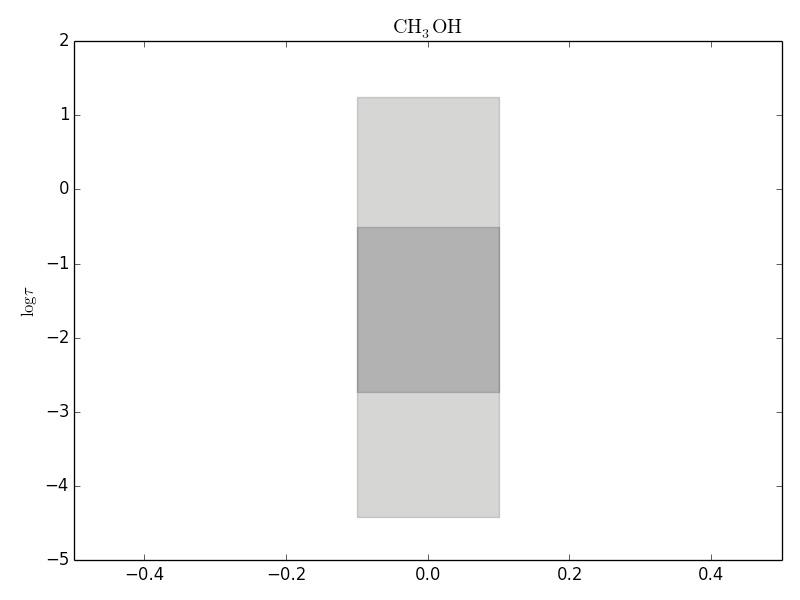}
\caption{\chem{CH_3OH}}
\label{CH3OH}
\end{figure*}
\clearpage
\begin{figure*}
\includegraphics[width=16cm]{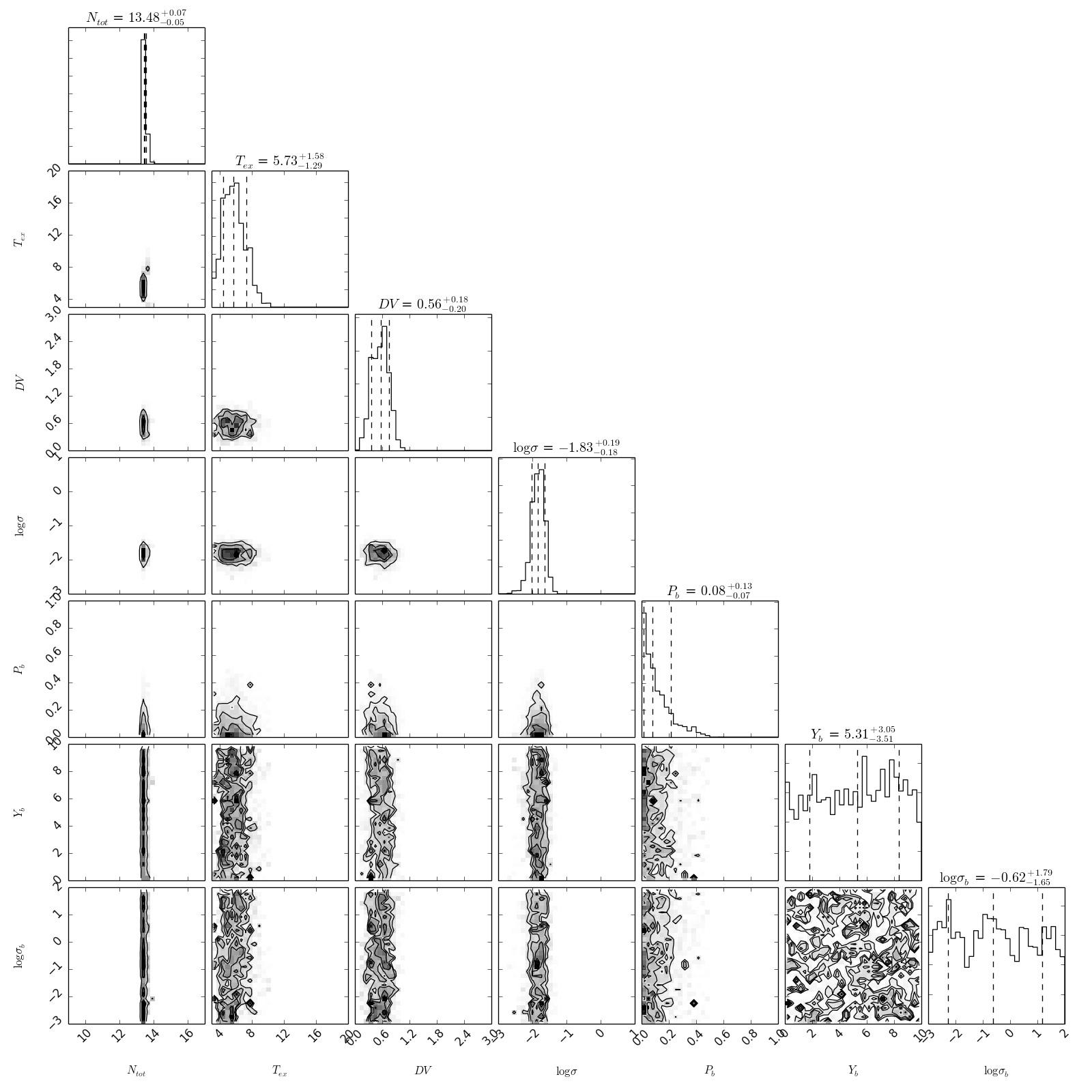}
\includegraphics[width=8cm]{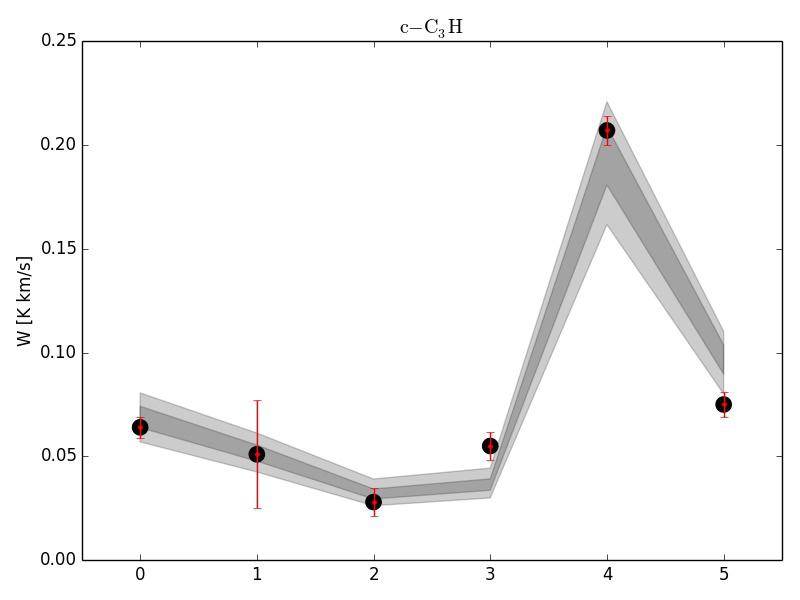}
\includegraphics[width=8cm]{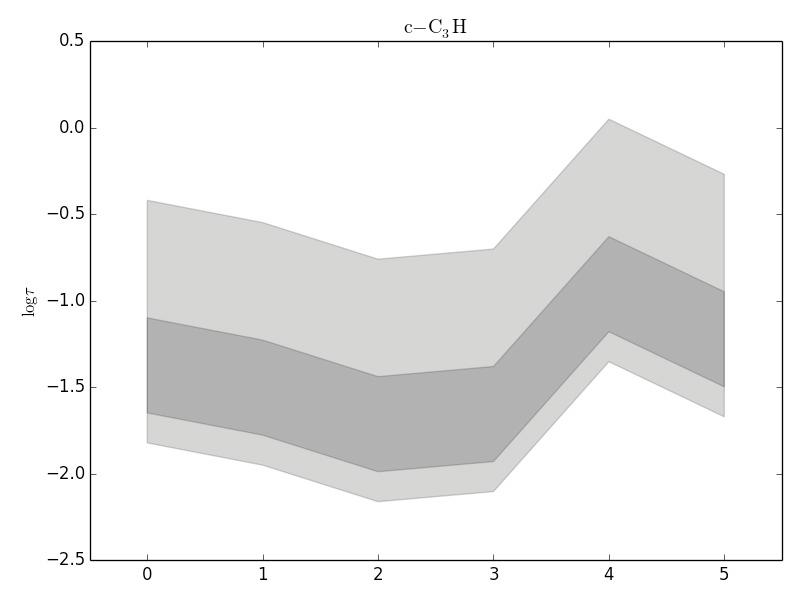}
\caption{\mbox{c-}\chem{C_3H}}
\label{c-C3H}
\end{figure*}
\clearpage
\begin{figure*}
\includegraphics[width=16cm]{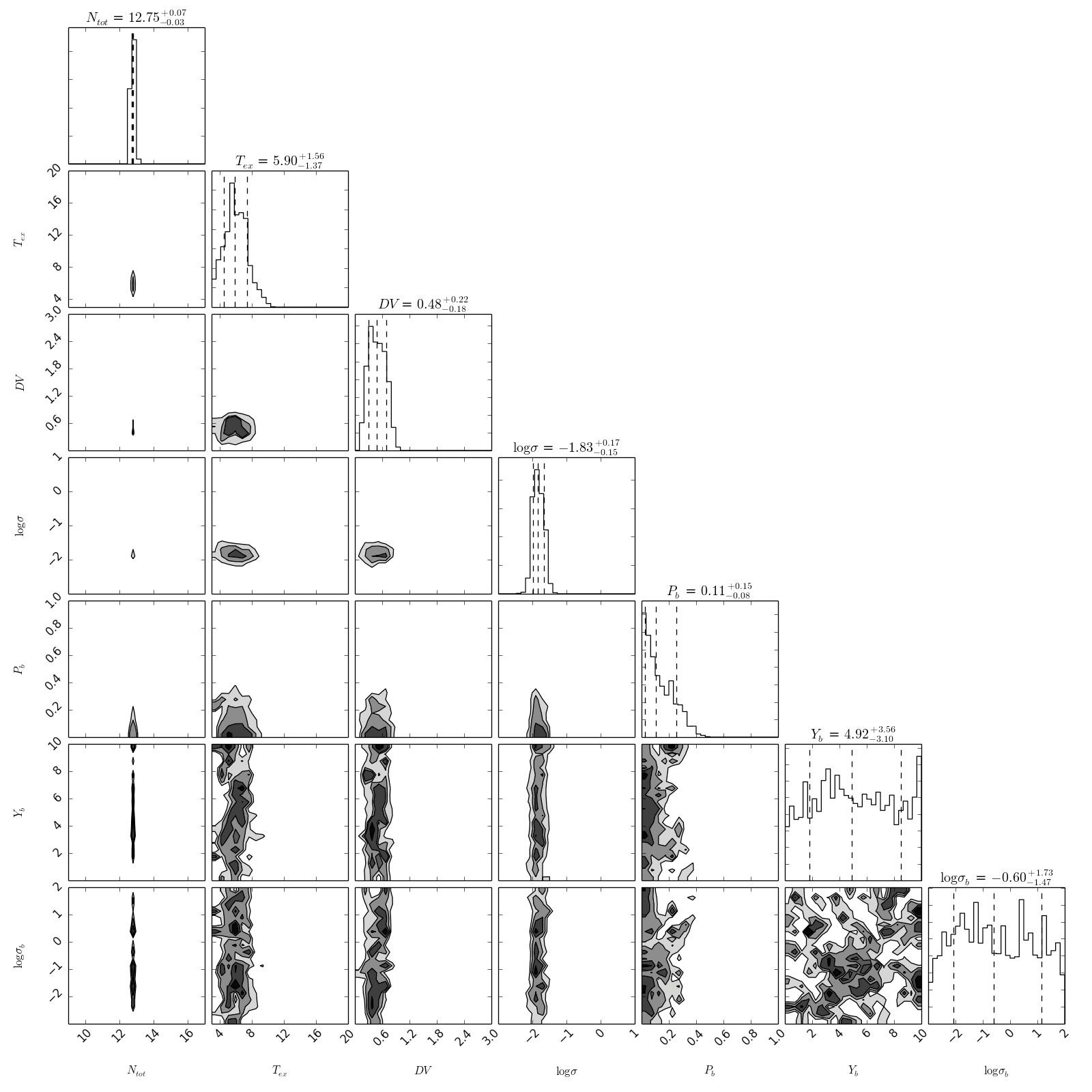}
\includegraphics[width=8cm]{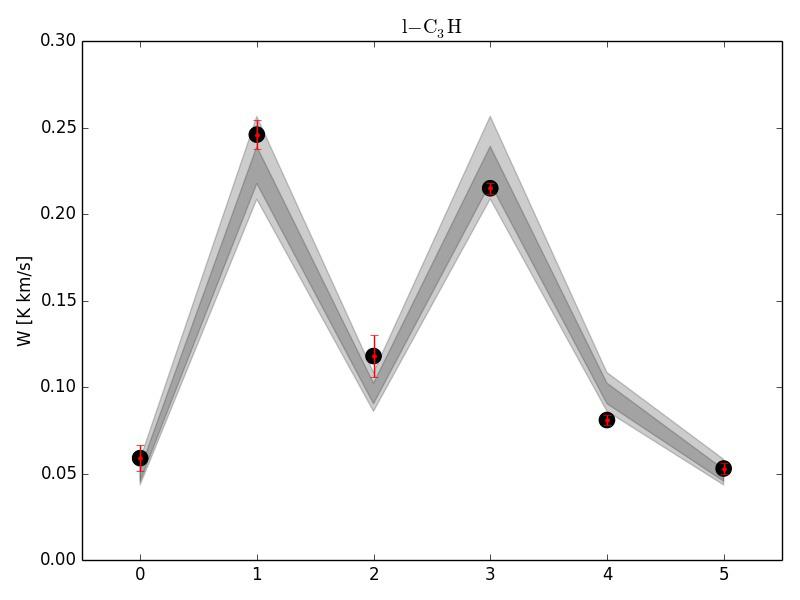}
\includegraphics[width=8cm]{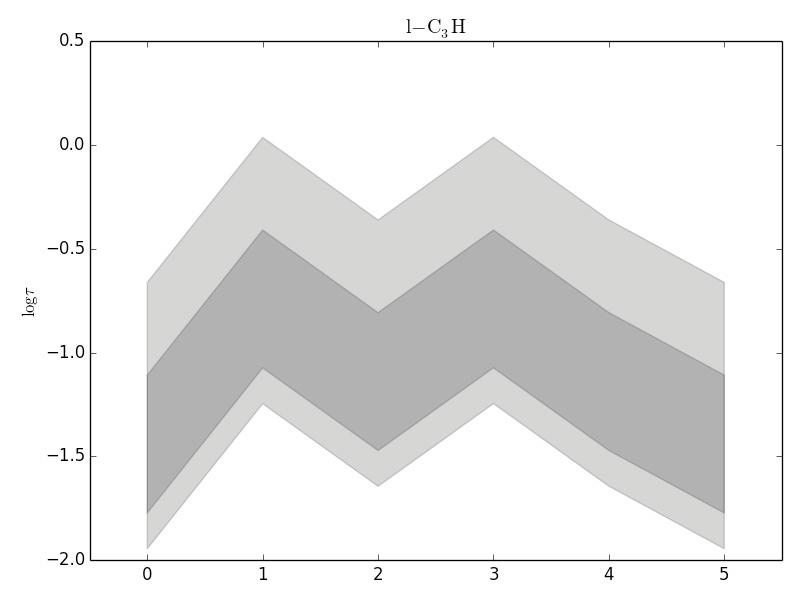}
\caption{\mbox{l-}\chem{C_3H}}
\label{C3H}
\end{figure*}
\clearpage
\begin{figure*}
\includegraphics[width=16cm]{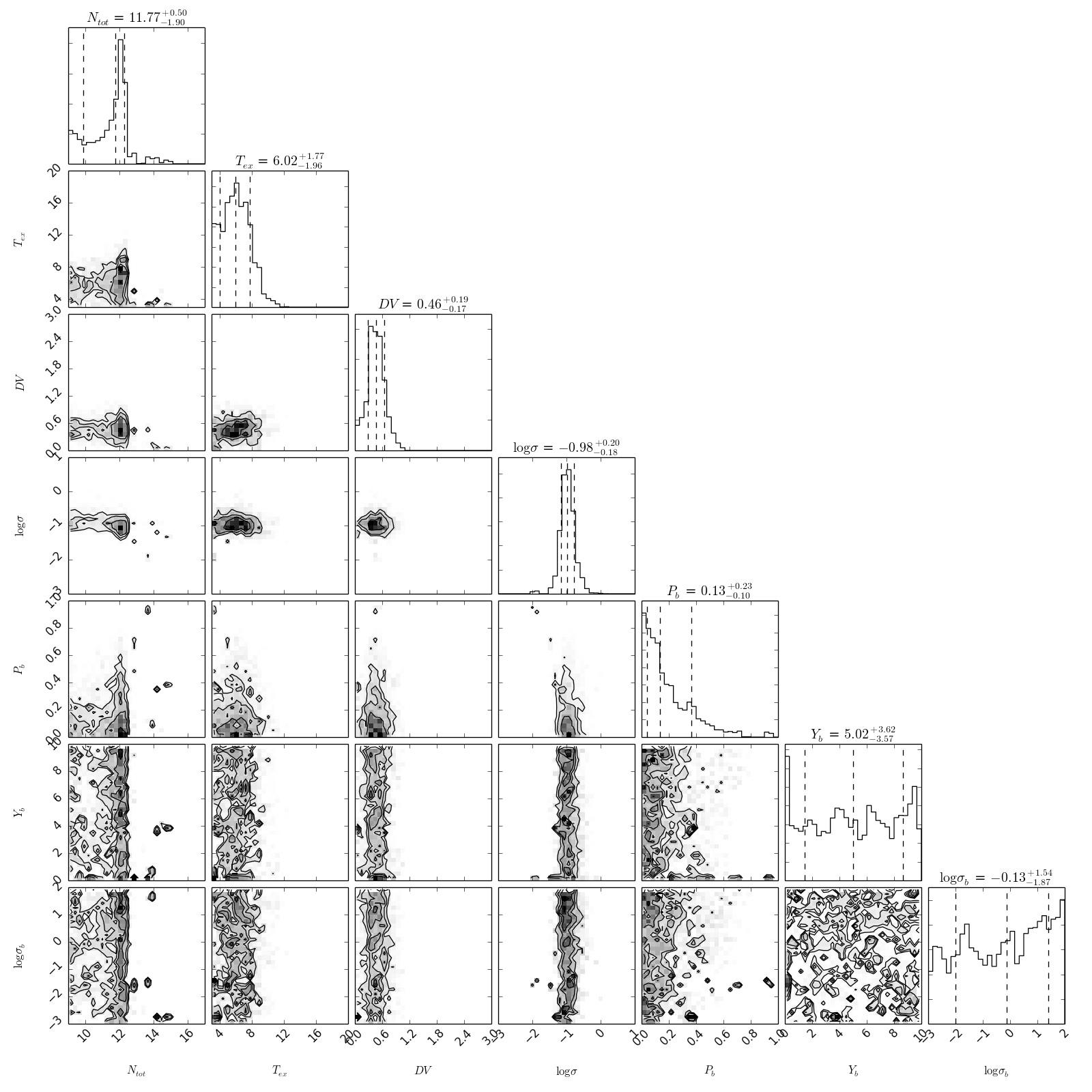}
\includegraphics[width=8cm]{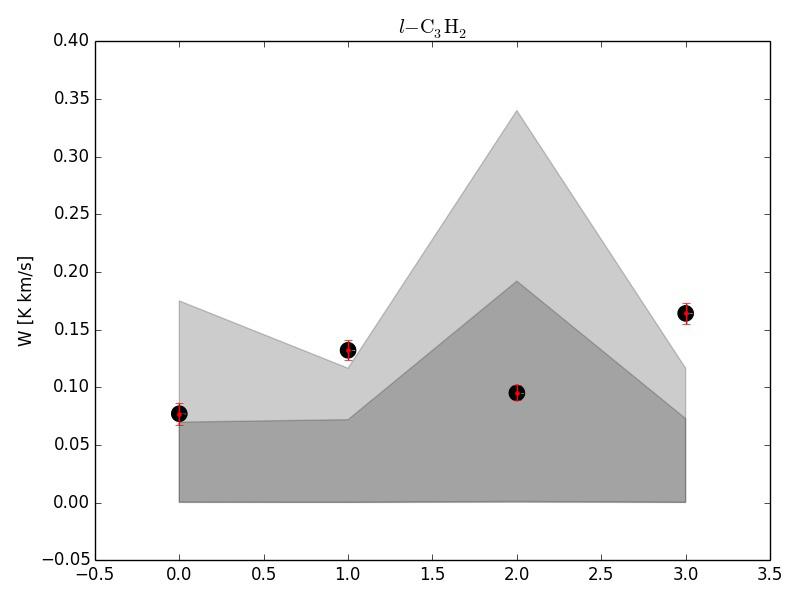}
\includegraphics[width=8cm]{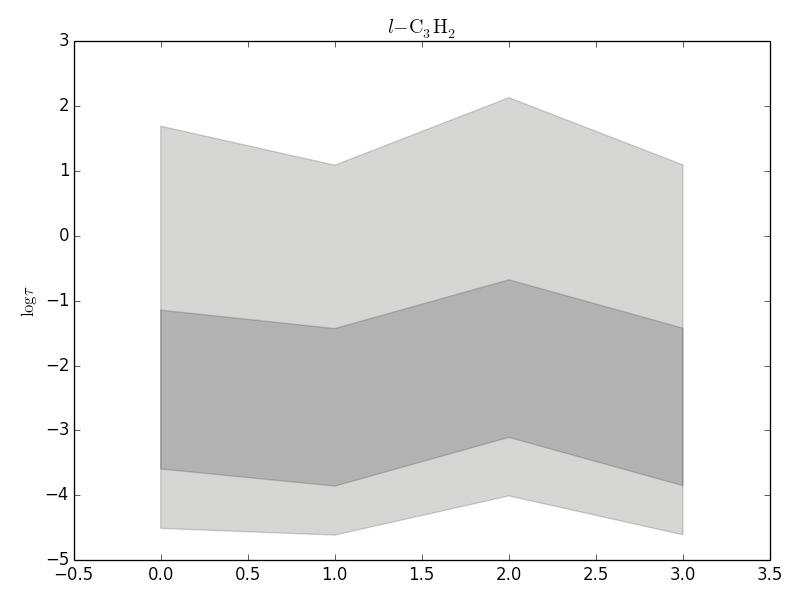}
\caption{\mbox{l-}\chem{C_3H_2}}
\label{H2C3}
\end{figure*}
\clearpage
\begin{figure*}
\includegraphics[width=16cm]{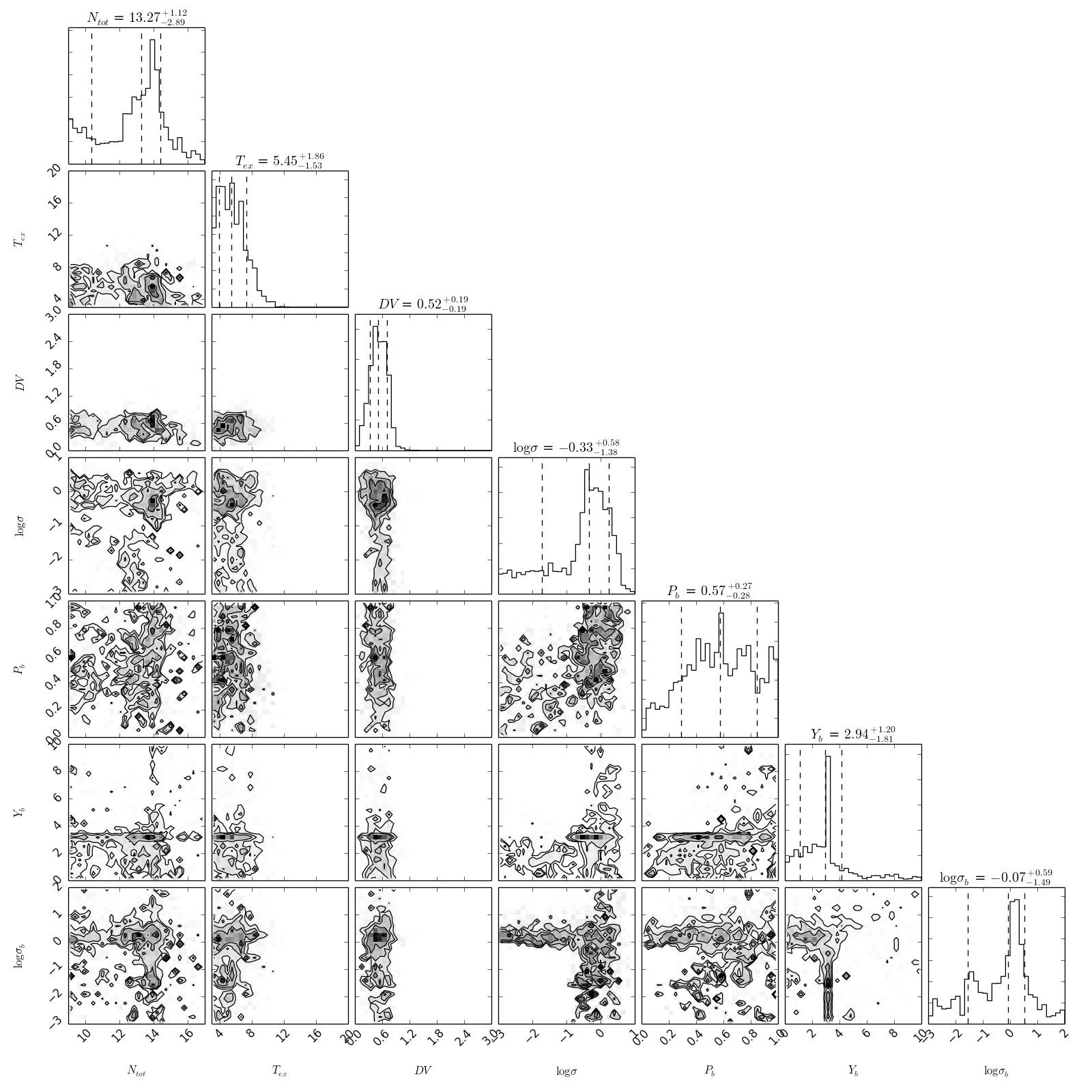}
\includegraphics[width=8cm]{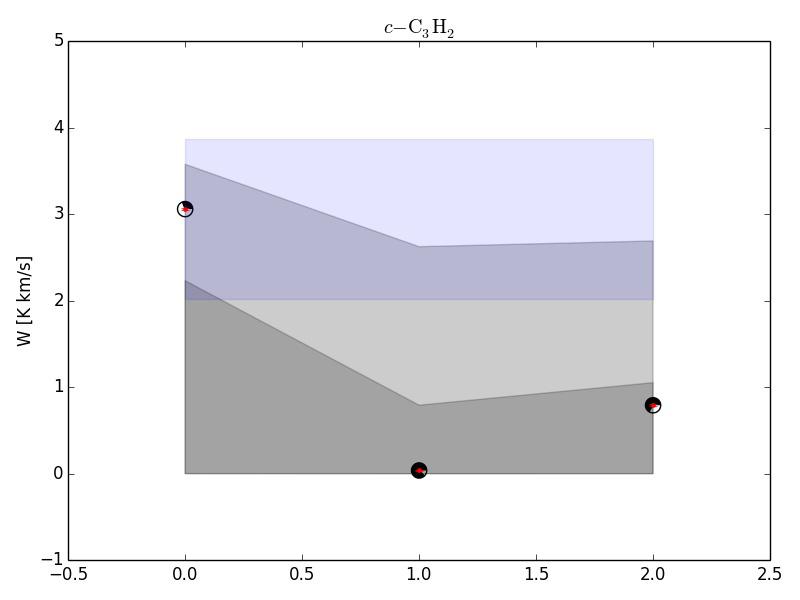}
\includegraphics[width=8cm]{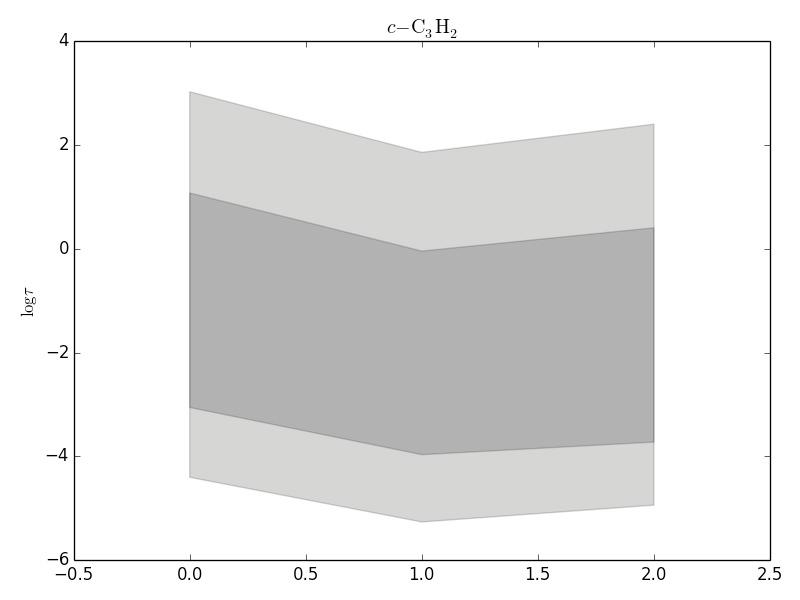}
\caption{\mbox{c-}\chem{C_3H_2}}
\label{c-C3H2}
\end{figure*}
\clearpage
\begin{figure*}
\includegraphics[width=16cm]{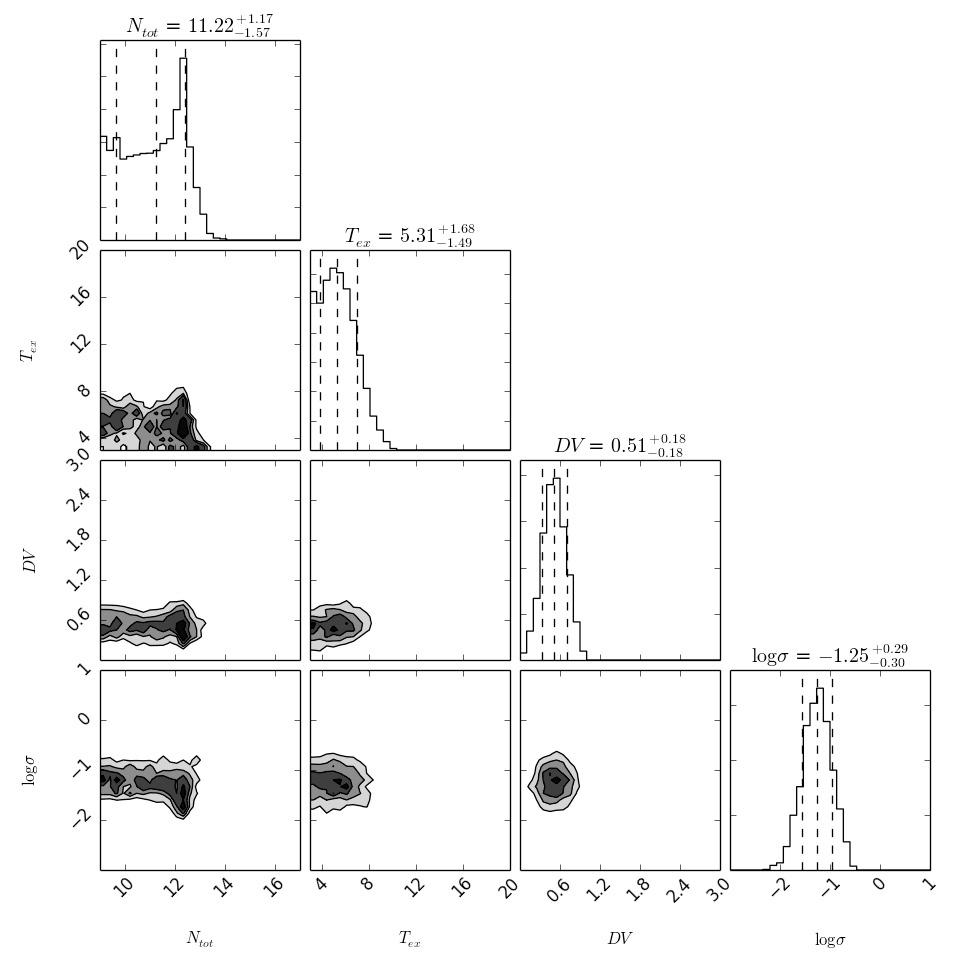}
\includegraphics[width=8cm]{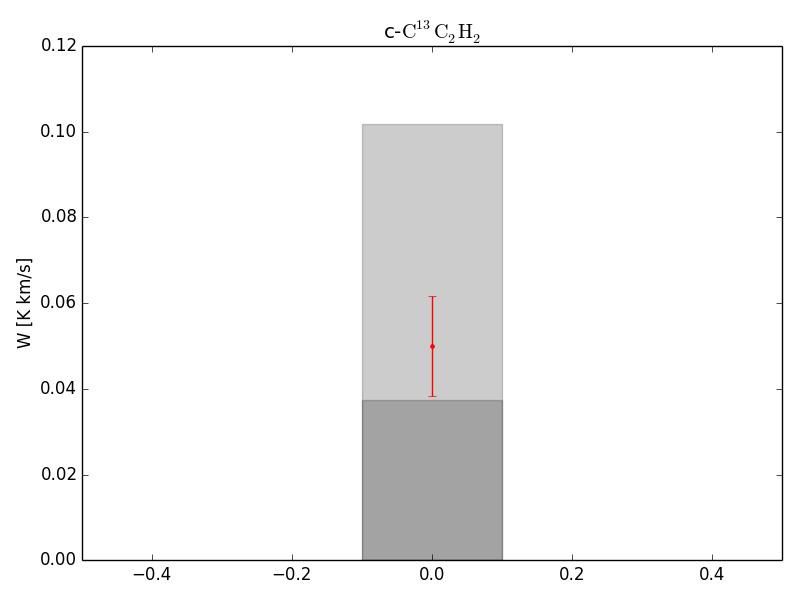}
\includegraphics[width=8cm]{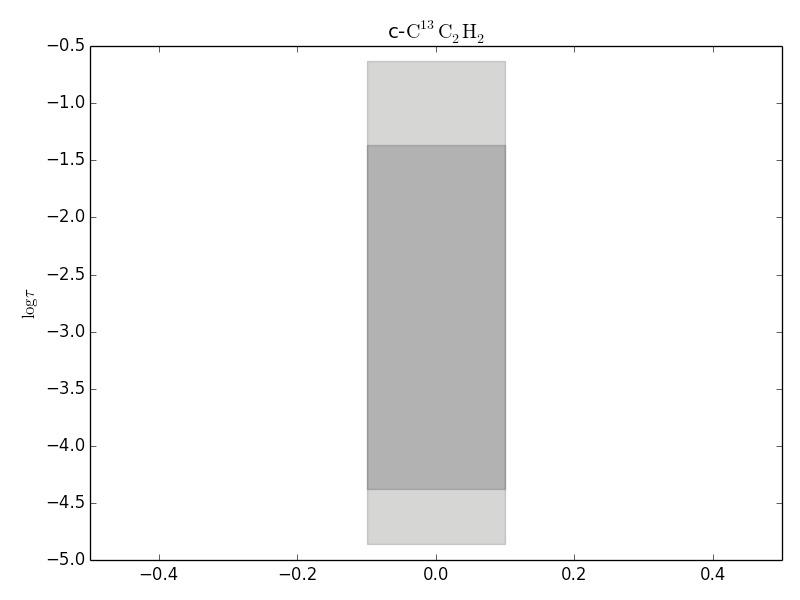}
\caption{\chem{^{13}CCH_2}}
\label{C13CCH2}
\end{figure*}
\clearpage
\begin{figure*}
\includegraphics[width=16cm]{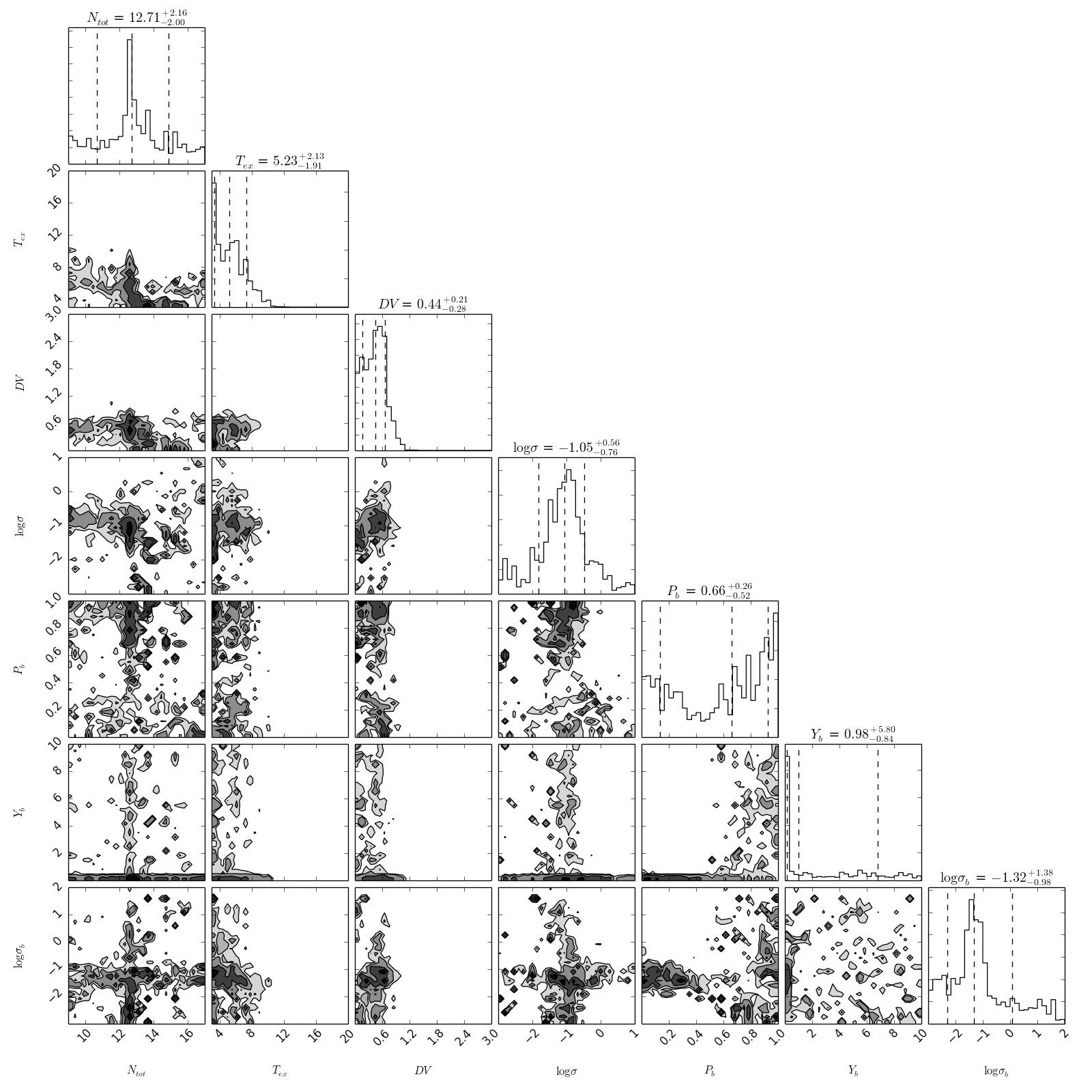}
\includegraphics[width=8cm]{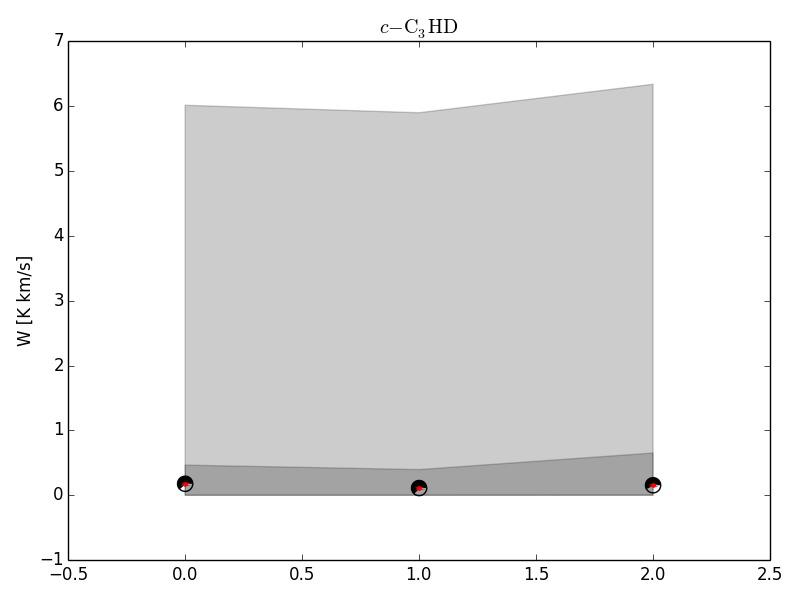}
\includegraphics[width=8cm]{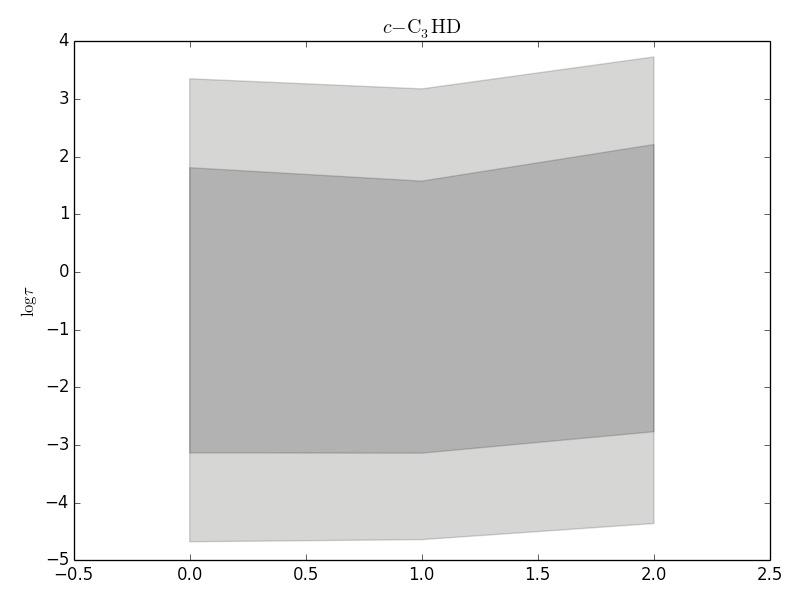}
\caption{\mbox{c-}\chem{C_3HD}}
\label{C3HD}
\end{figure*}
\clearpage
\begin{figure*}
\includegraphics[width=16cm]{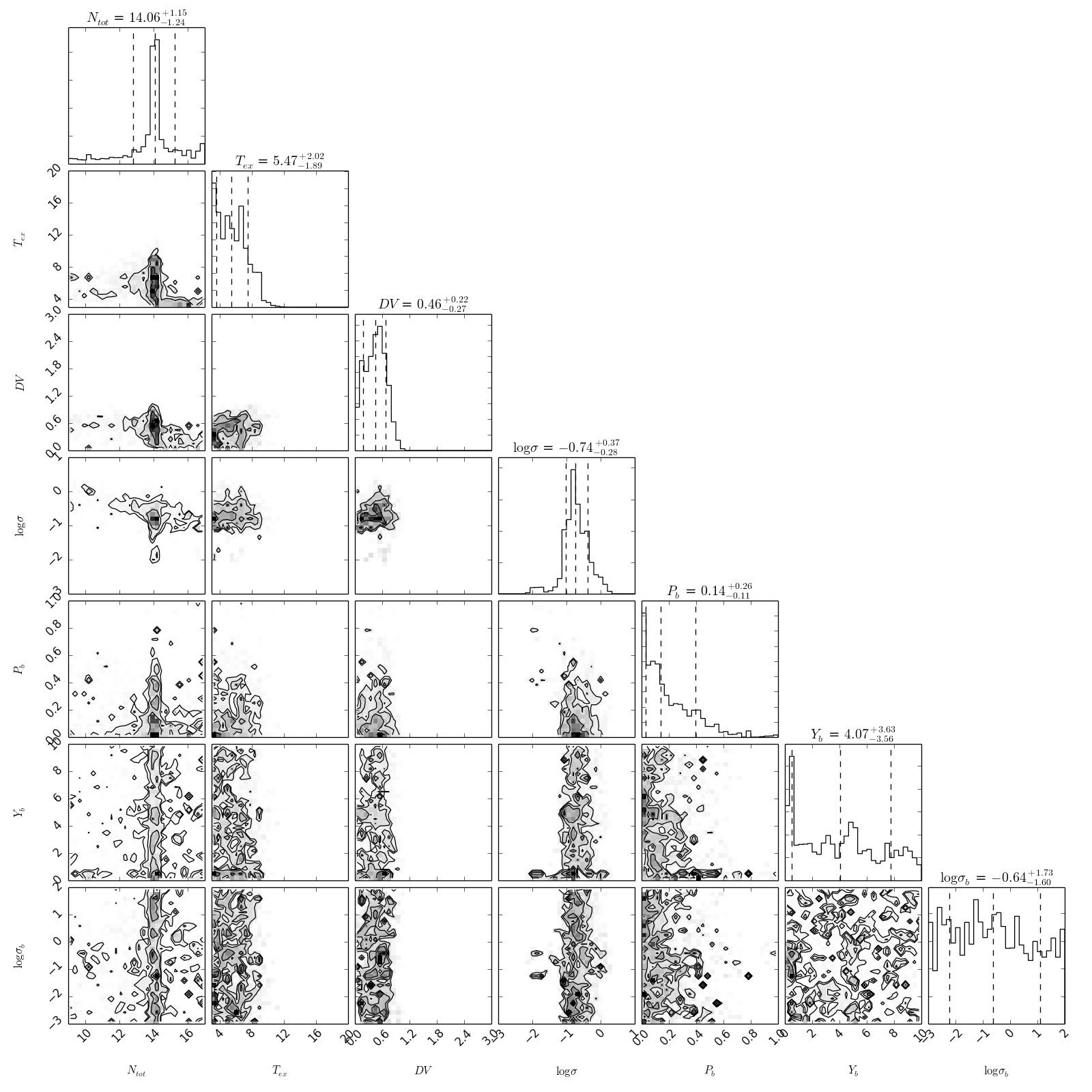}
\includegraphics[width=8cm]{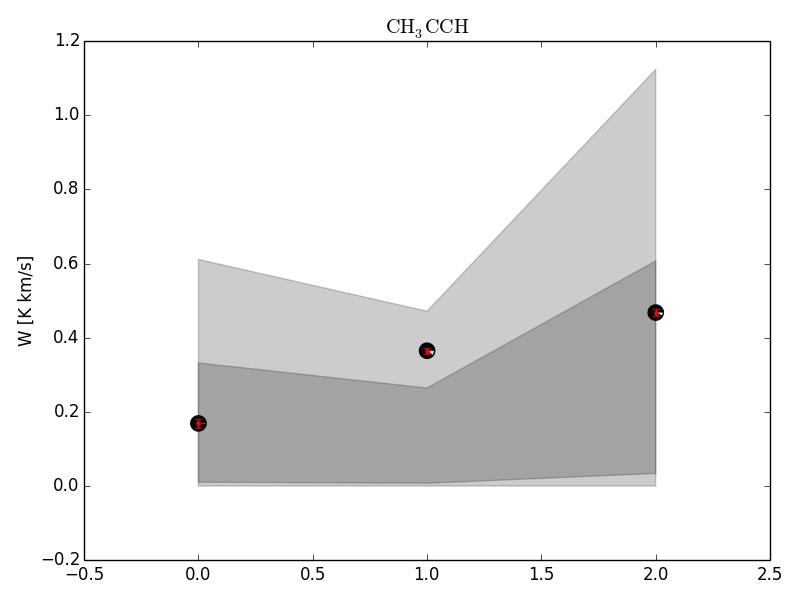}
\includegraphics[width=8cm]{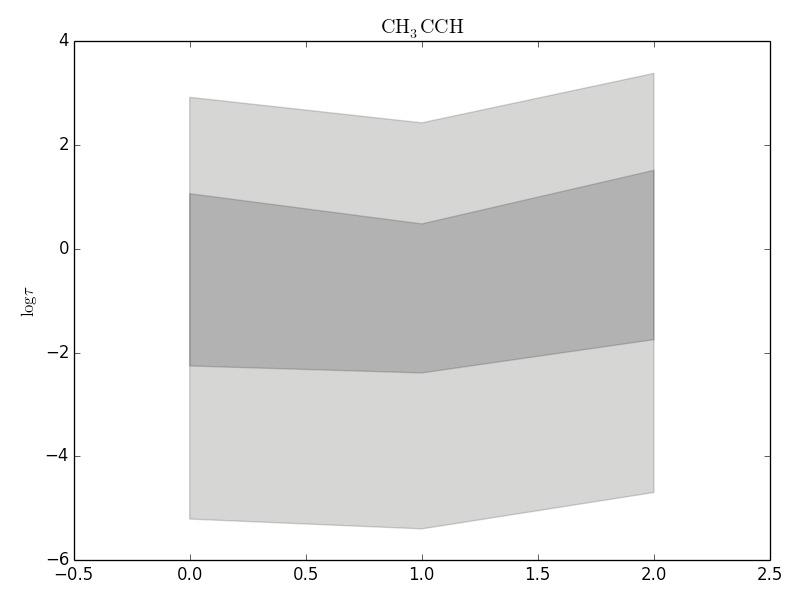}
\caption{\chem{CH_3CCH}}
\label{CH3CCH}
\end{figure*}
\clearpage
\begin{figure*}
\includegraphics[width=16cm]{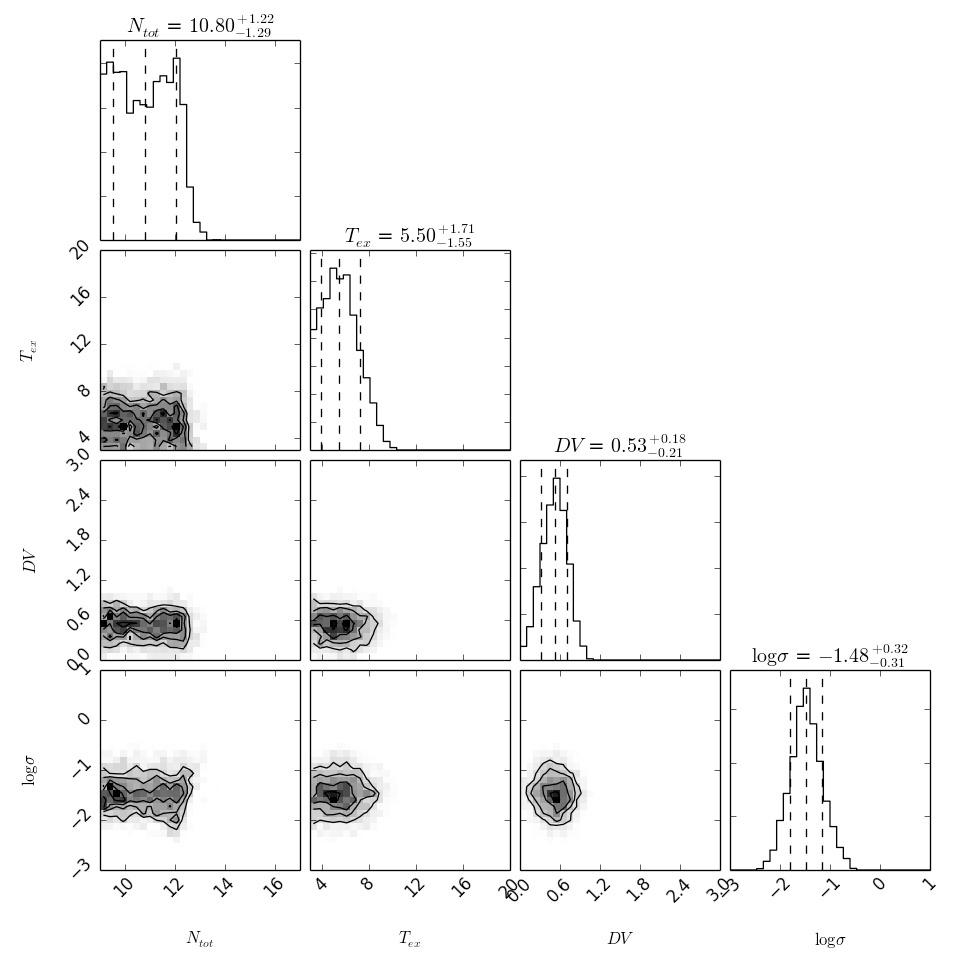}
\includegraphics[width=8cm]{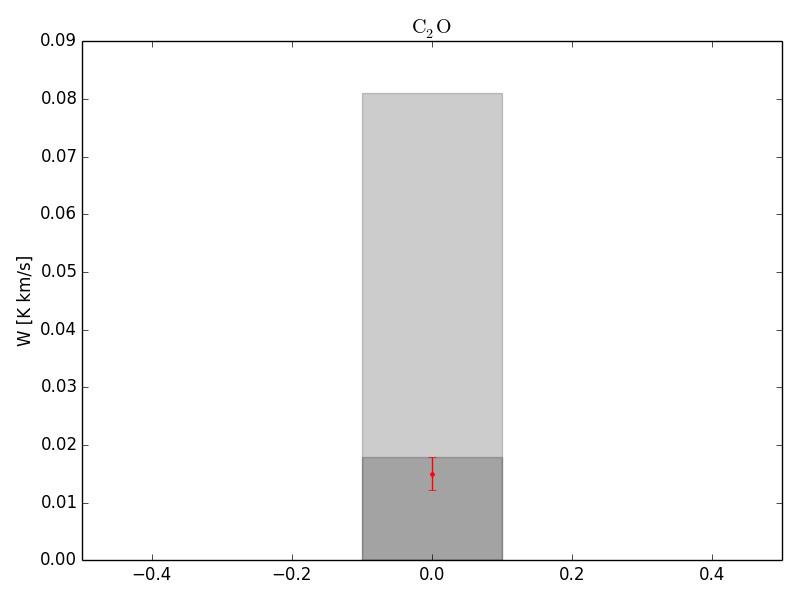}
\includegraphics[width=8cm]{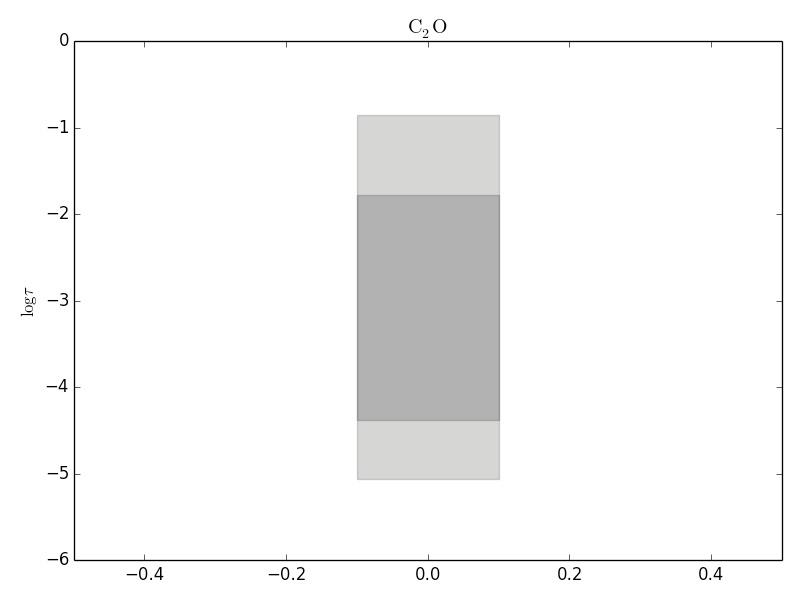}
\caption{\chem{C_2O}}
\label{C2O}
\end{figure*}
\clearpage
\clearpage
\begin{figure*}
\includegraphics[width=16cm]{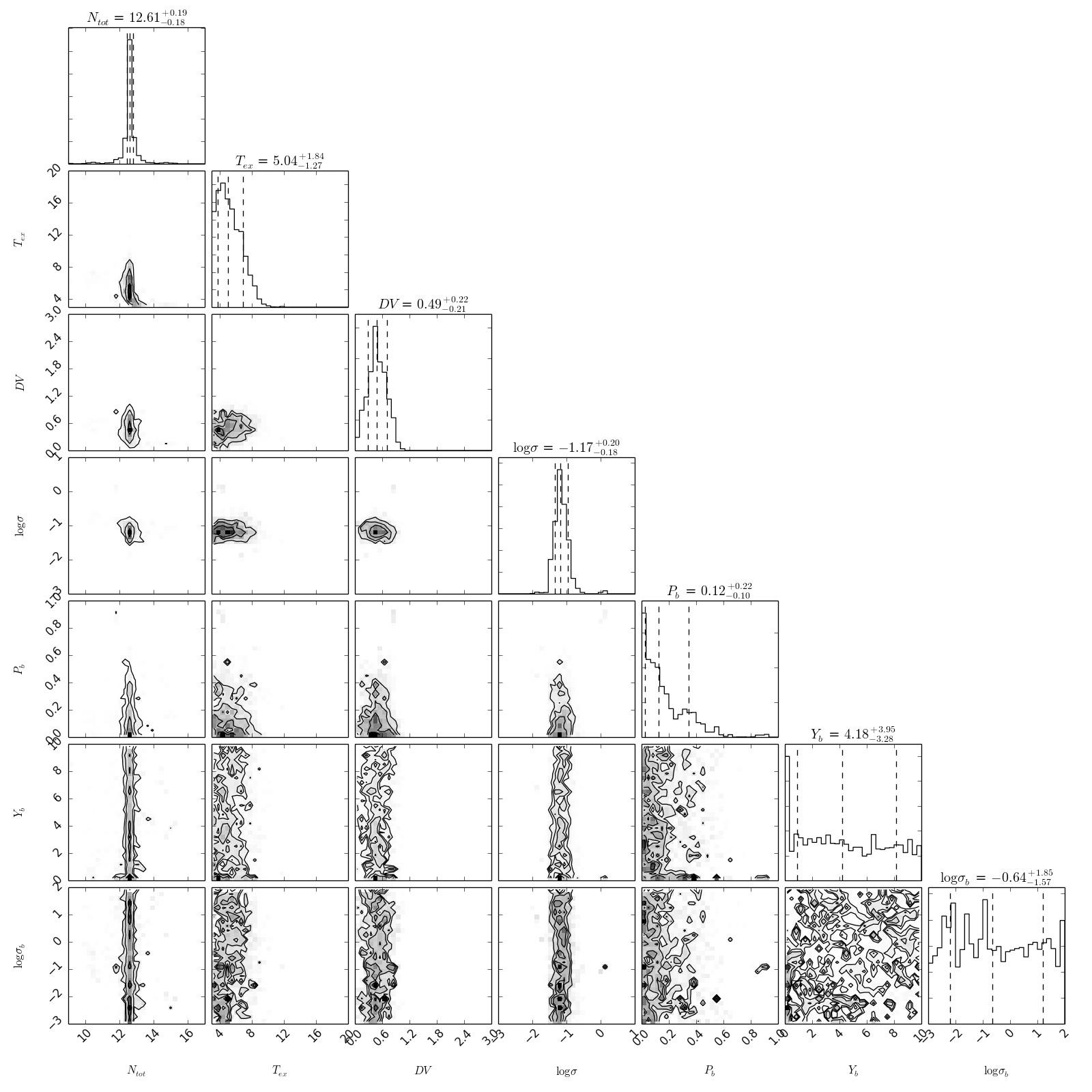}
\includegraphics[width=8cm]{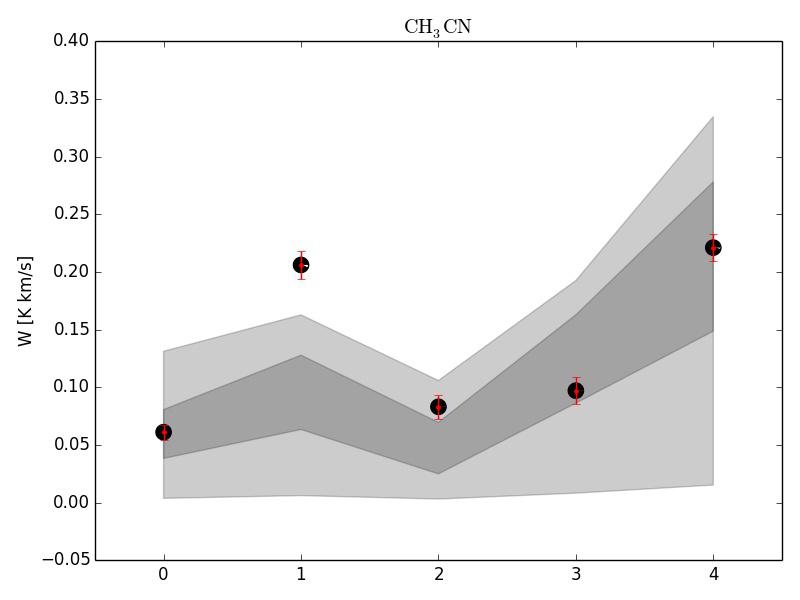}
\includegraphics[width=8cm]{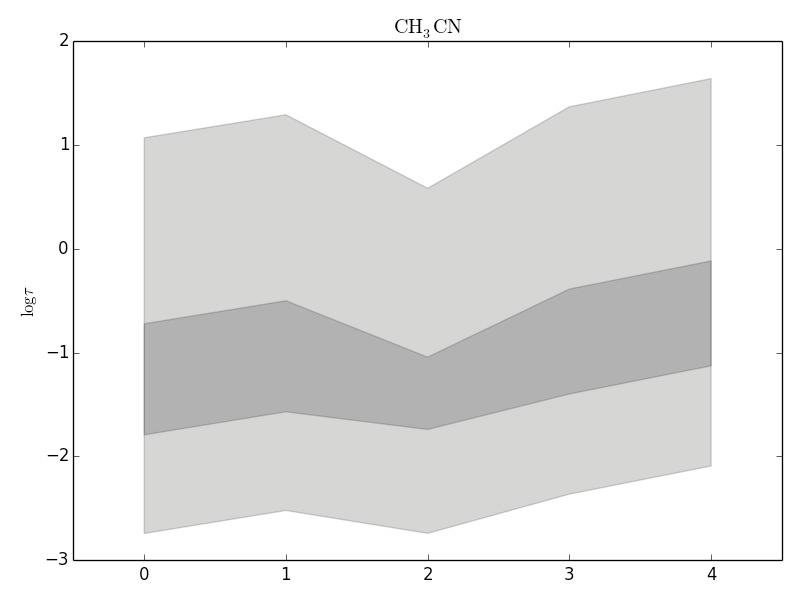}
\caption{\chem{CH_3CN}}
\label{CH3CN}
\end{figure*}
\clearpage
\begin{figure*}
\includegraphics[width=16cm]{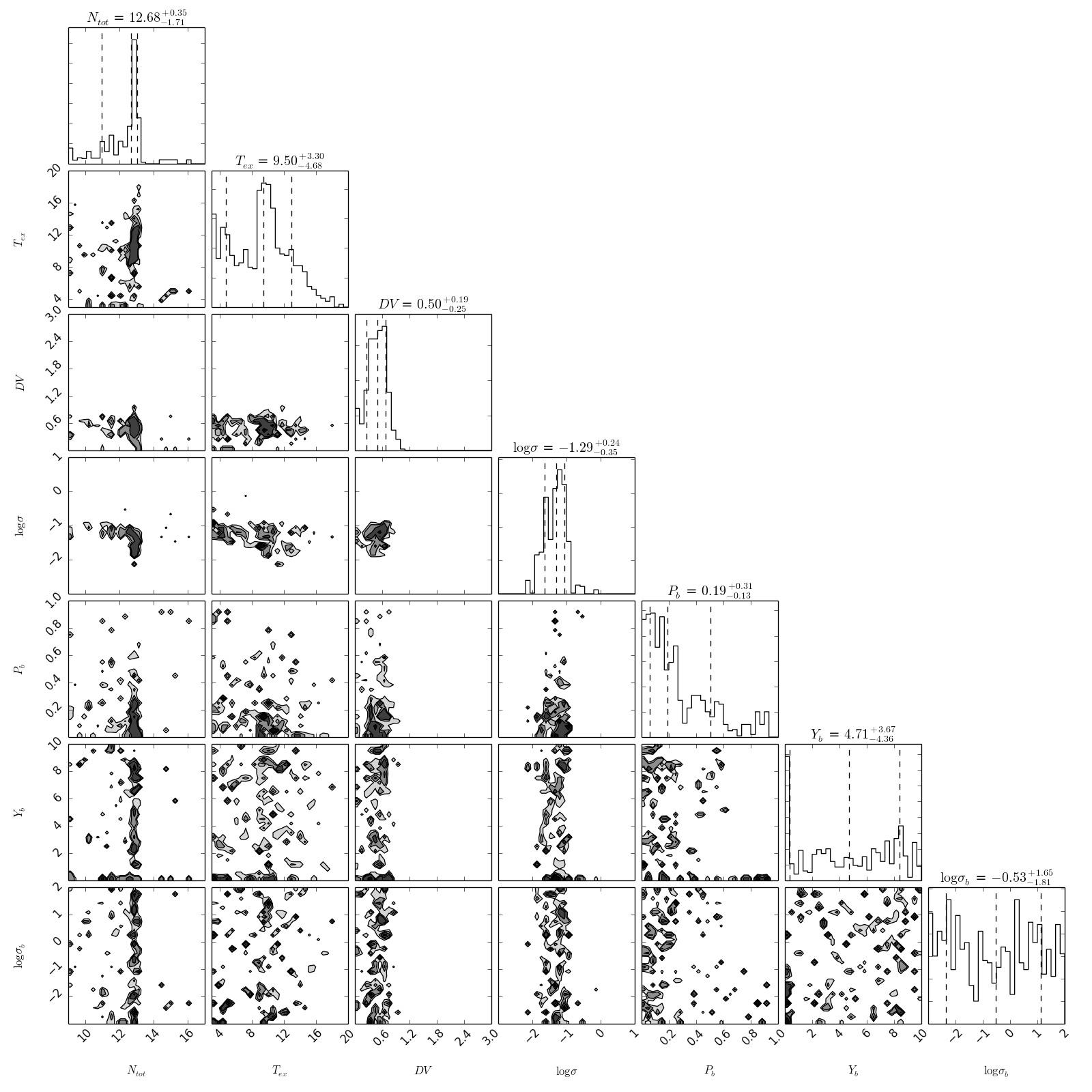}
\includegraphics[width=8cm]{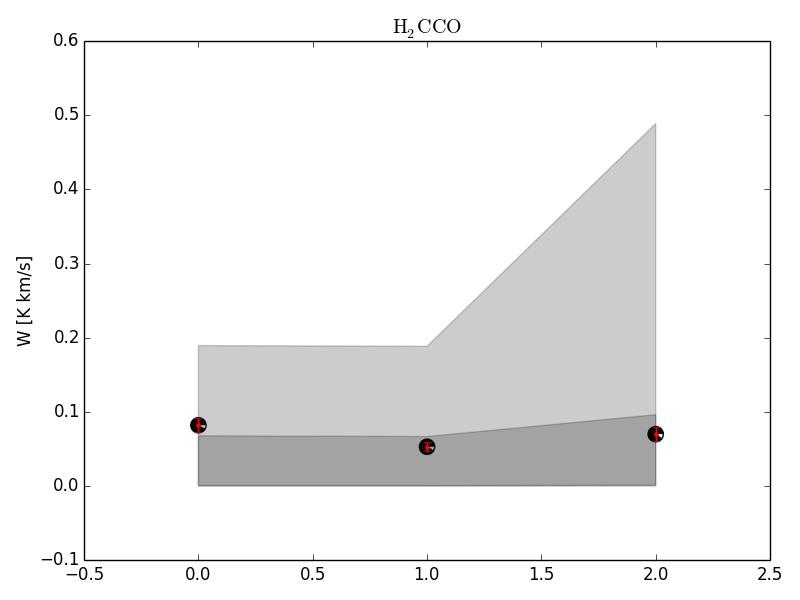}
\includegraphics[width=8cm]{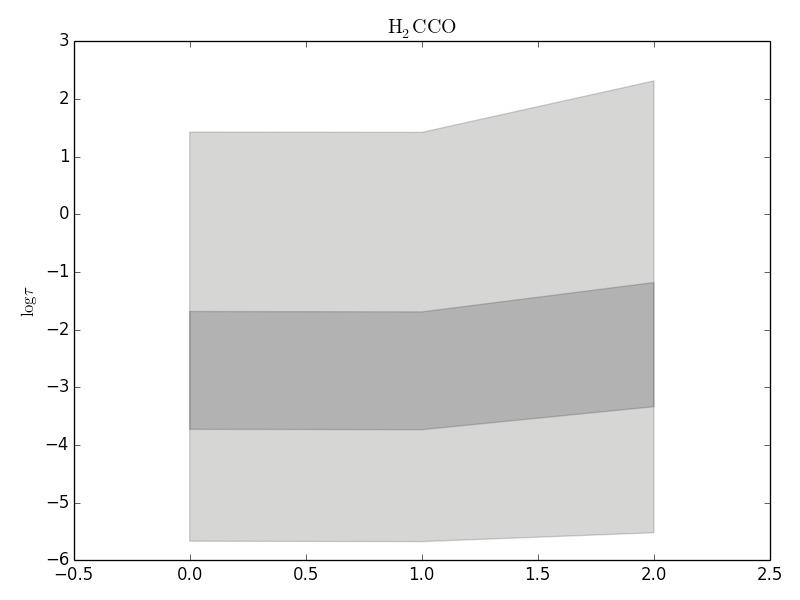}
\caption{\chem{H_2CCO}}
\label{H2CCO}
\end{figure*}
\clearpage
\begin{figure*}
\includegraphics[width=16cm]{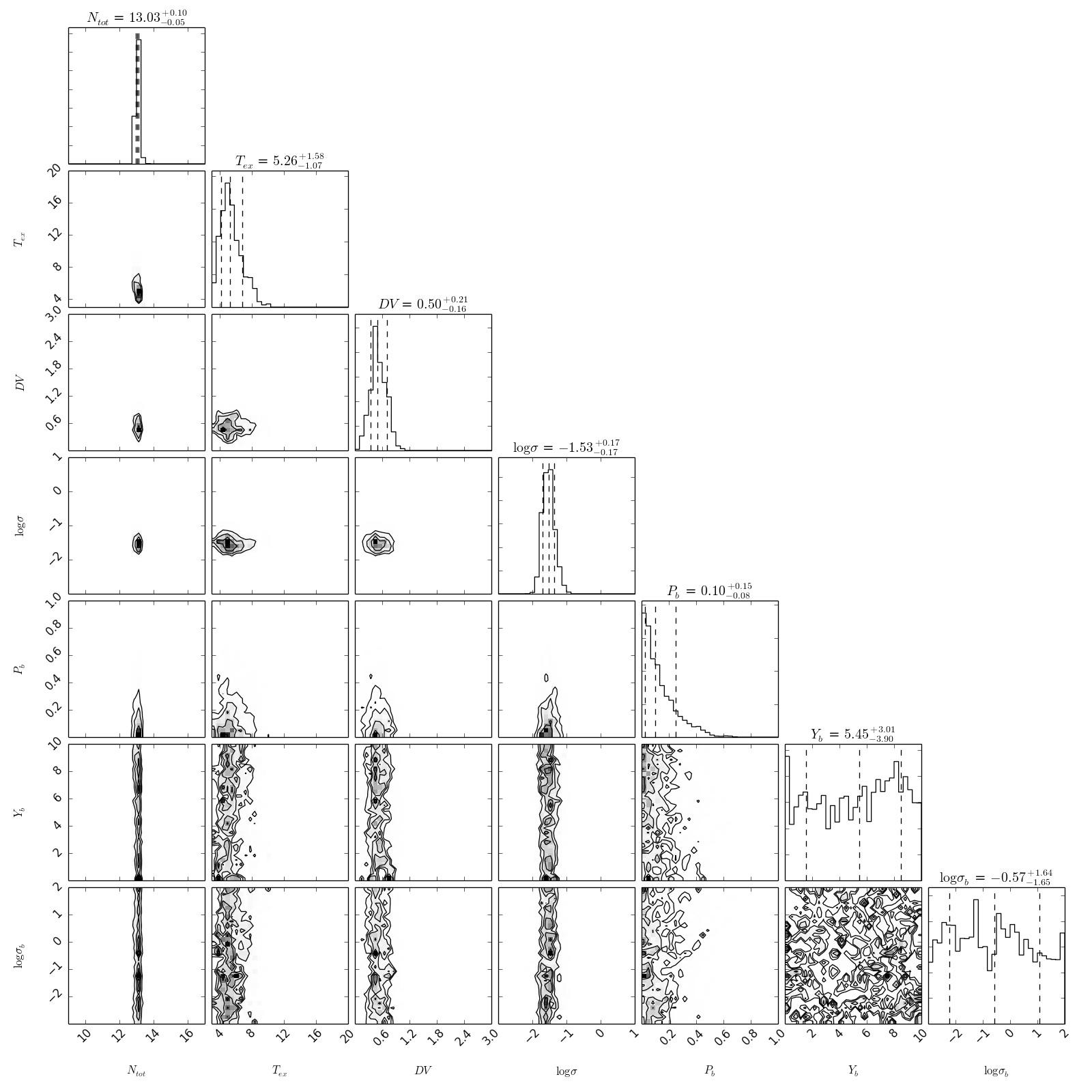}
\includegraphics[width=8cm]{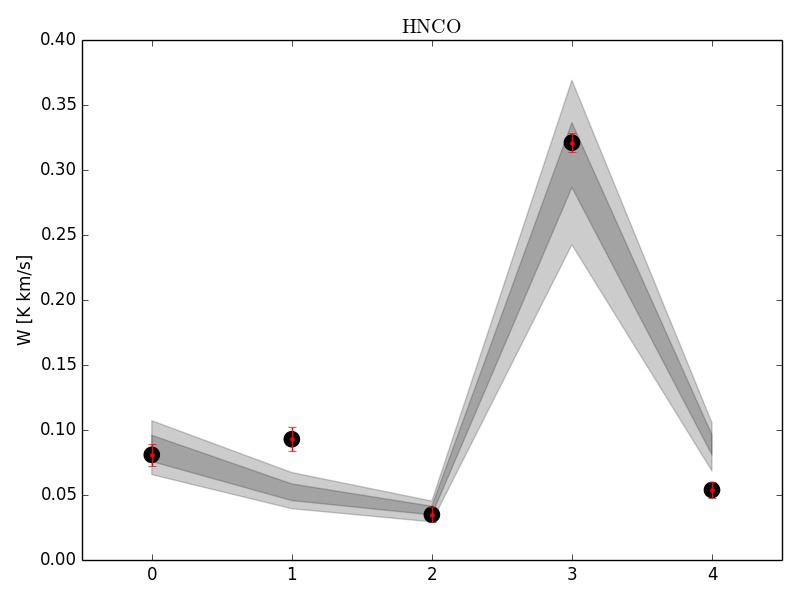}
\includegraphics[width=8cm]{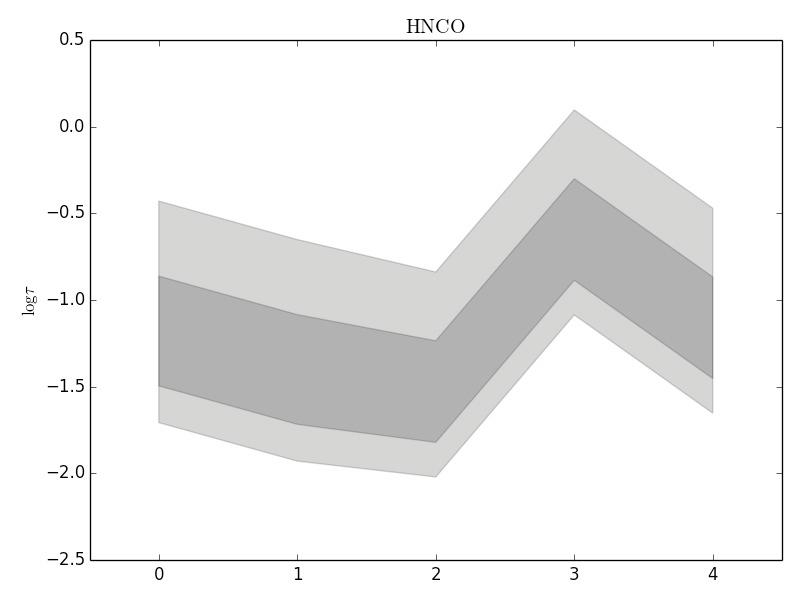}
\caption{\chem{HNCO}}
\label{HNCO}
\end{figure*}
\clearpage
\begin{figure*}
\includegraphics[width=16cm]{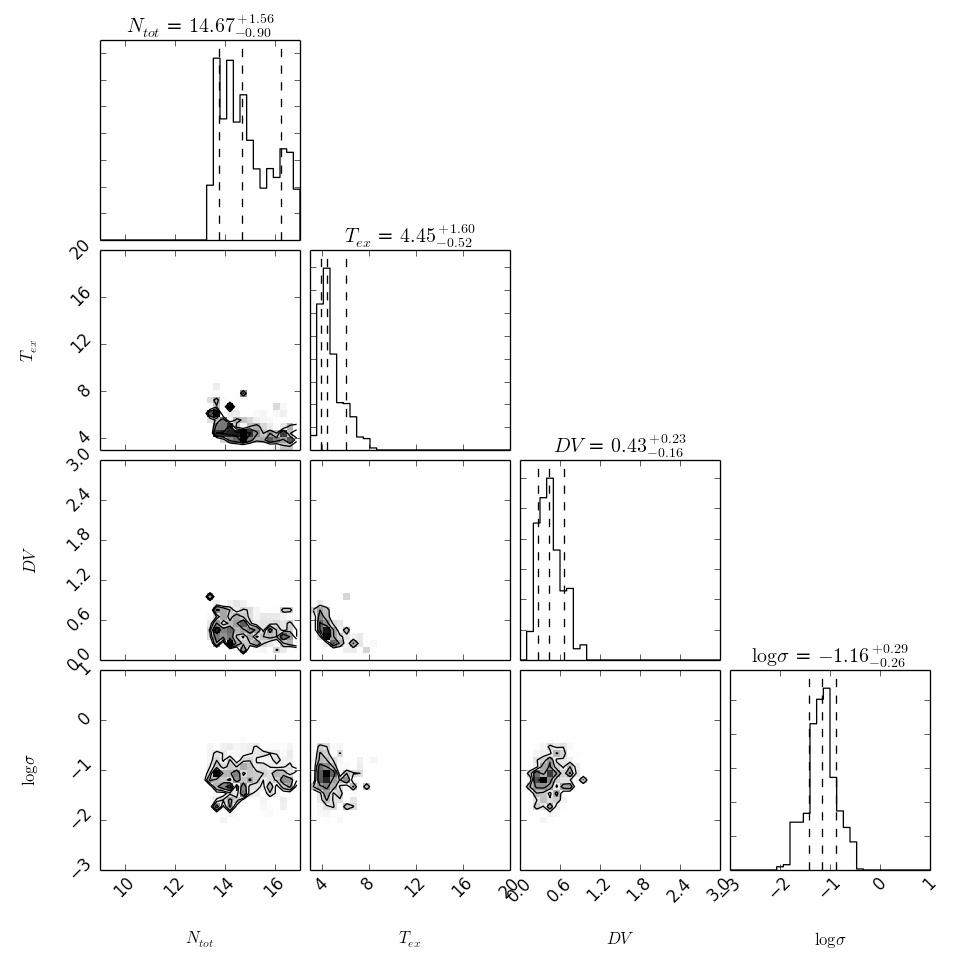}
\includegraphics[width=8cm]{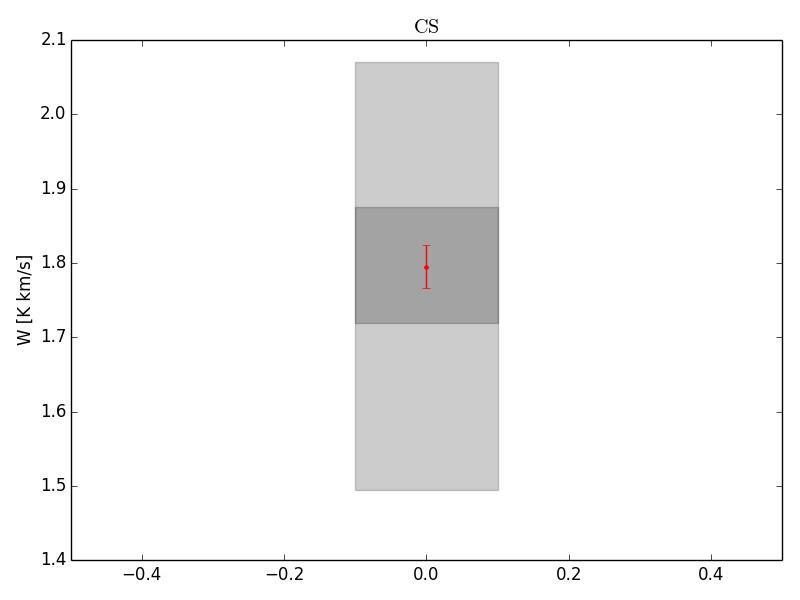}
\includegraphics[width=8cm]{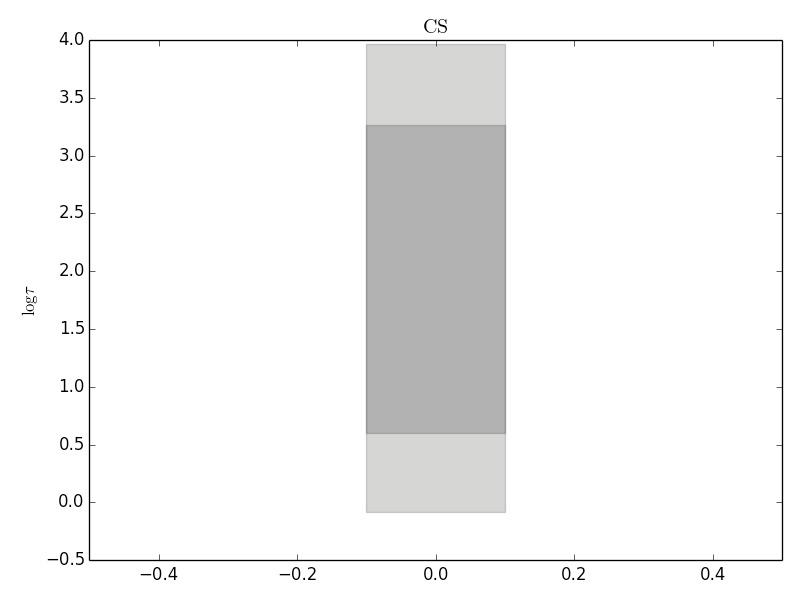}
\caption{\chem{CS}}
\label{CS}
\end{figure*}
\clearpage
\begin{figure*}
\includegraphics[width=16cm]{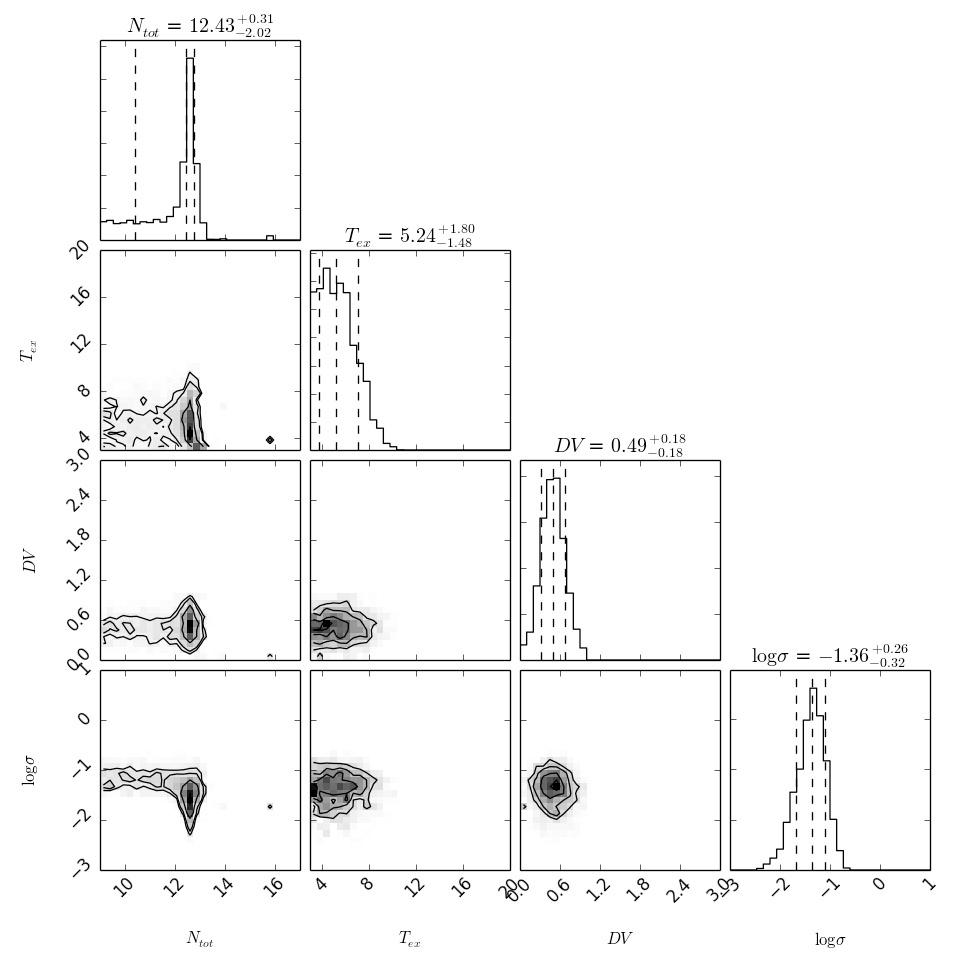}
\includegraphics[width=8cm]{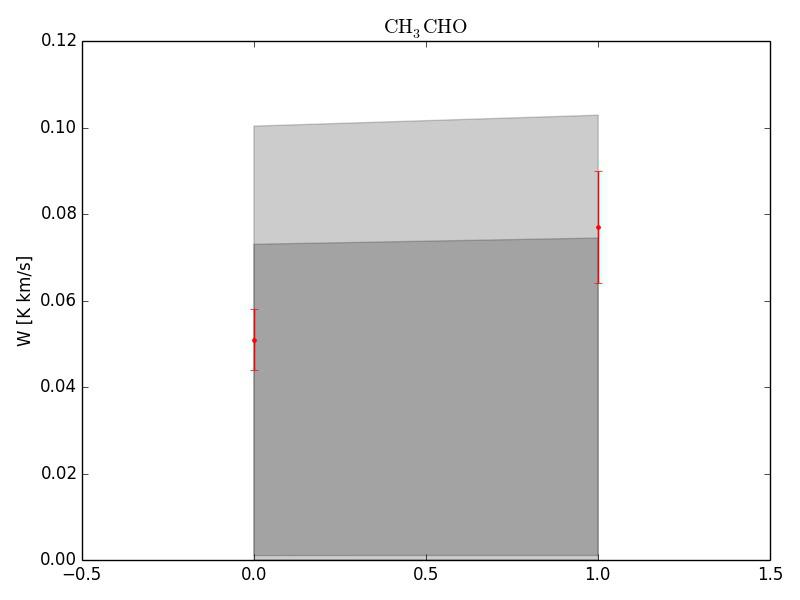}
\includegraphics[width=8cm]{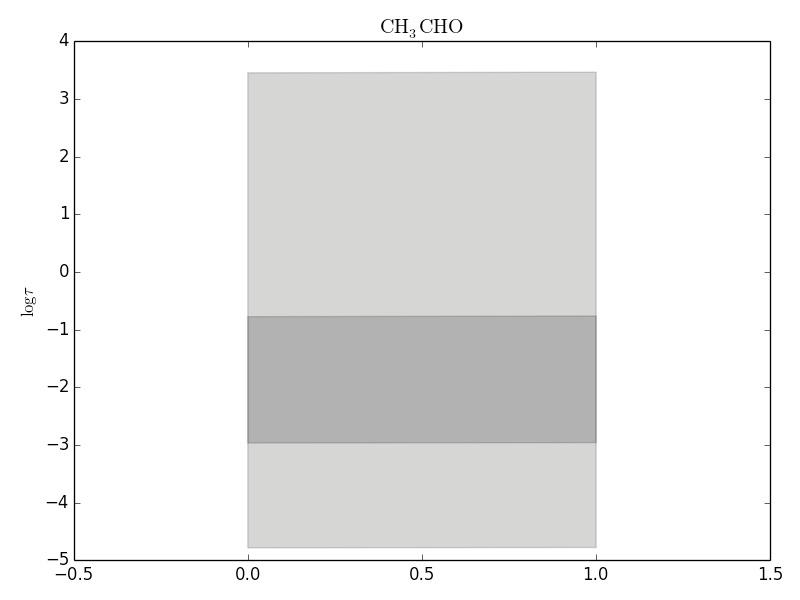}
\caption{\chem{CH_3CHO}}
\label{CH3CHO}
\end{figure*}
\clearpage
\begin{figure*}
\includegraphics[width=16cm]{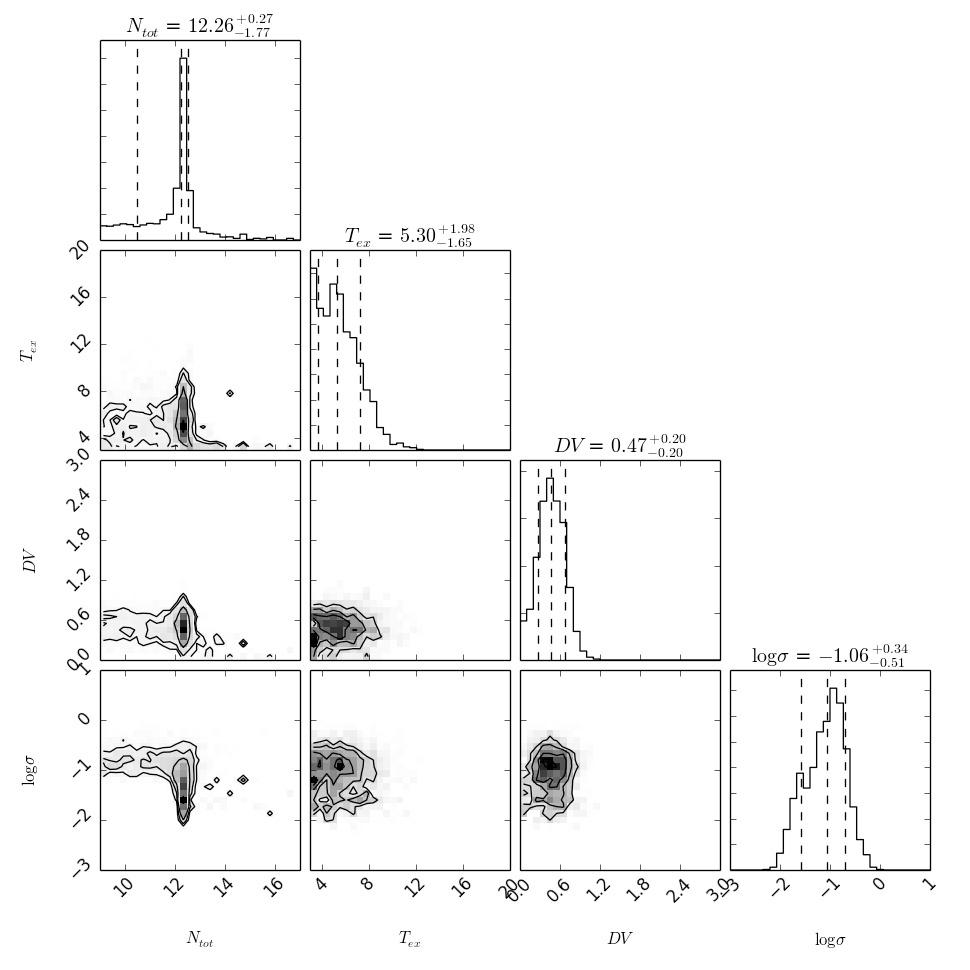}
\includegraphics[width=8cm]{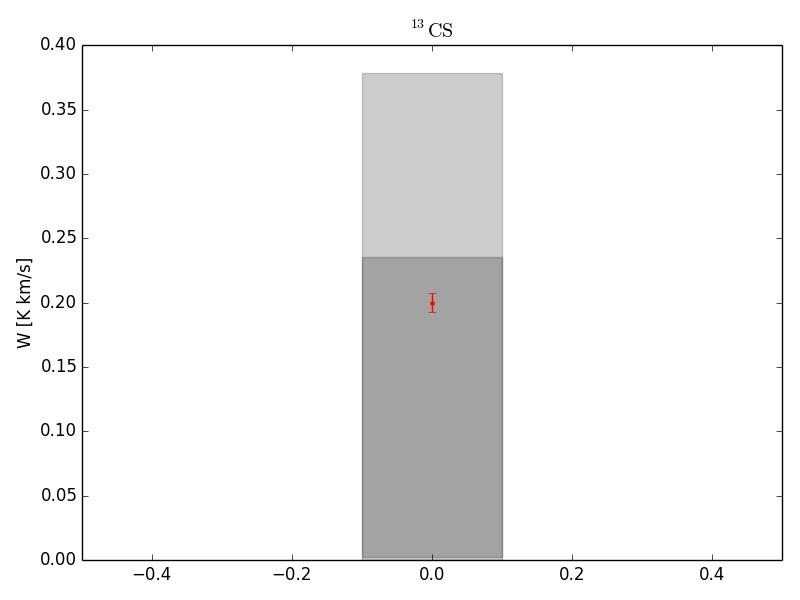}
\includegraphics[width=8cm]{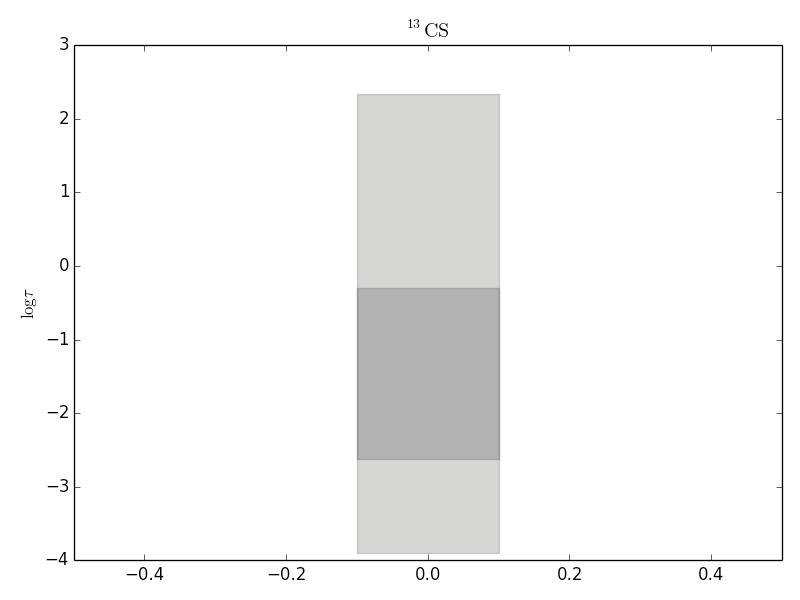}
\caption{\chem{^{13}CS}}
\label{thCS}
\end{figure*}
\clearpage
\begin{figure*}
\includegraphics[width=16cm]{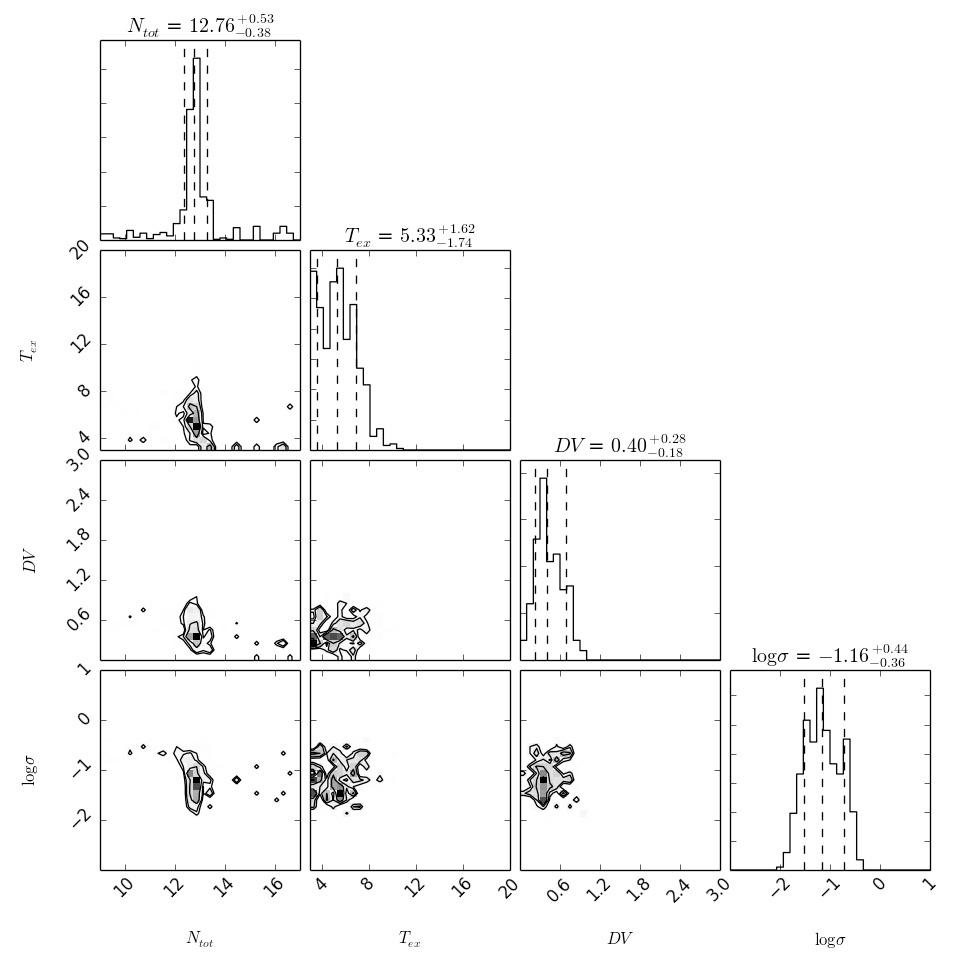}
\includegraphics[width=8cm]{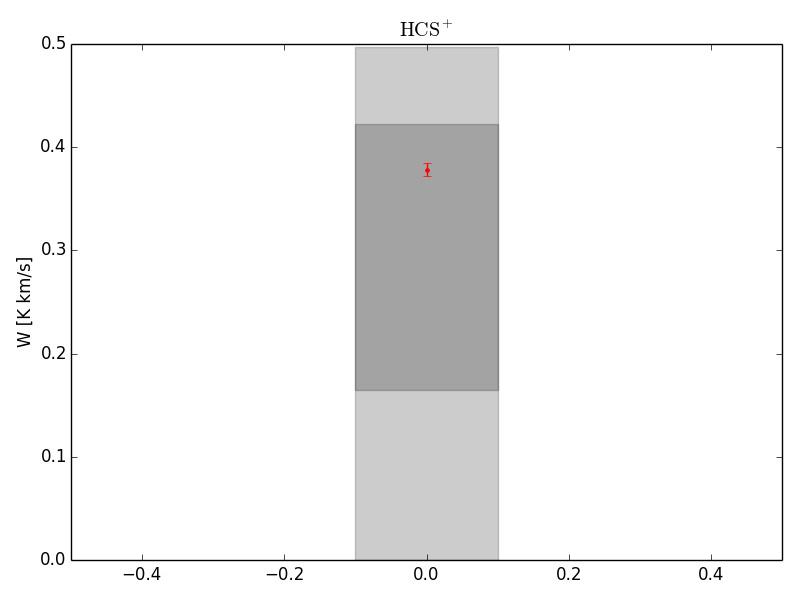}
\includegraphics[width=8cm]{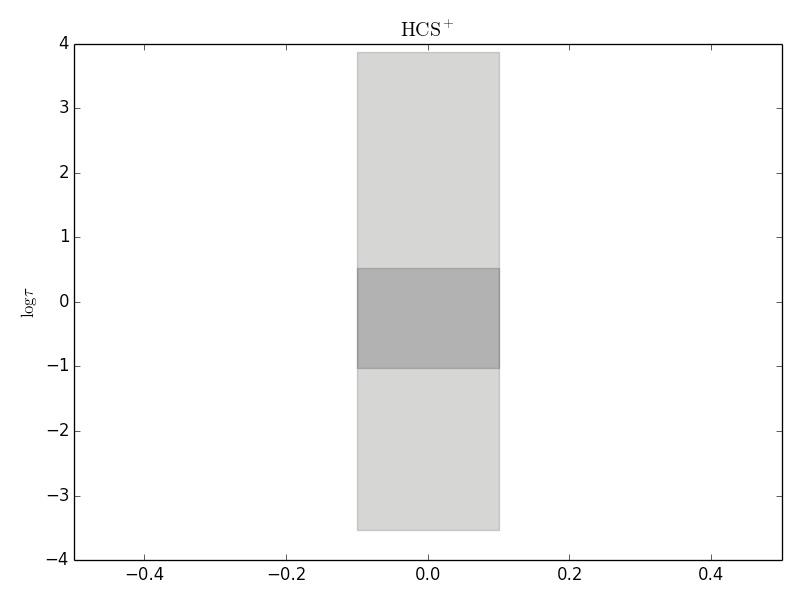}
\caption{\chem{HCS^+}}
\label{HCSp}
\end{figure*}
\clearpage
\begin{figure*}
\includegraphics[width=16cm]{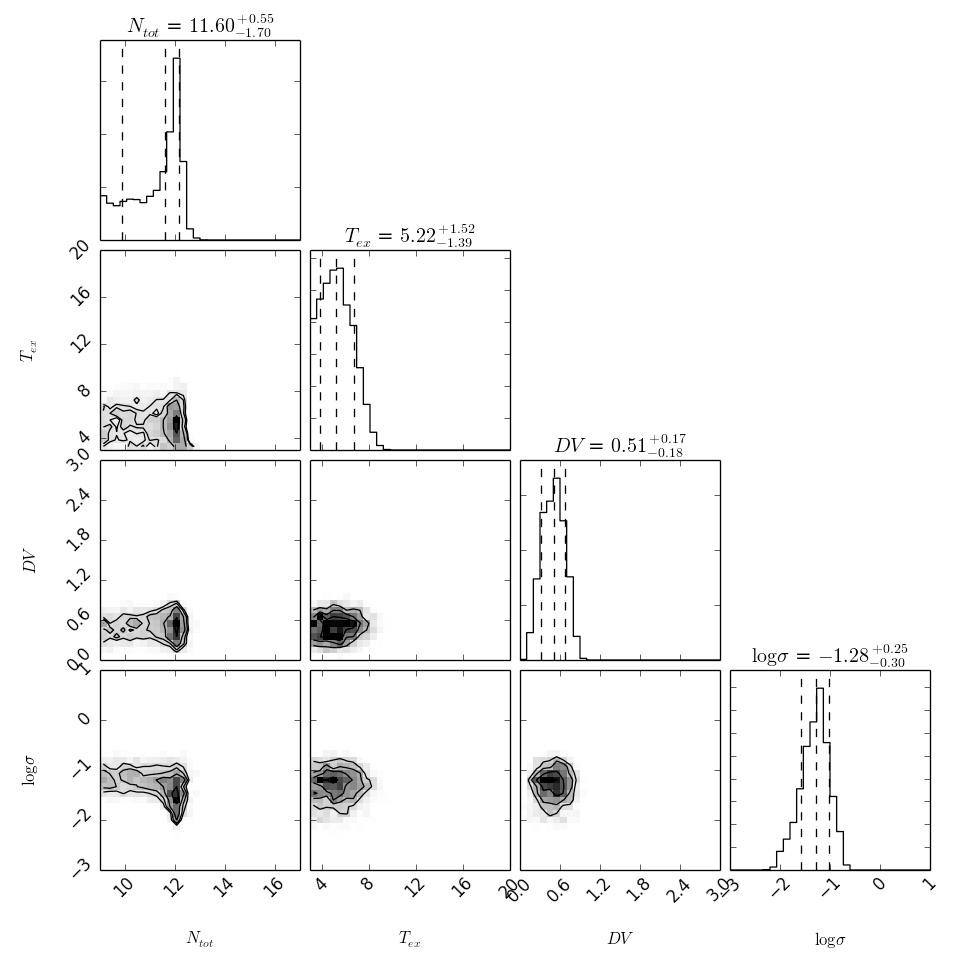}
\includegraphics[width=8cm]{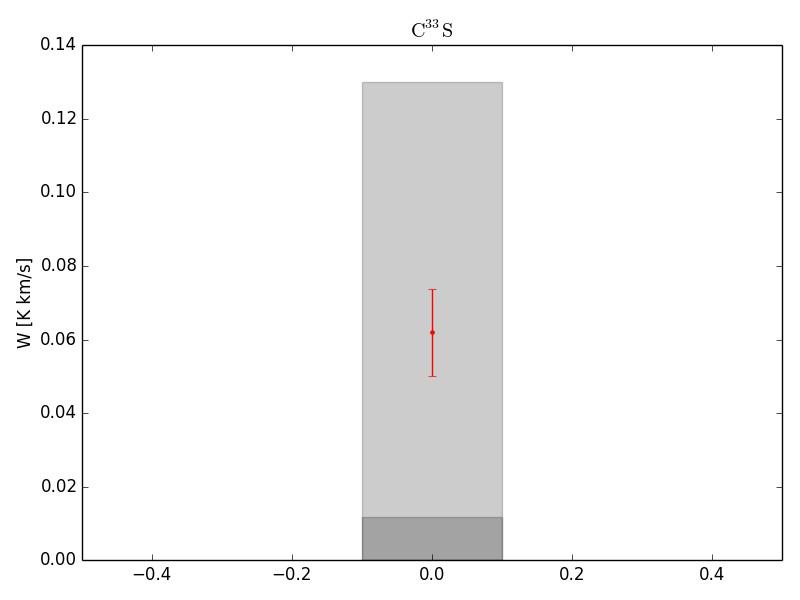}
\includegraphics[width=8cm]{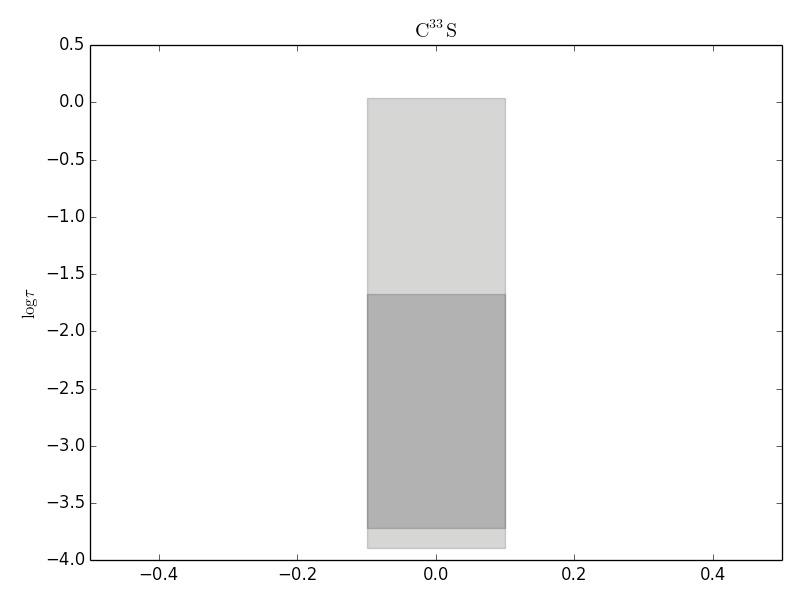}
\caption{\chem{C^{33}S}}
\label{C33S}
\end{figure*}
\clearpage
\begin{figure*}
\includegraphics[width=16cm]{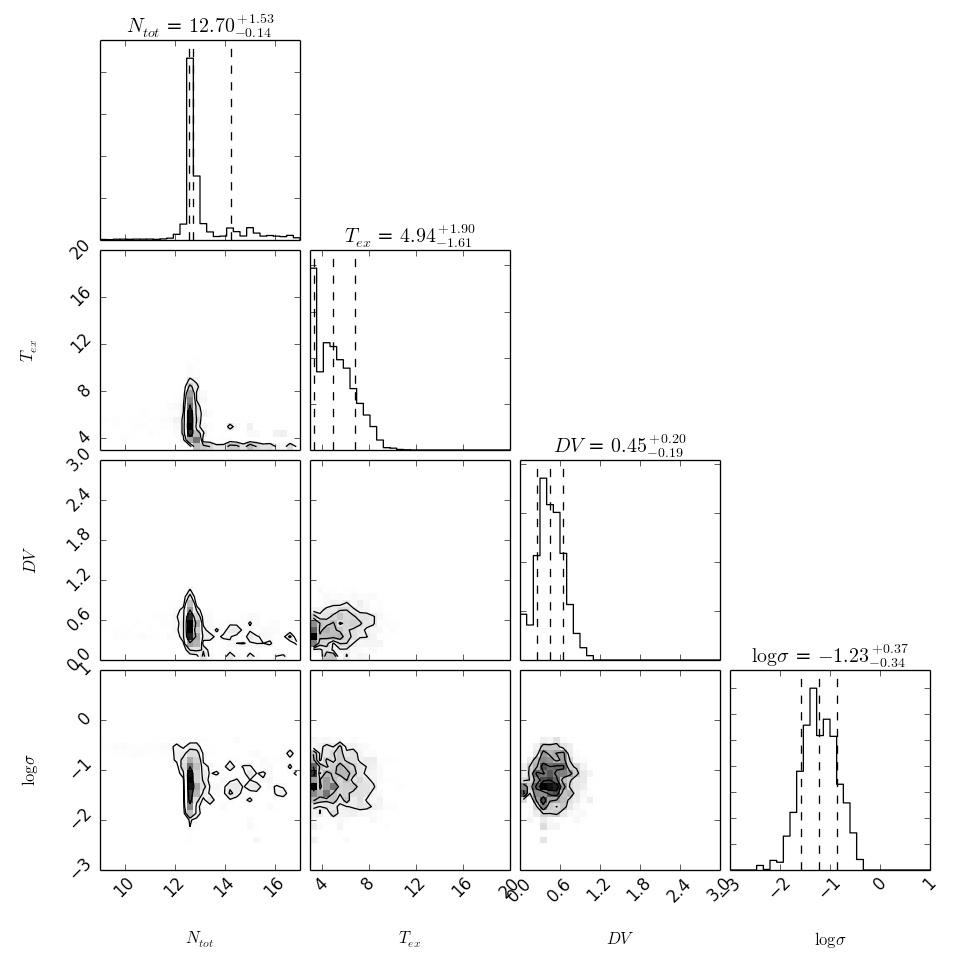}
\includegraphics[width=8cm]{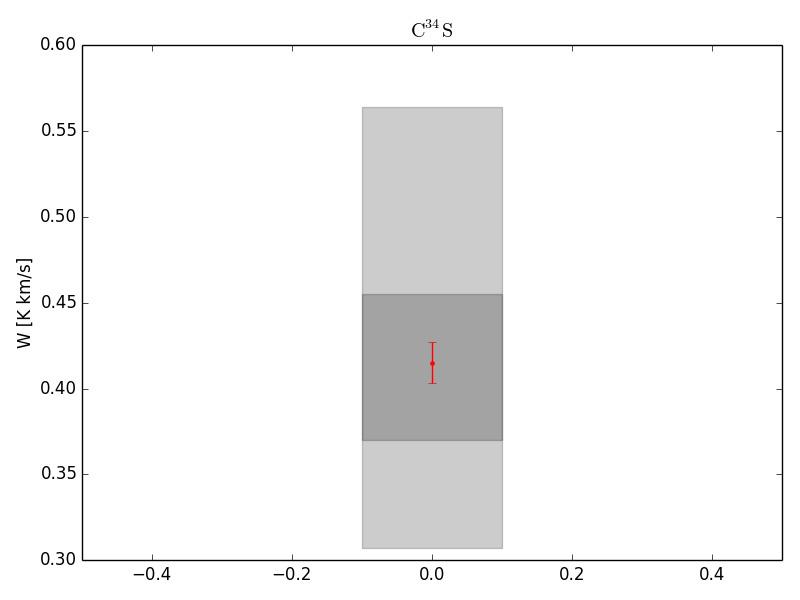}
\includegraphics[width=8cm]{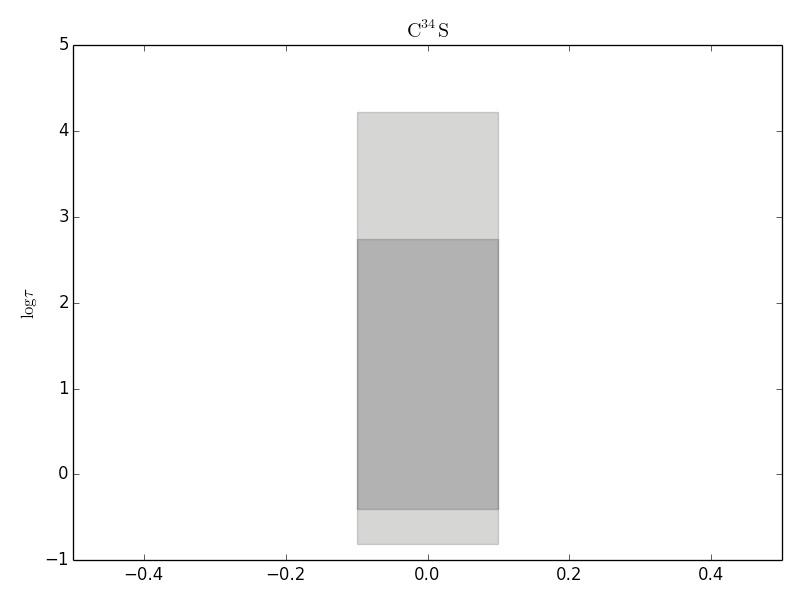}
\caption{\chem{C^{34}S}}
\label{C34S}
\end{figure*}
\clearpage
\begin{figure*}
\includegraphics[width=16cm]{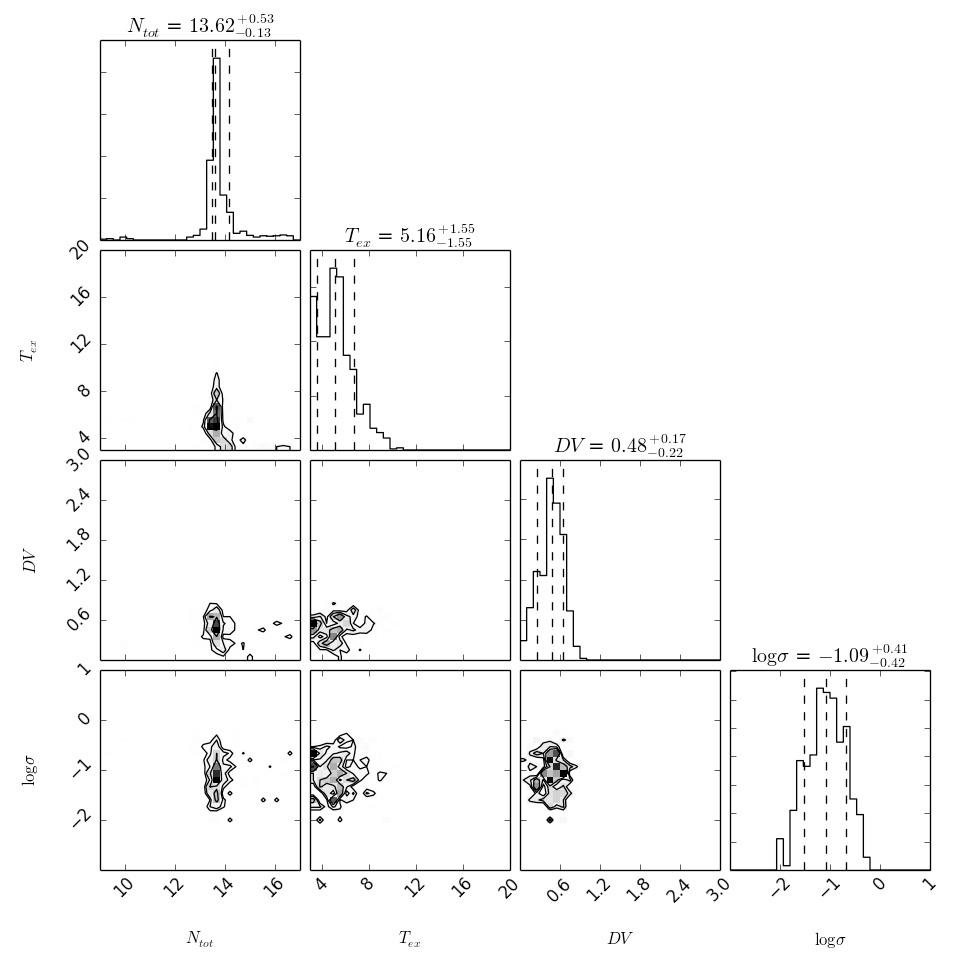}
\includegraphics[width=8cm]{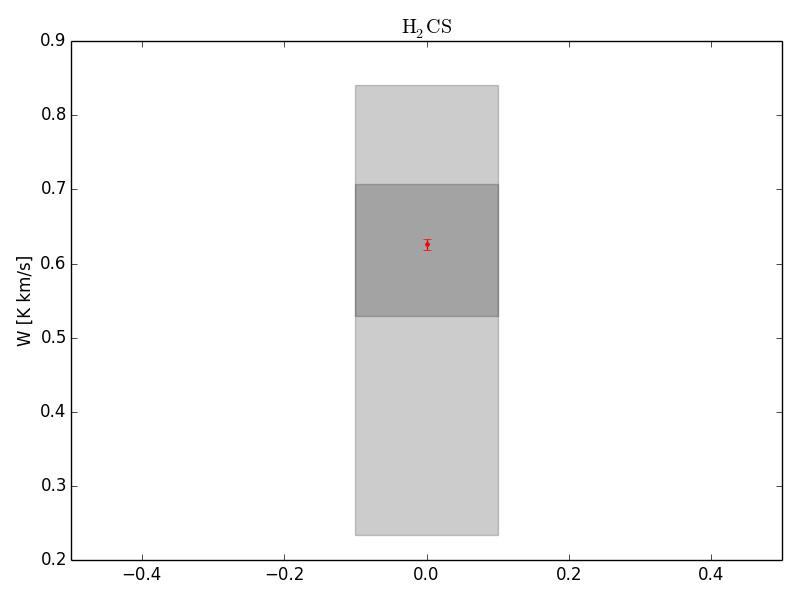}
\includegraphics[width=8cm]{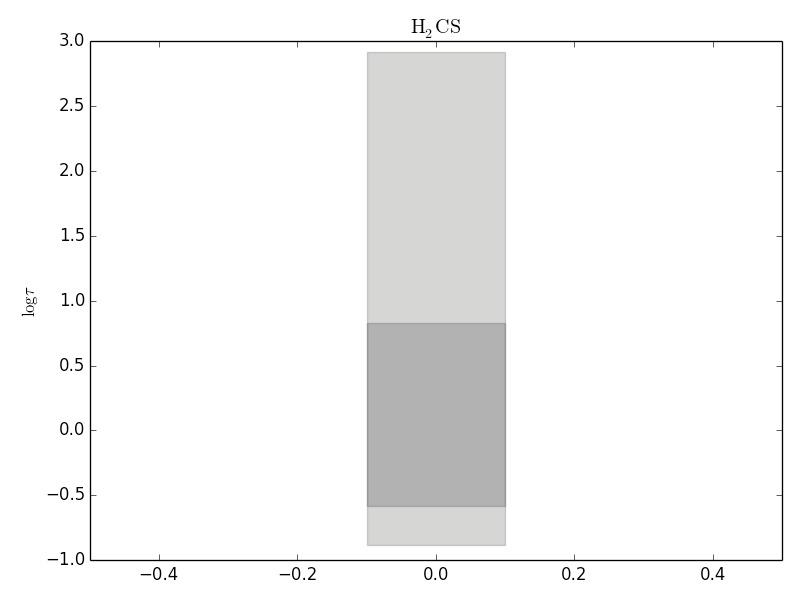}
\caption{\chem{H_2CS}}
\label{H2CS}
\end{figure*}
\clearpage
\begin{figure*}
\includegraphics[width=16cm]{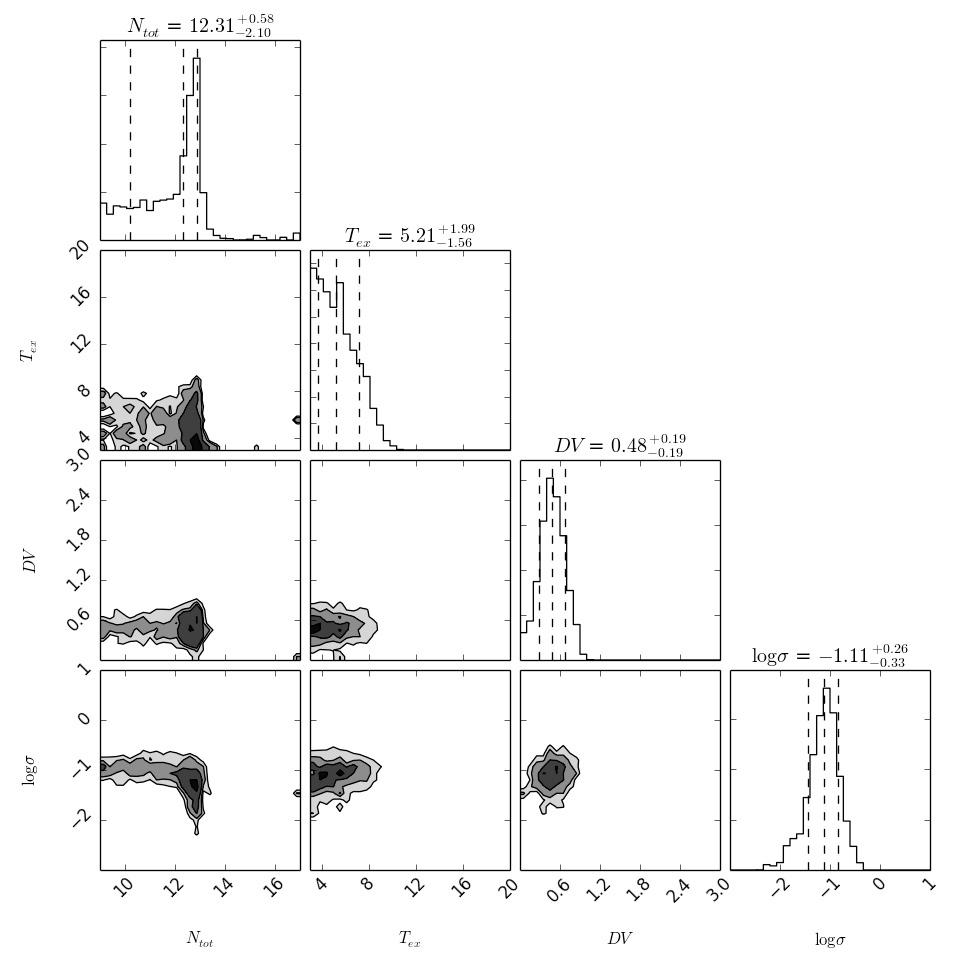}
\includegraphics[width=8cm]{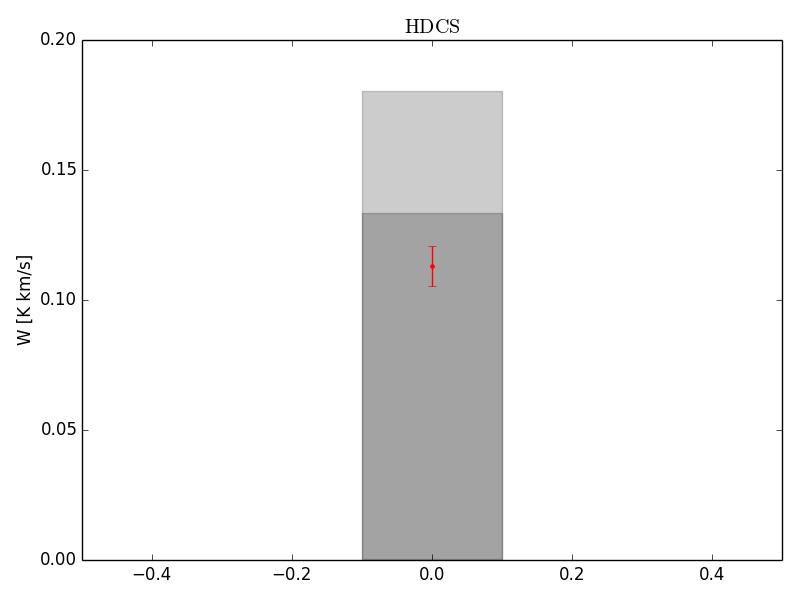}
\includegraphics[width=8cm]{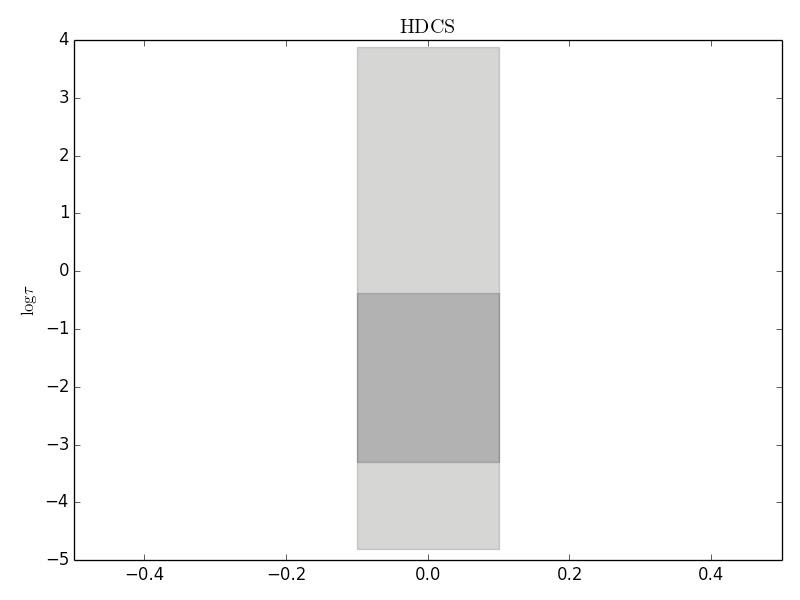}
\caption{\chem{HDCS}}
\label{HDCS}
\end{figure*}
\clearpage
\begin{figure*}
\includegraphics[width=16cm]{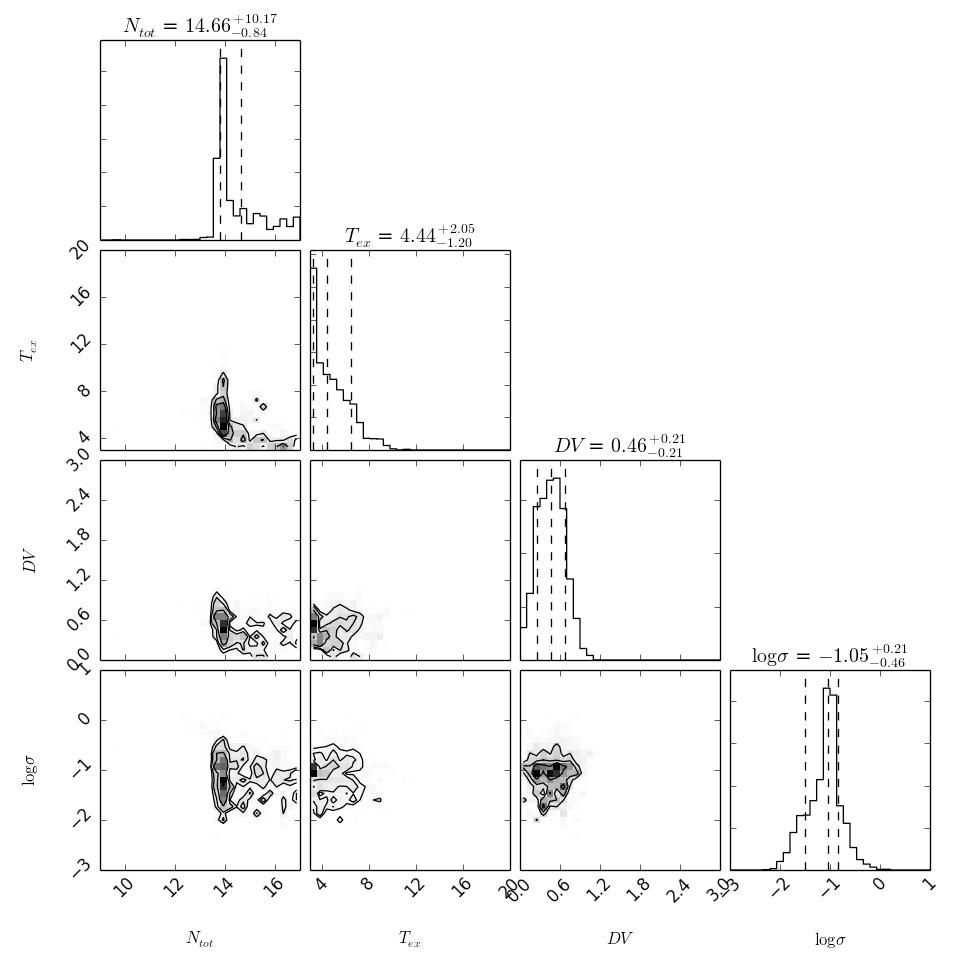}
\includegraphics[width=8cm]{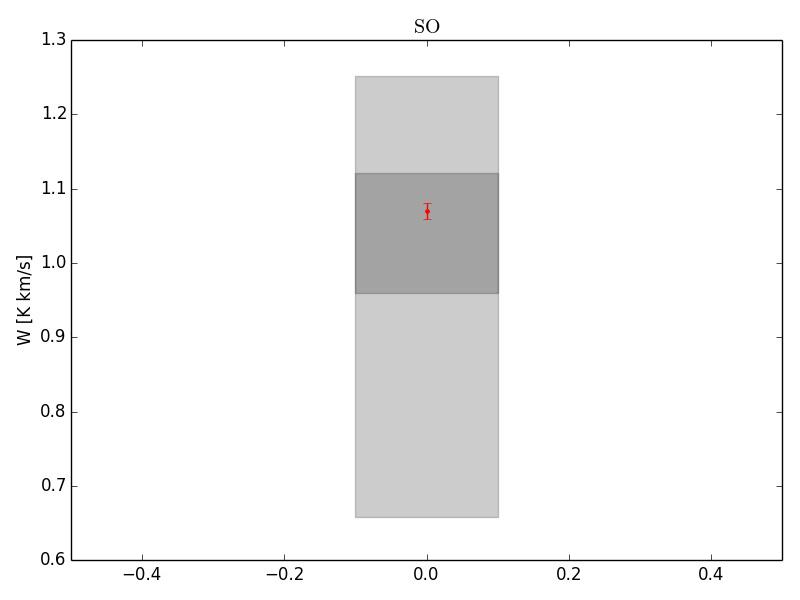}
\includegraphics[width=8cm]{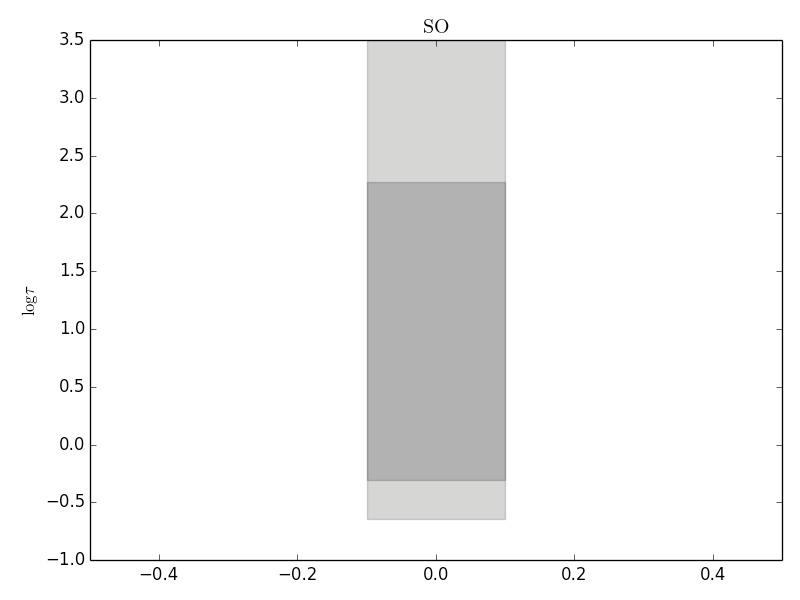}
\caption{\chem{SO}}
\label{SO}
\end{figure*}
\clearpage
\begin{figure*}
\includegraphics[width=16cm]{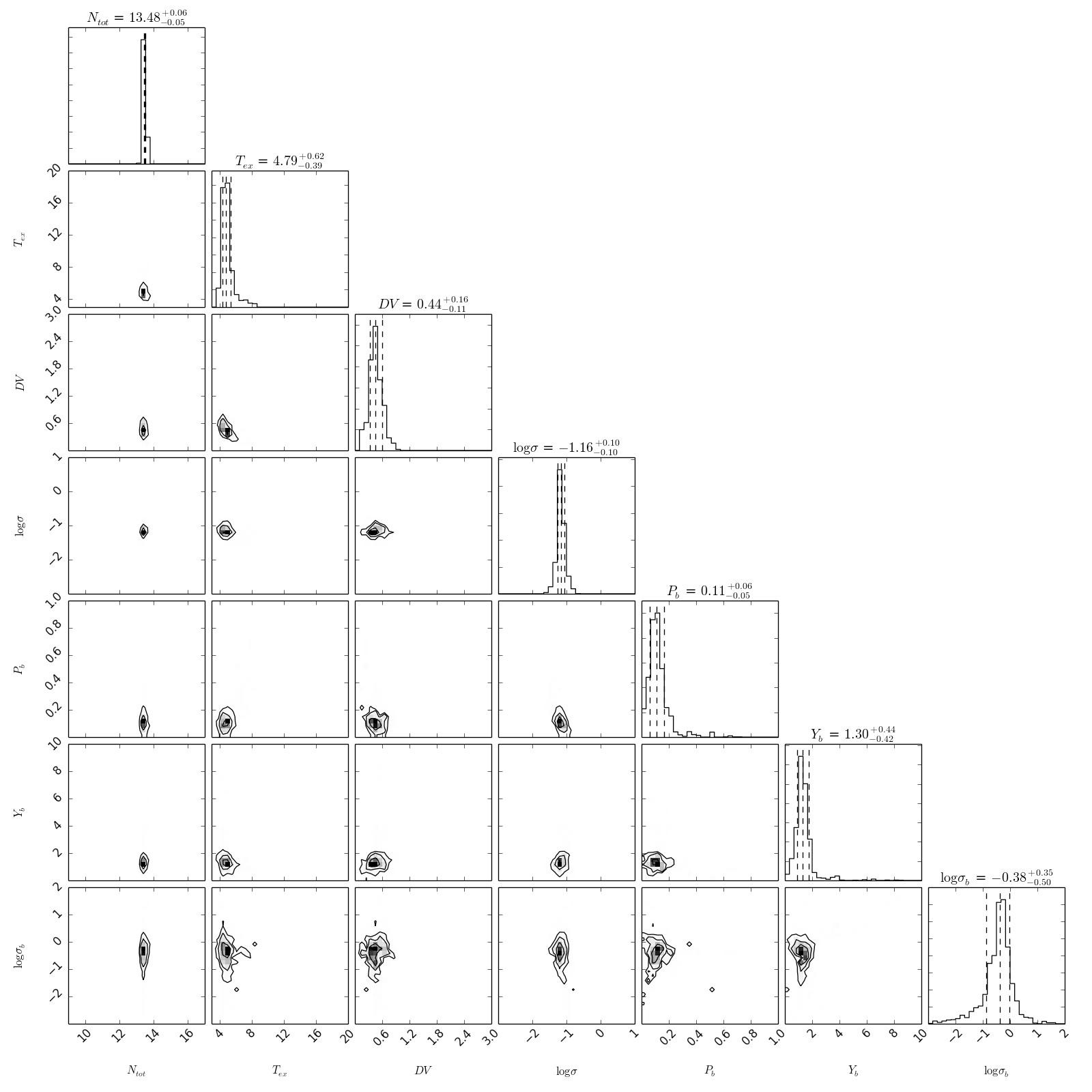}
\includegraphics[width=8cm]{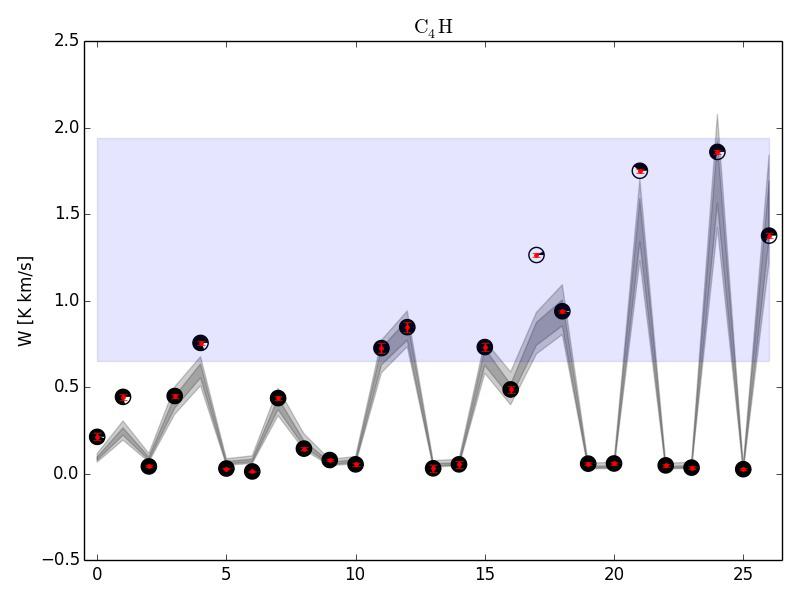}
\includegraphics[width=8cm]{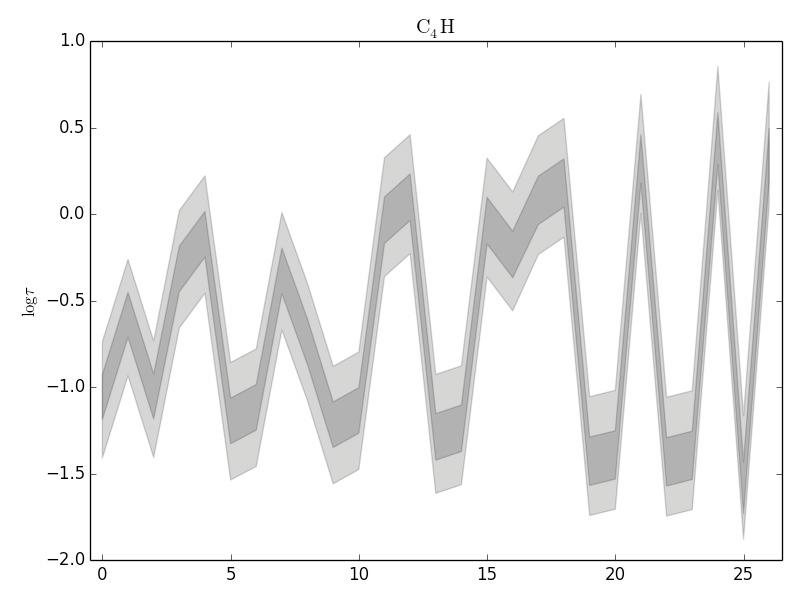}
\caption{\chem{C_4H}}
\label{C4H}
\end{figure*}
\clearpage
\begin{figure*}
\includegraphics[width=16cm]{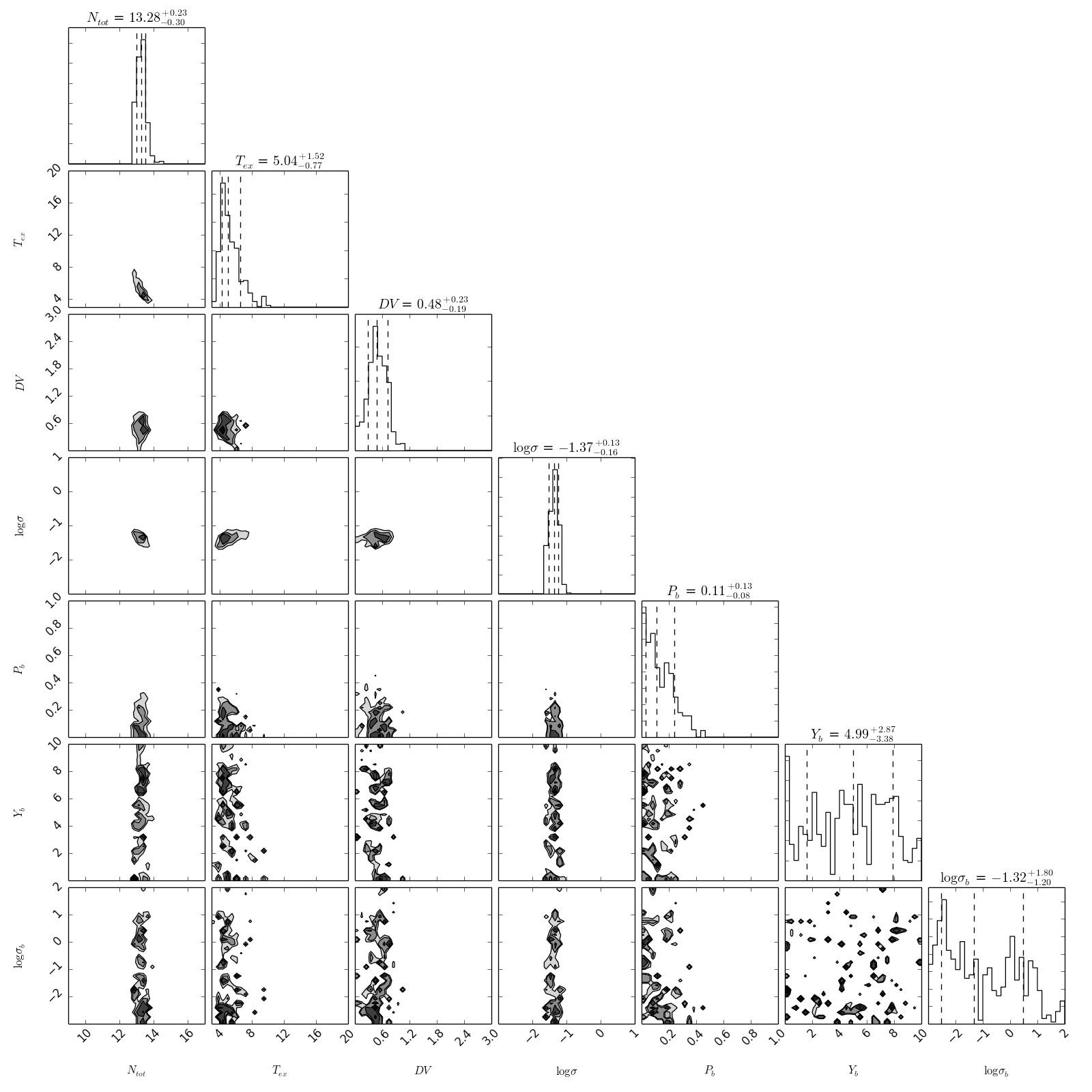}
\includegraphics[width=8cm]{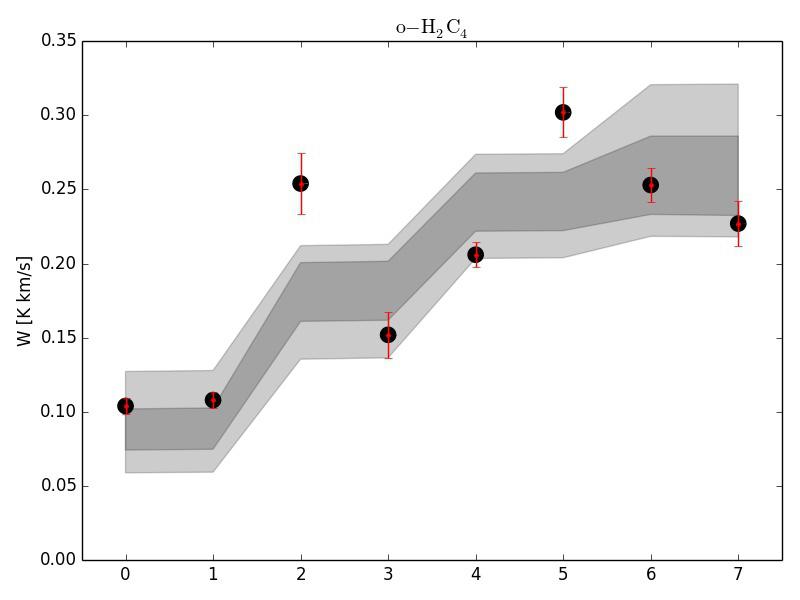}
\includegraphics[width=8cm]{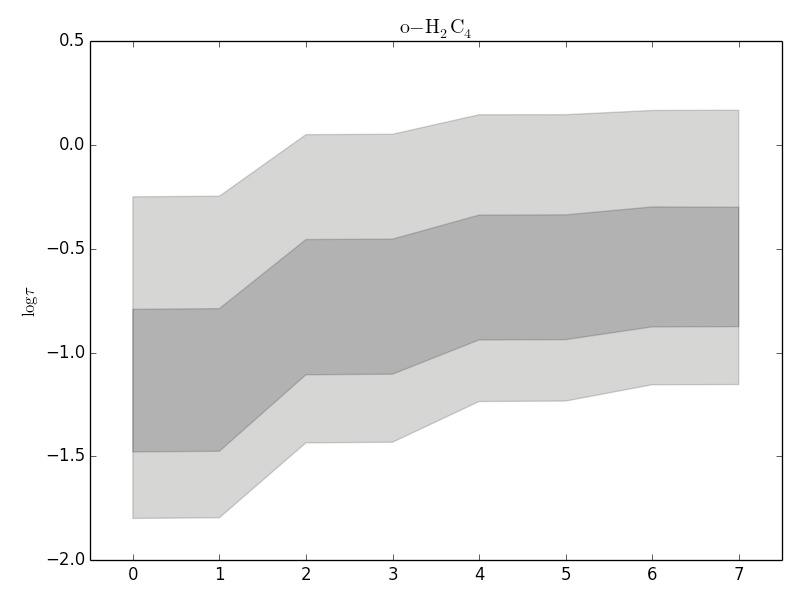}
\caption{\mbox{o-}\chem{C_4H_2}}
\label{o-H2C4}
\end{figure*}
\clearpage
\begin{figure*}
\includegraphics[width=16cm]{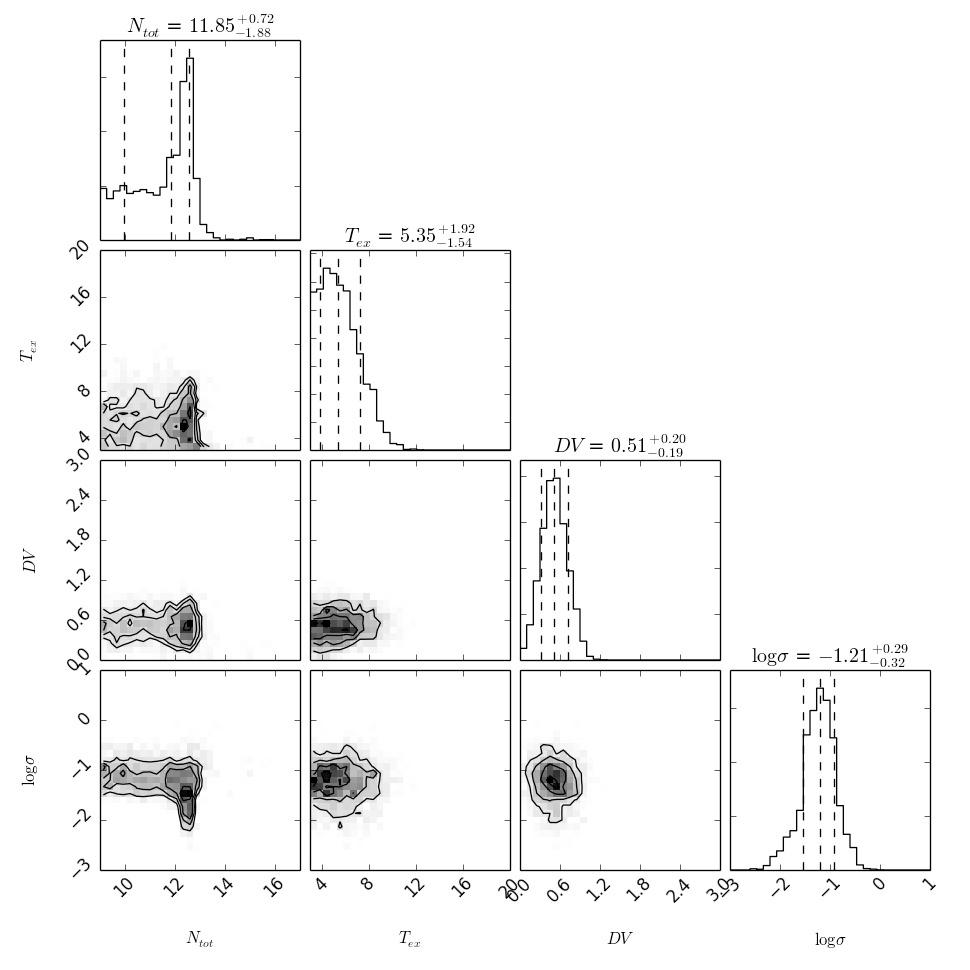}
\includegraphics[width=8cm]{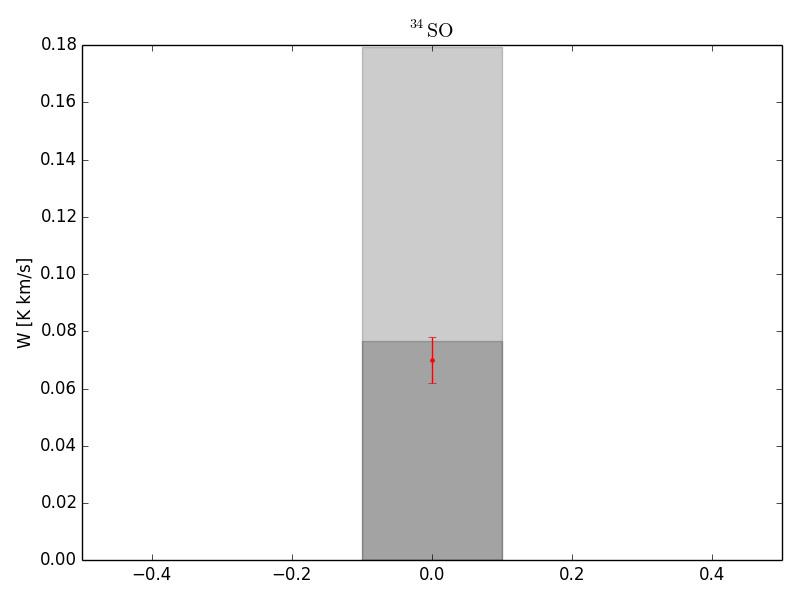}
\includegraphics[width=8cm]{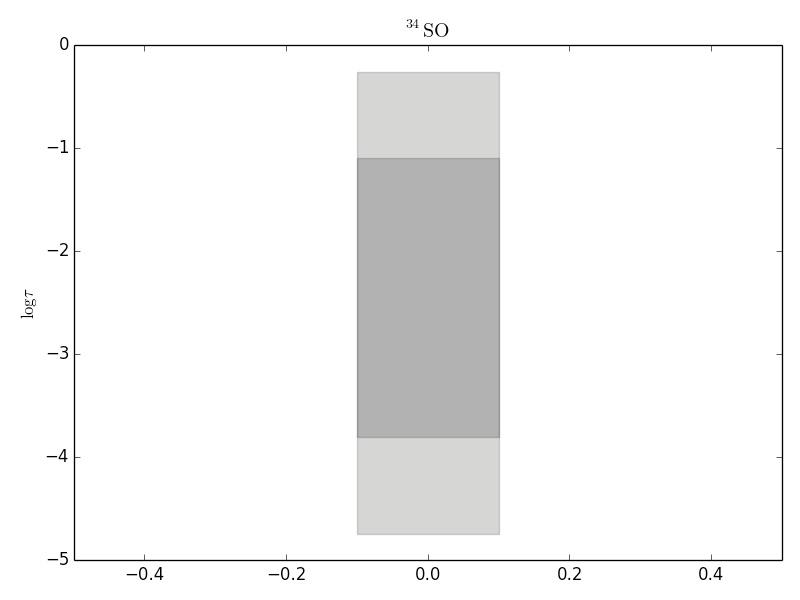}
\caption{\chem{^{34}SO}}
\label{34SO}
\end{figure*}
\clearpage
\begin{figure*}
\includegraphics[width=16cm]{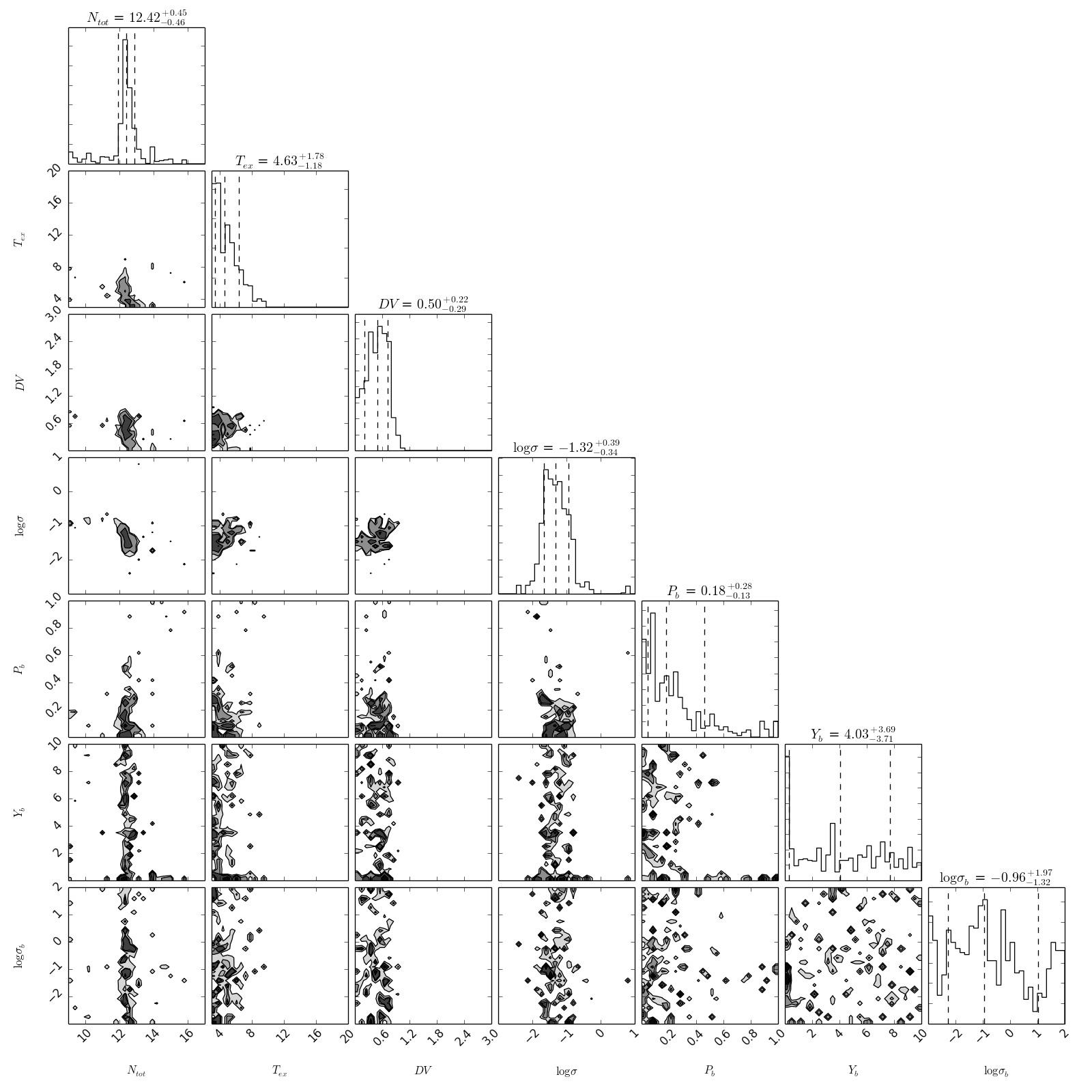}
\includegraphics[width=8cm]{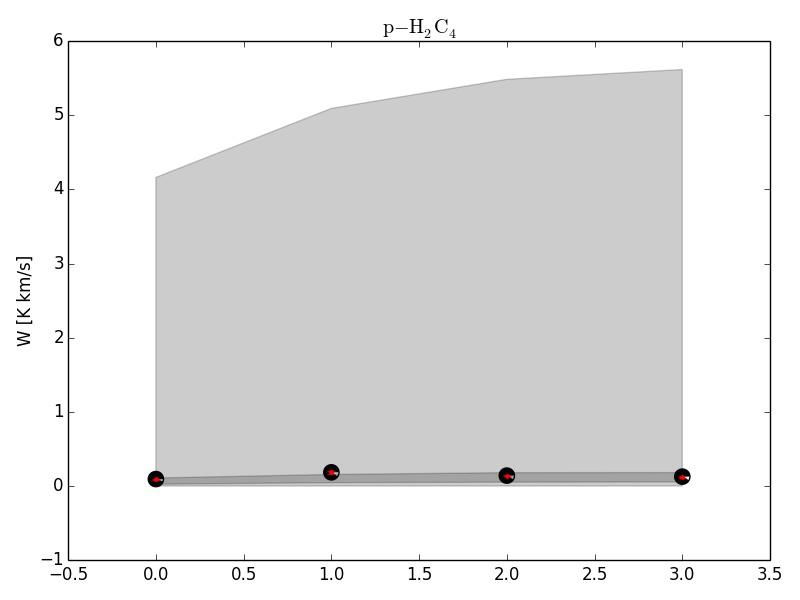}
\includegraphics[width=8cm]{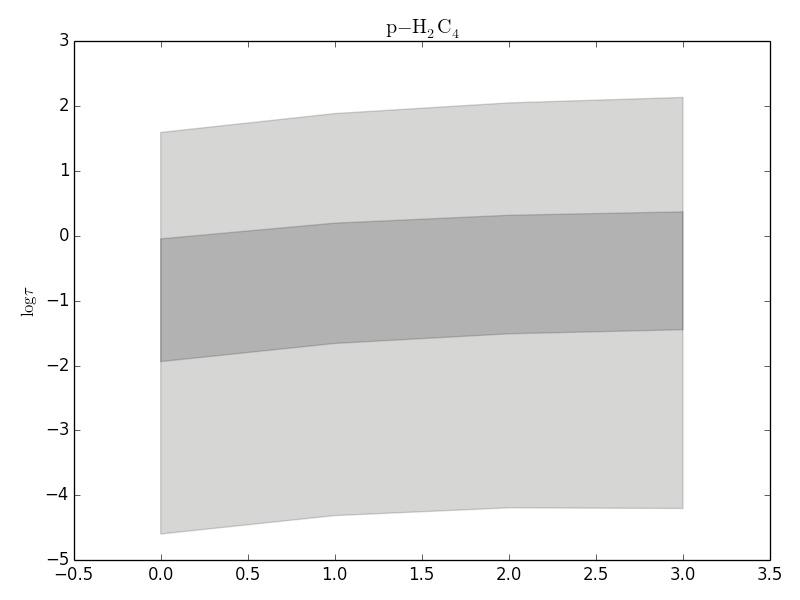}
\caption{\mbox{p-}\chem{C_4H_2}}
\label{p-H2C4}
\end{figure*}
\clearpage
\clearpage
\begin{figure*}
\includegraphics[width=16cm]{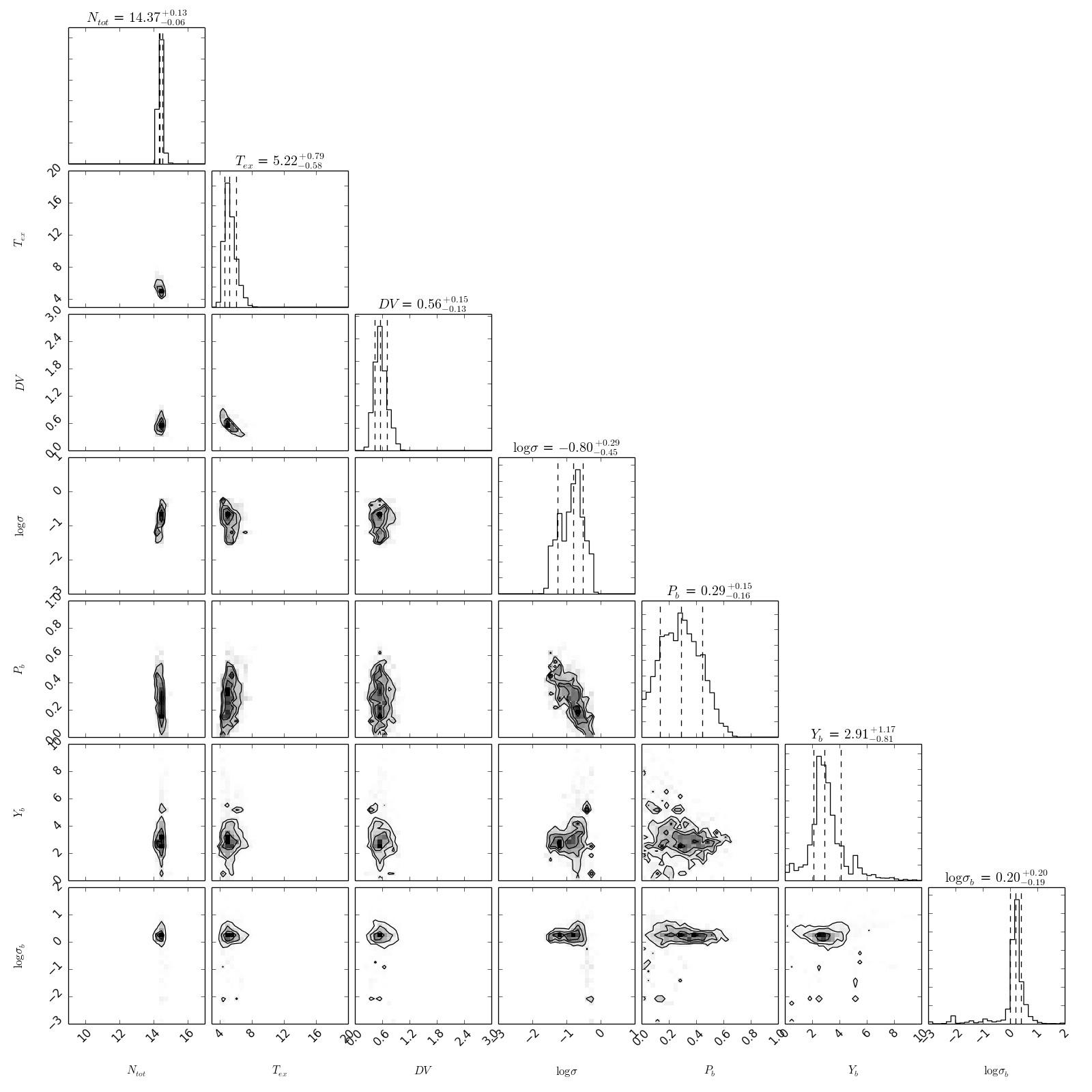}
\includegraphics[width=8cm]{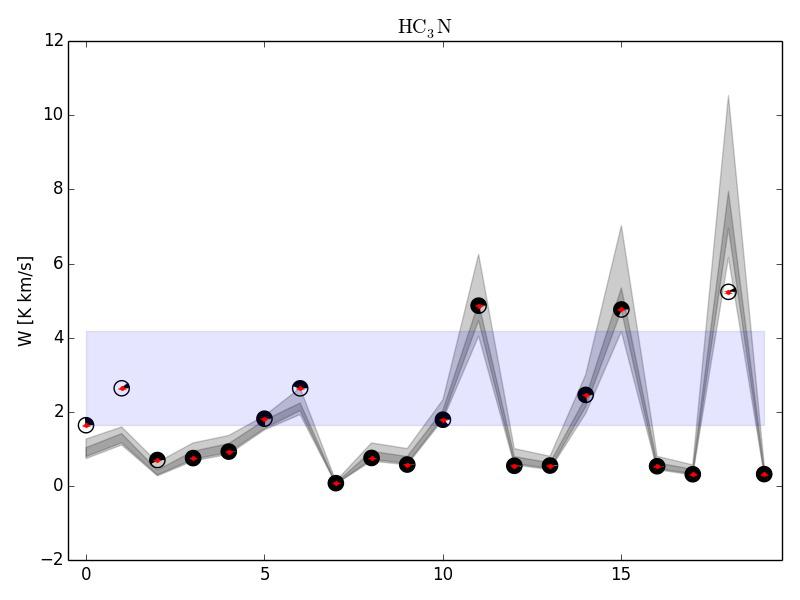}
\includegraphics[width=8cm]{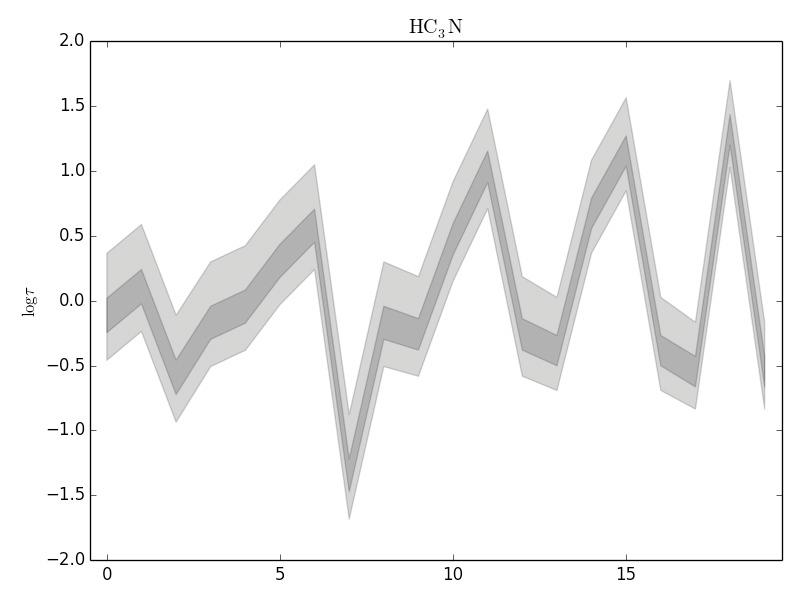}
\caption{\chem{HC_3N}}
\label{HC3N}
\end{figure*}
\clearpage
\begin{figure*}
\includegraphics[width=16cm]{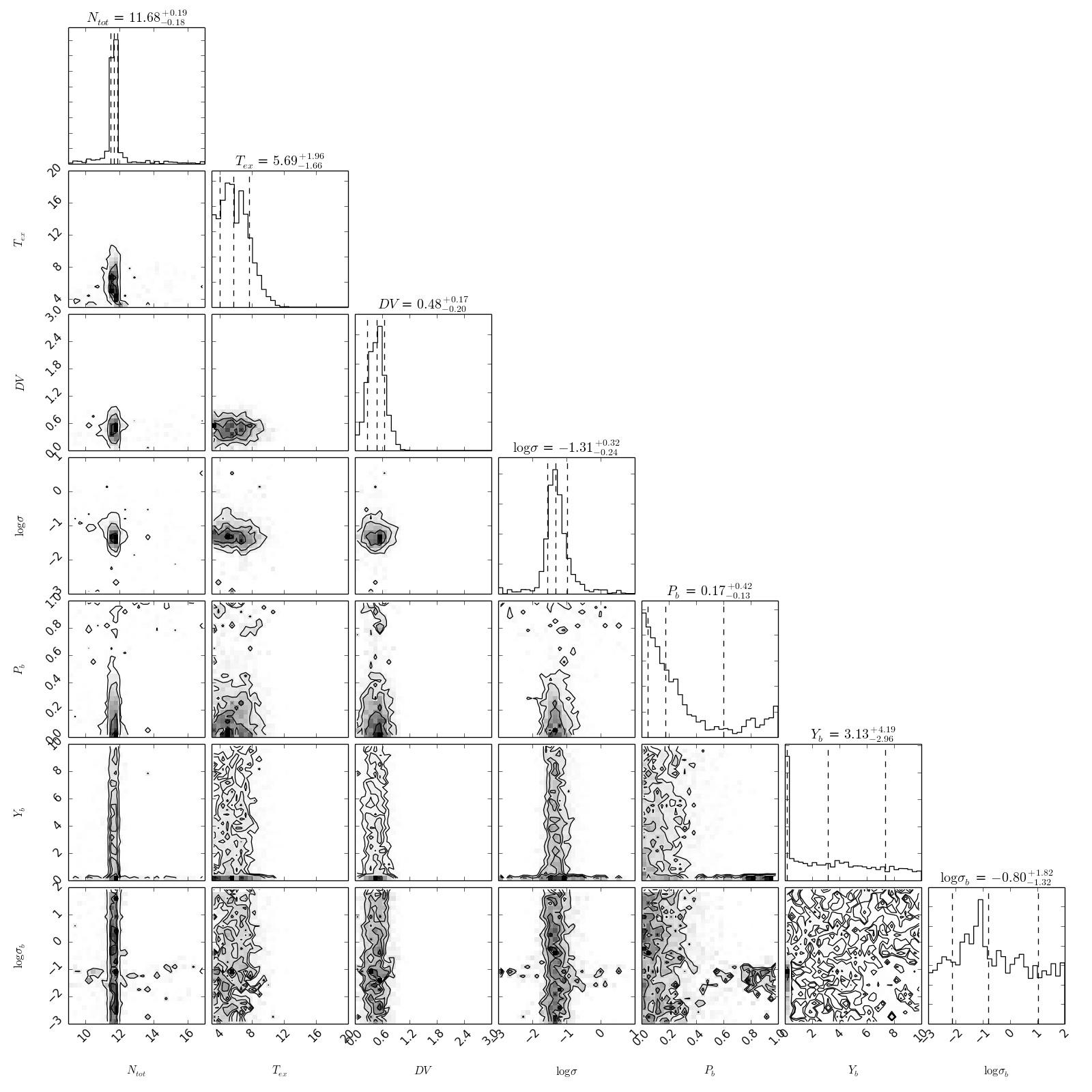}
\includegraphics[width=8cm]{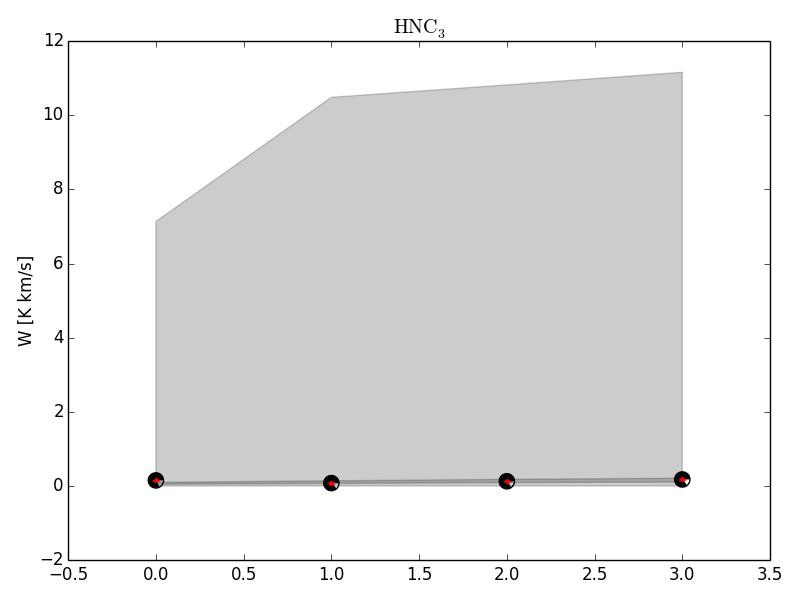}
\includegraphics[width=8cm]{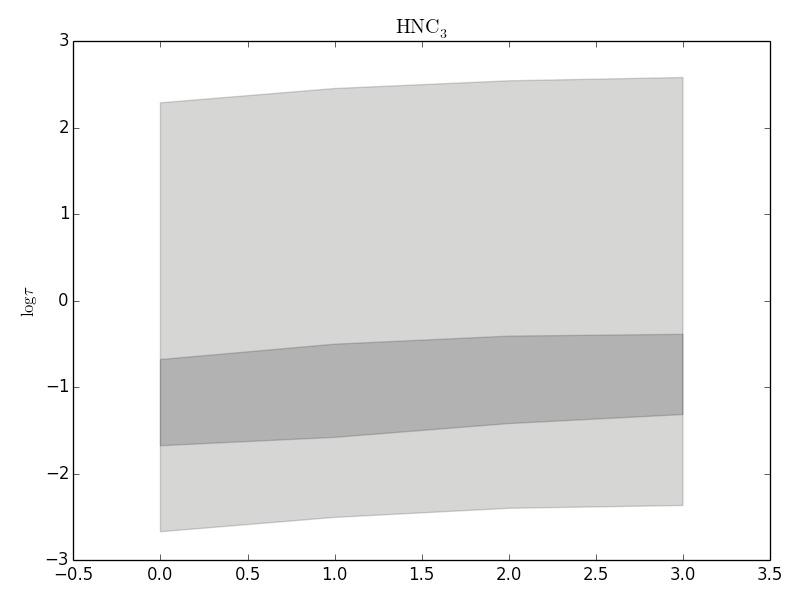}
\caption{\chem{HNC_3}}
\label{HNCCC}
\end{figure*}
\clearpage
\begin{figure*}
\includegraphics[width=16cm]{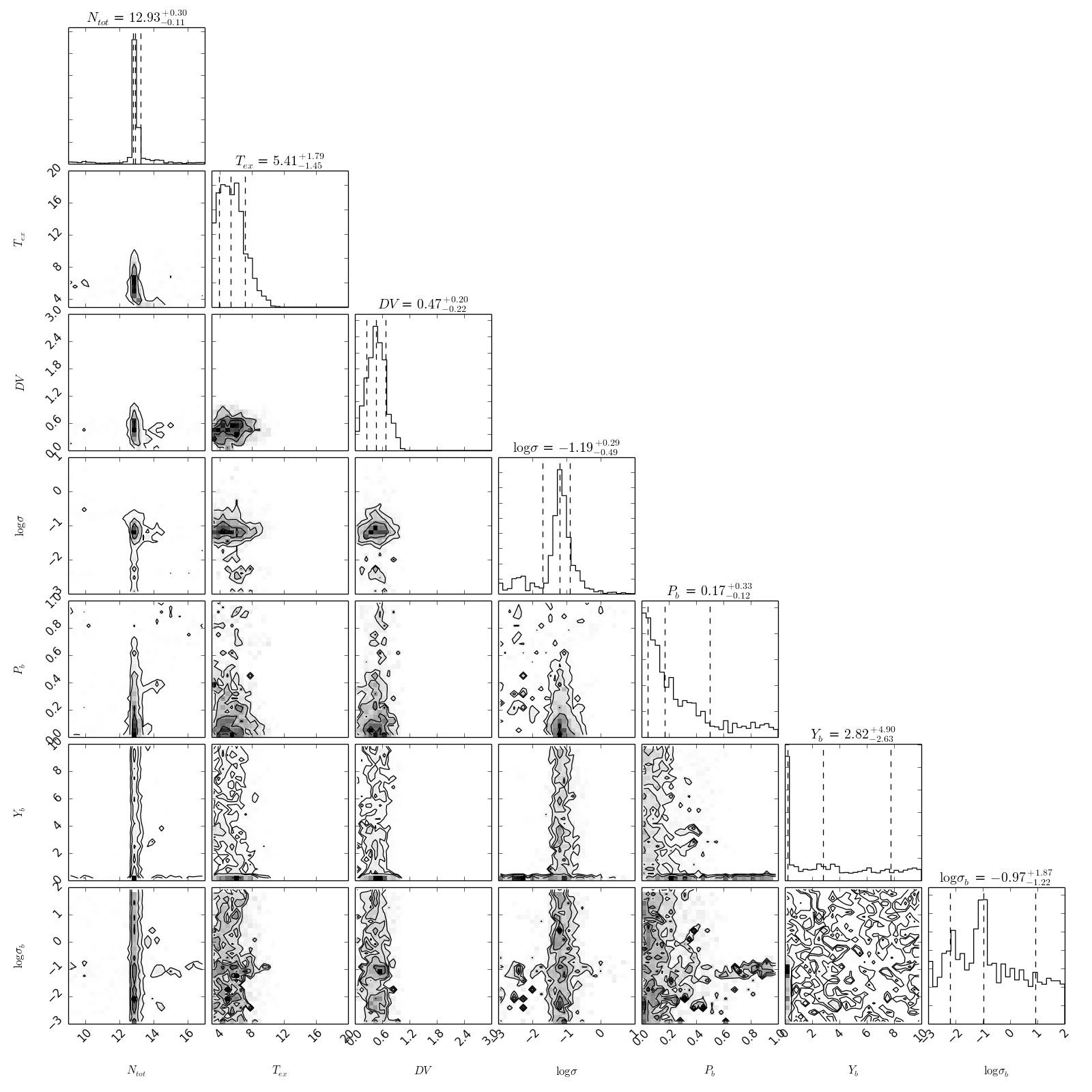}
\includegraphics[width=8cm]{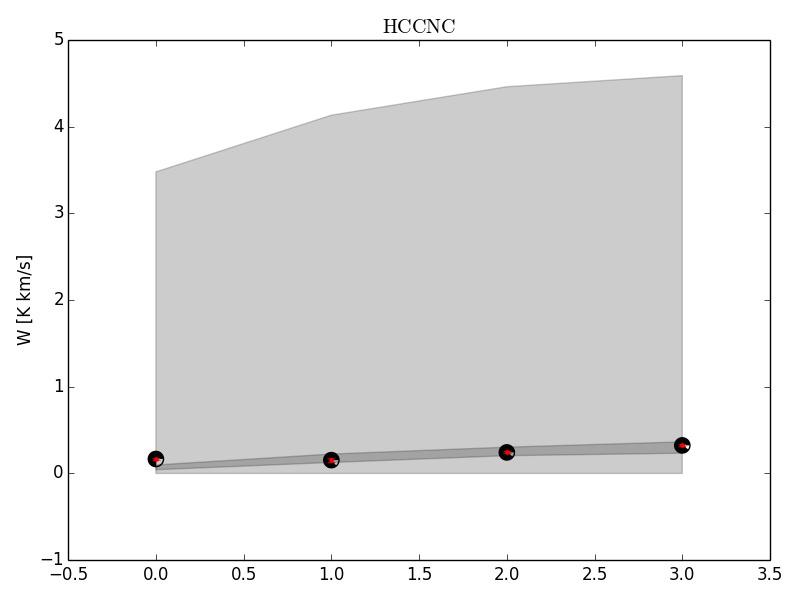}
\includegraphics[width=8cm]{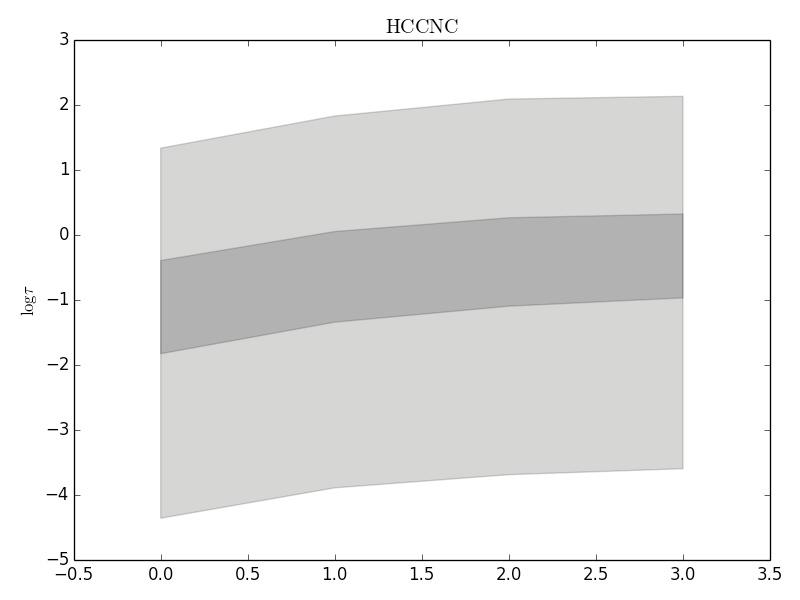}
\caption{\chem{HCCNC}}
\label{HCCNC}
\end{figure*}
\clearpage
\begin{figure*}
\includegraphics[width=16cm]{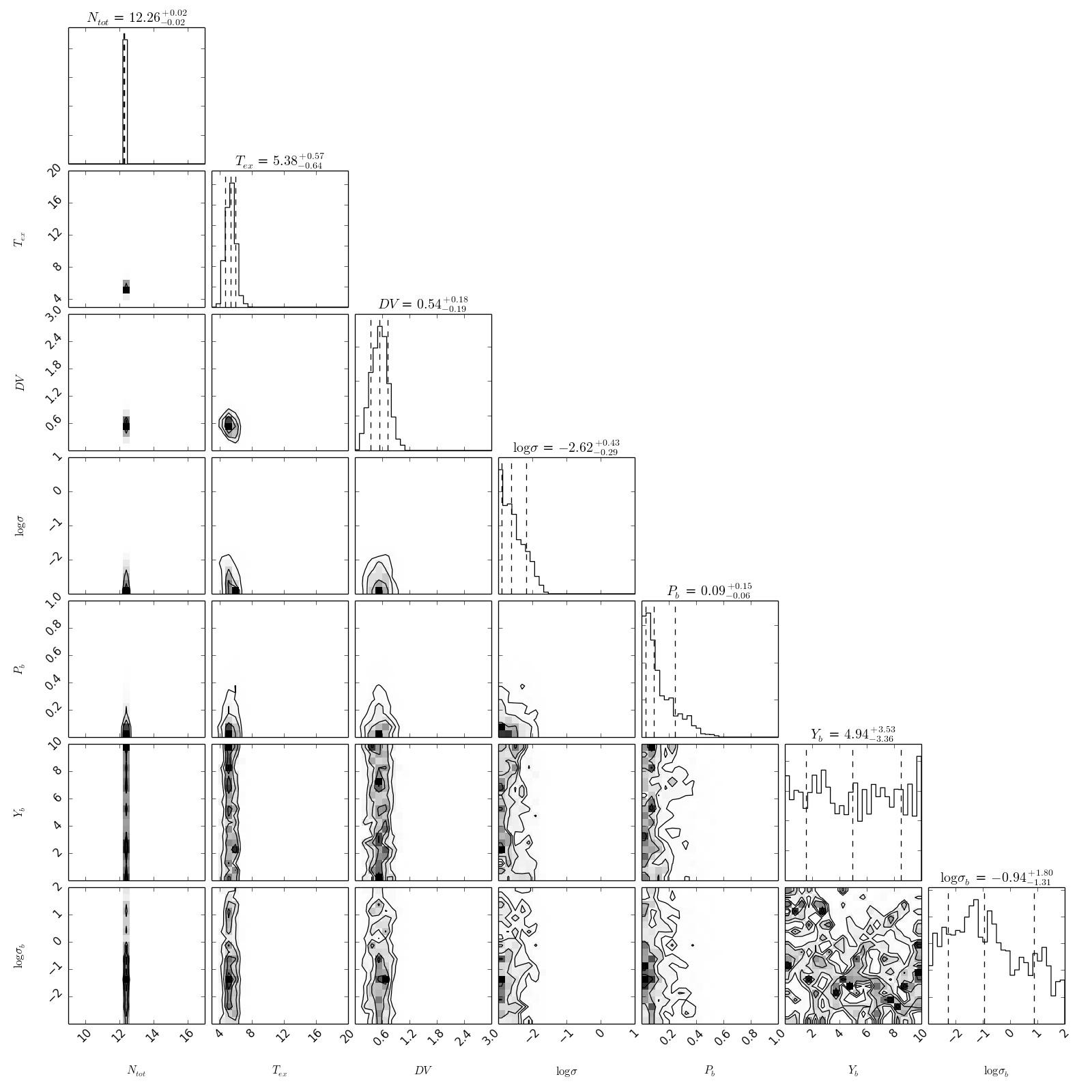}
\includegraphics[width=8cm]{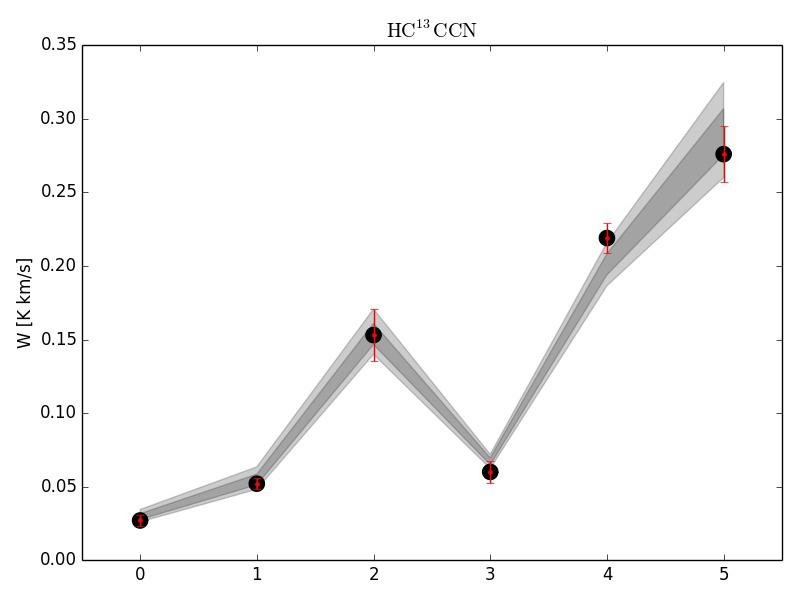}
\includegraphics[width=8cm]{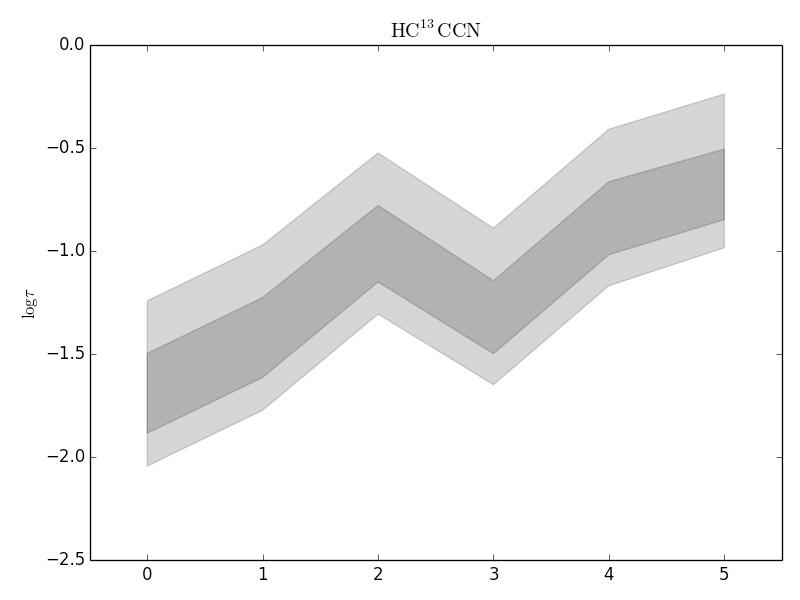}
\caption{\chem{HC^{13}CCN}}
\label{HC13CCN}
\end{figure*}
\clearpage
\begin{figure*}
\includegraphics[width=16cm]{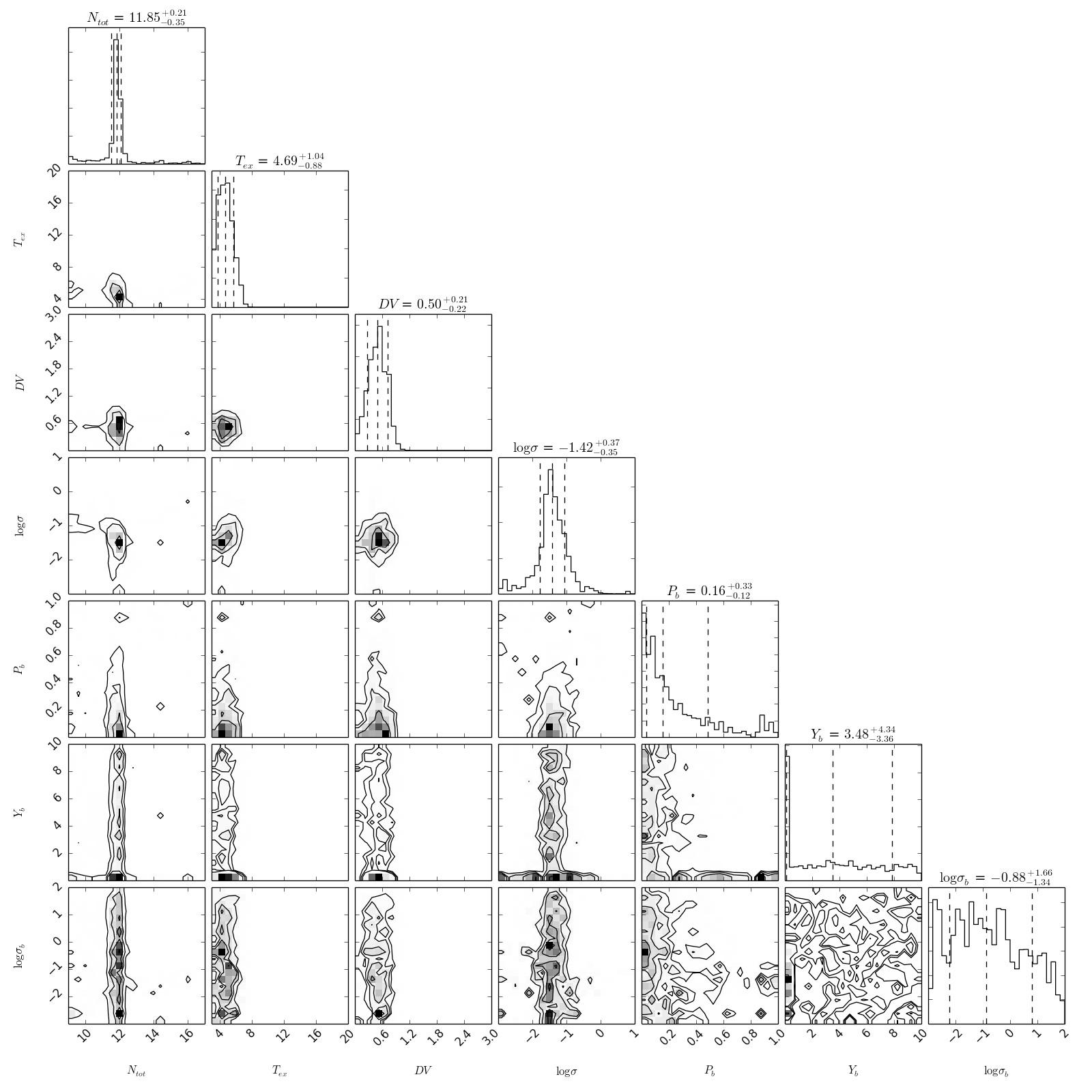}
\includegraphics[width=8cm]{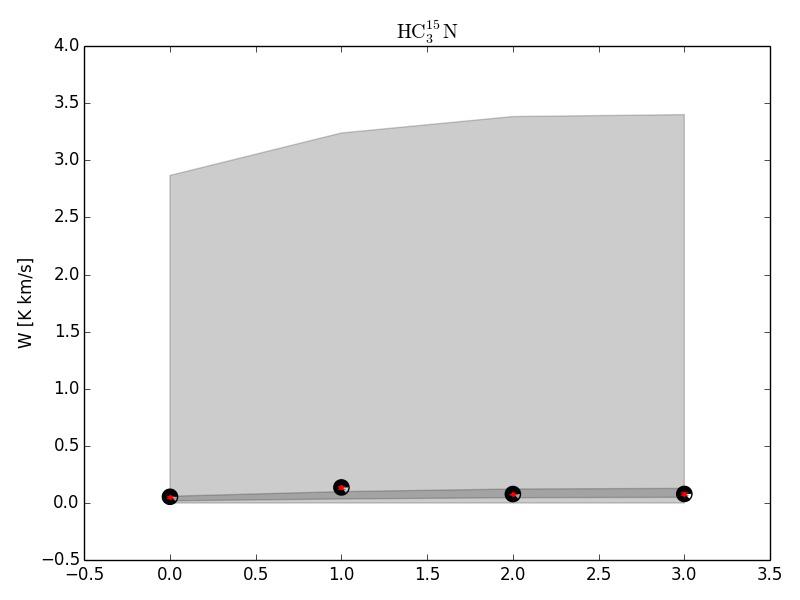}
\includegraphics[width=8cm]{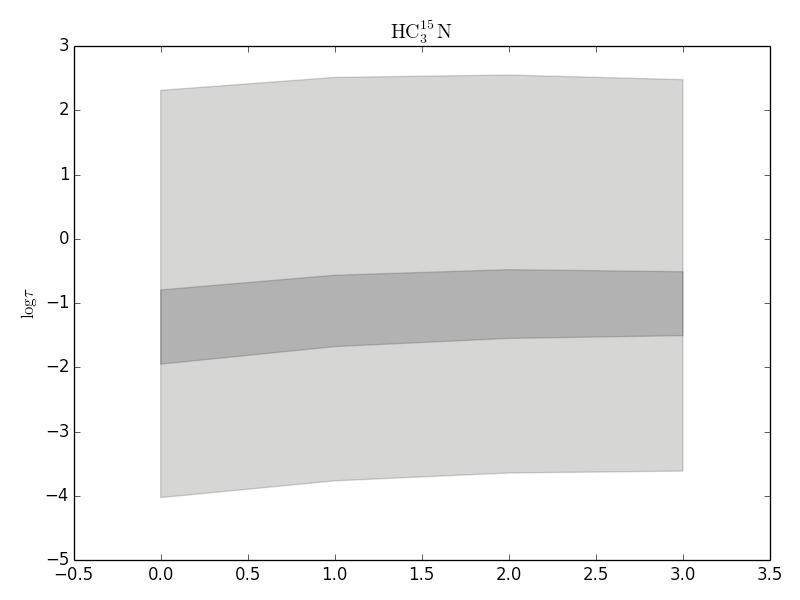}
\caption{\chem{HC_3^{15}N}}
\label{HC315N}
\end{figure*}
\clearpage
\begin{figure*}
\includegraphics[width=16cm]{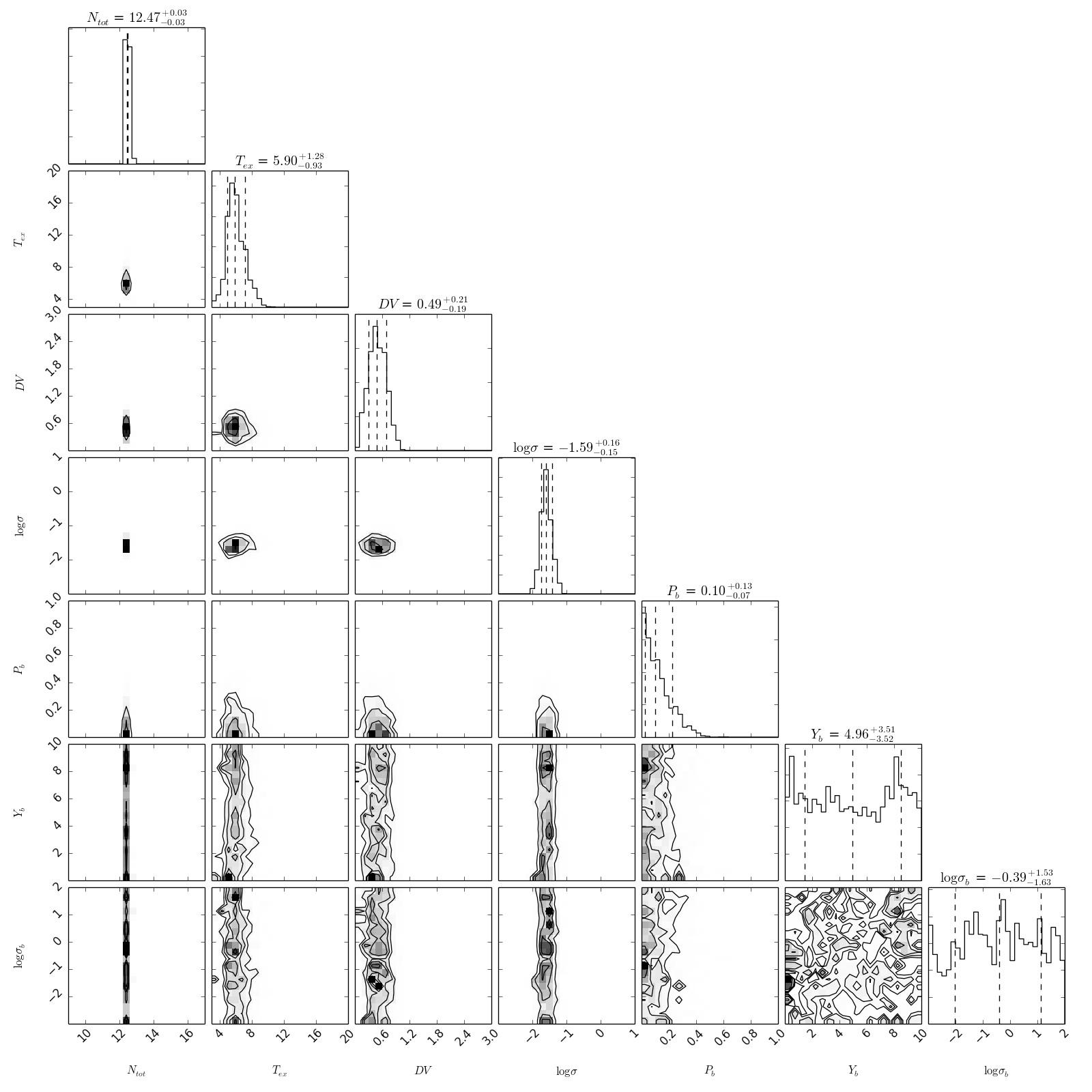}
\includegraphics[width=8cm]{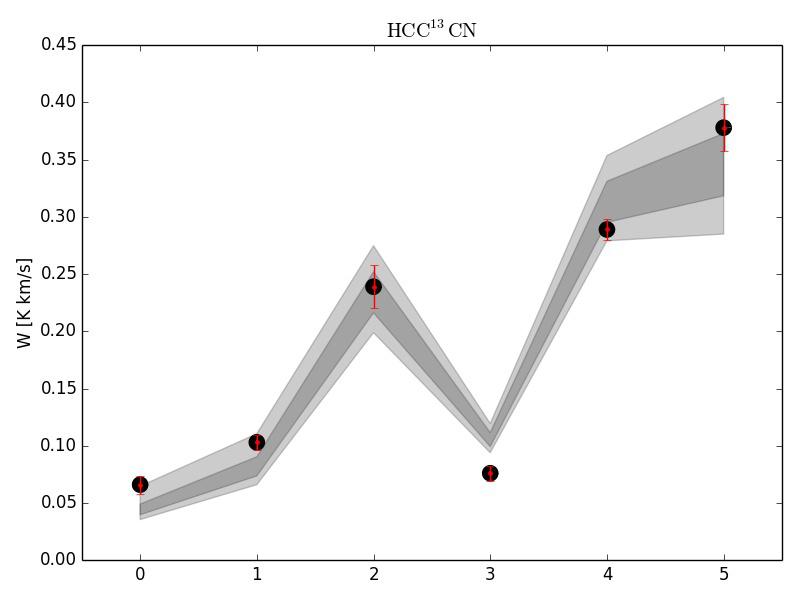}
\includegraphics[width=8cm]{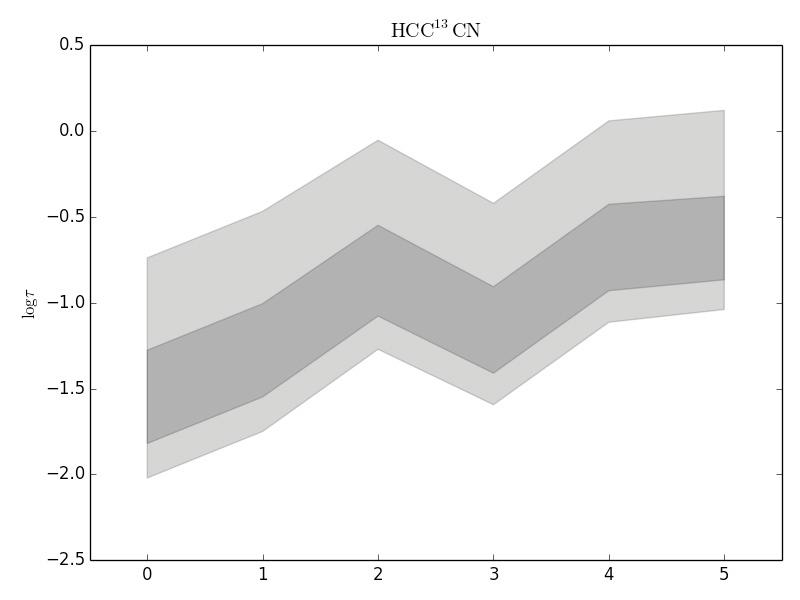}
\caption{\chem{HC_2^{13}CN}}
\label{HCC13CN}
\end{figure*}
\clearpage
\begin{figure*}
\includegraphics[width=16cm]{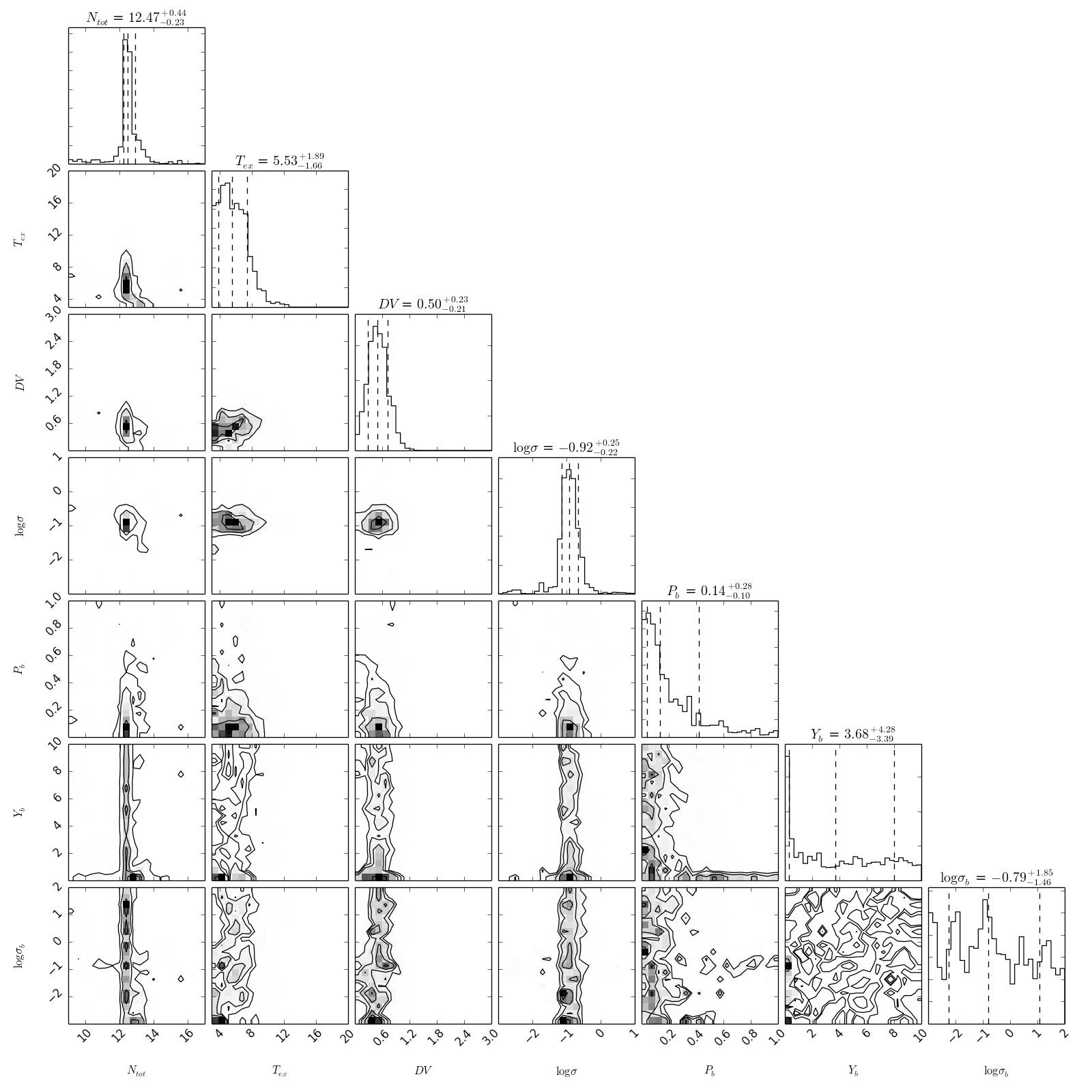}
\includegraphics[width=8cm]{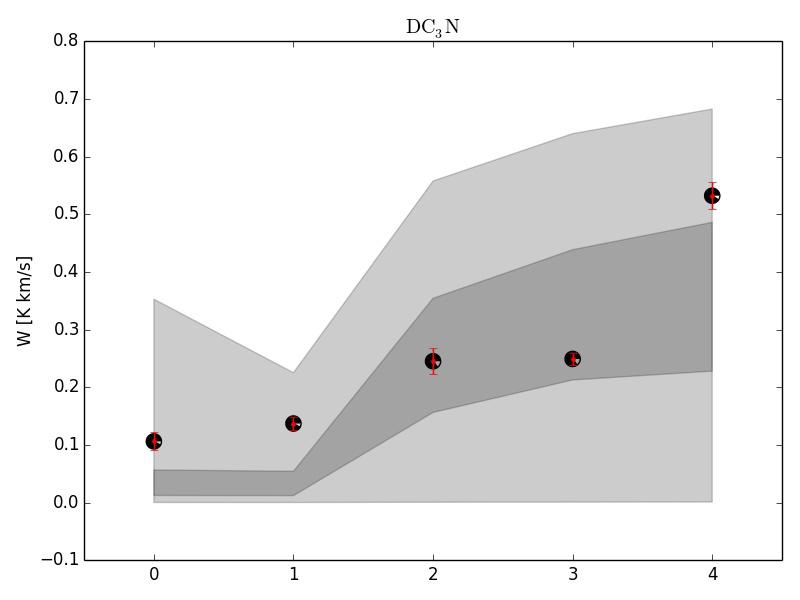}
\includegraphics[width=8cm]{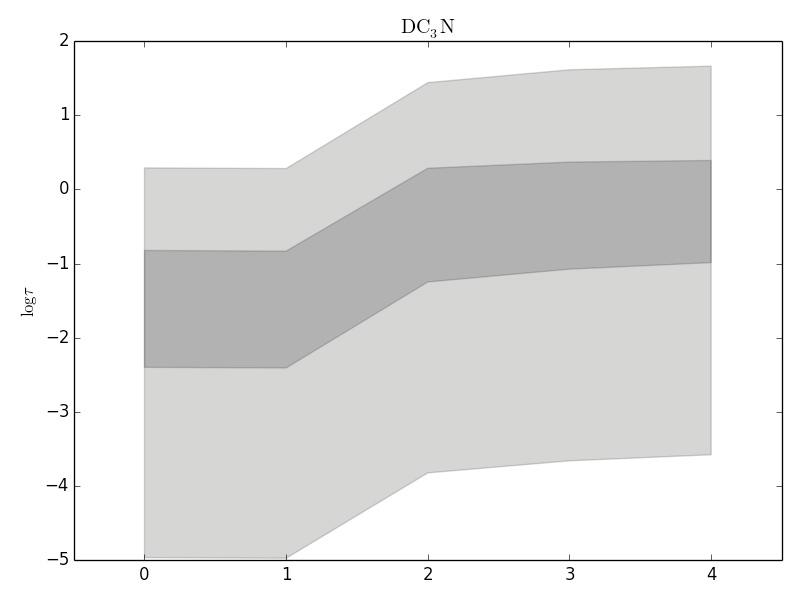}
\caption{\chem{DC_3N}}
\label{DC3N}
\end{figure*}
\clearpage
\begin{figure*}
\includegraphics[width=16cm]{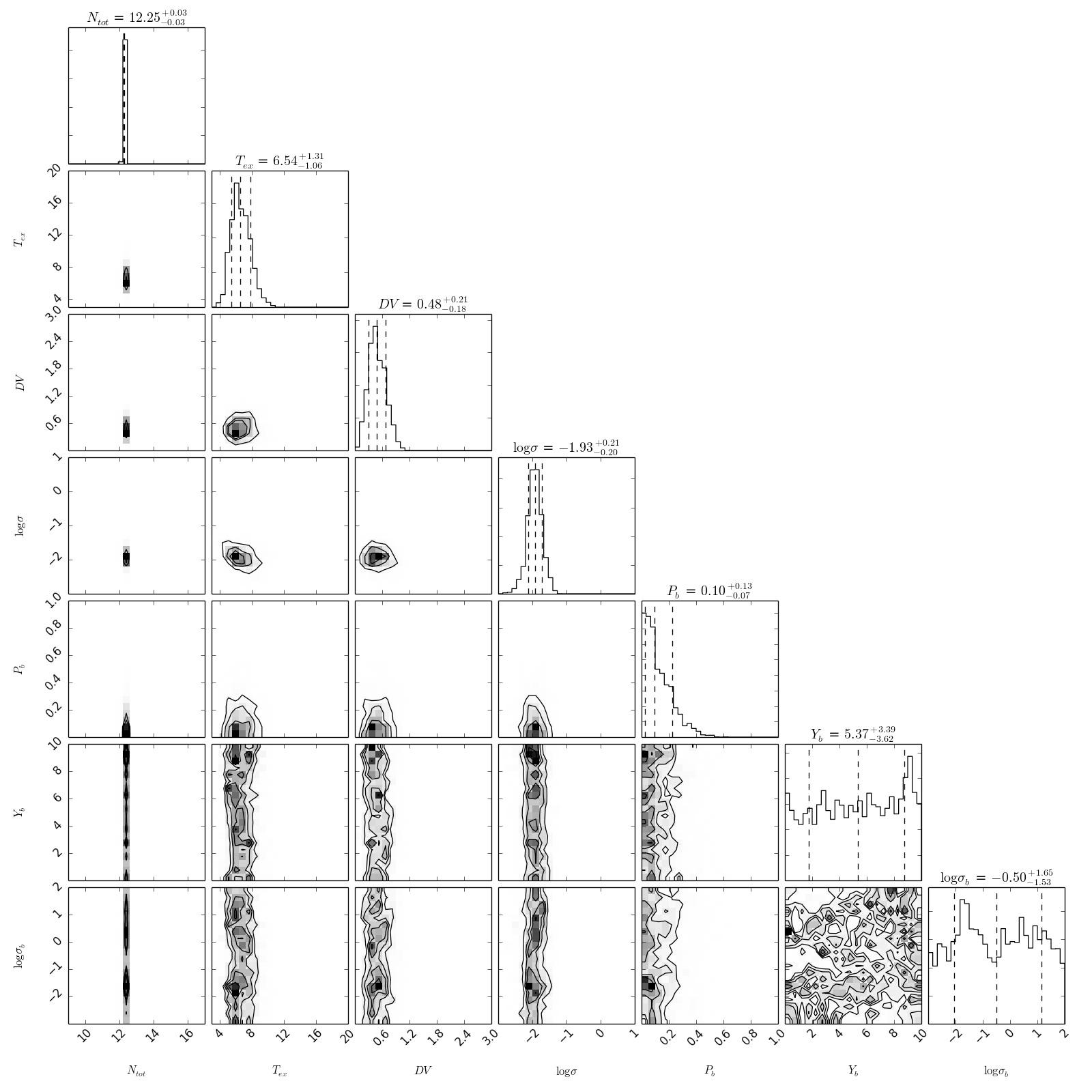}
\includegraphics[width=8cm]{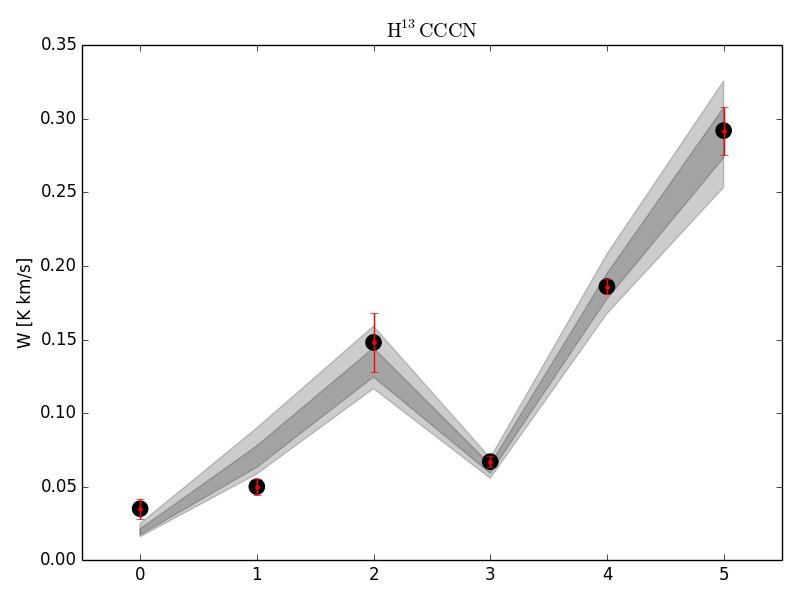}
\includegraphics[width=8cm]{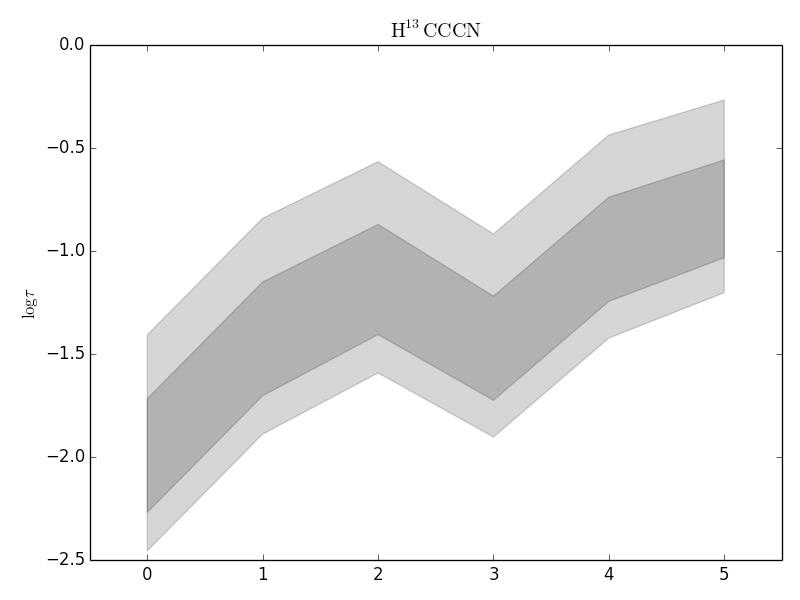}
\caption{\chem{H^{13}CC_2N}}
\label{H13CCCN}
\end{figure*}
\clearpage
\begin{figure*}
\includegraphics[width=16cm]{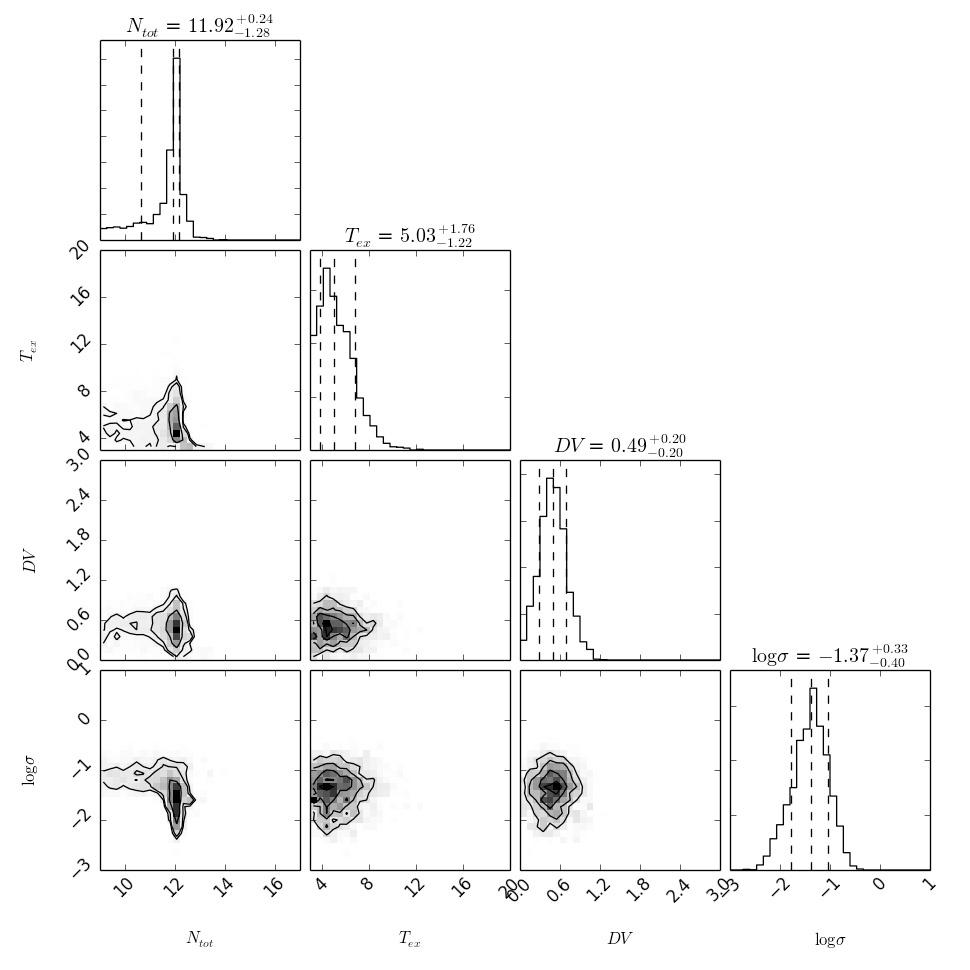}
\includegraphics[width=8cm]{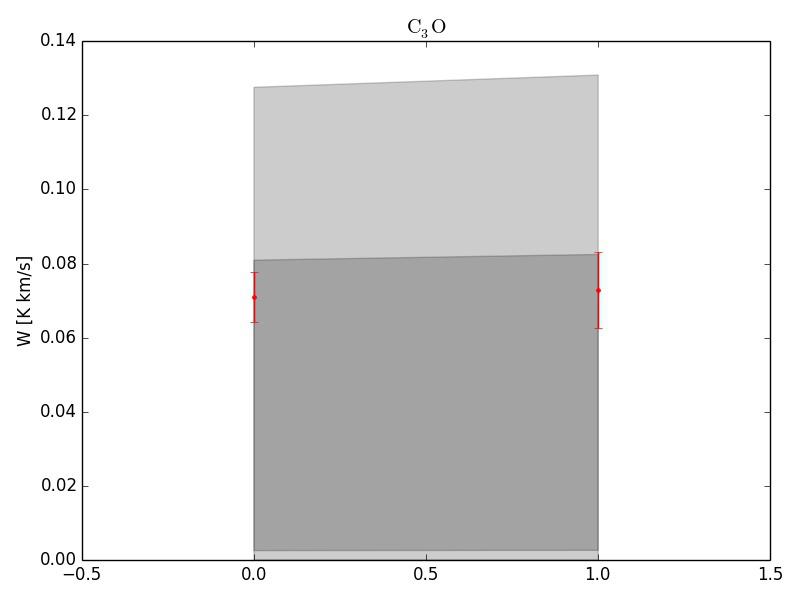}
\includegraphics[width=8cm]{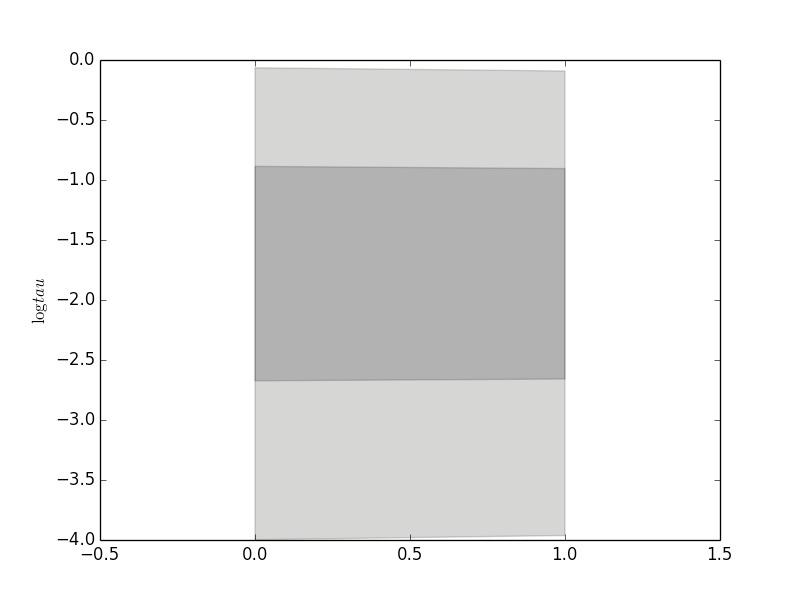}
\caption{\chem{C_3O}}
\label{C3O}
\end{figure*}
\clearpage
\begin{figure*}
\includegraphics[width=16cm]{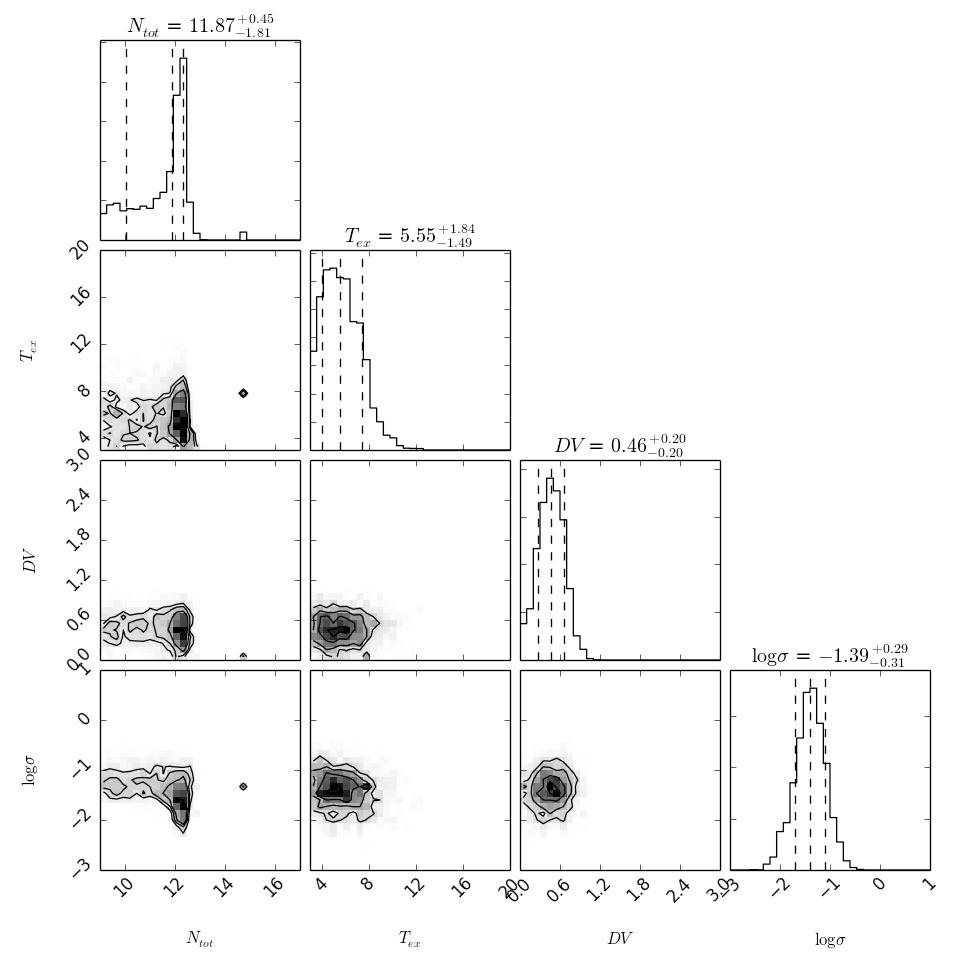}
\includegraphics[width=8cm]{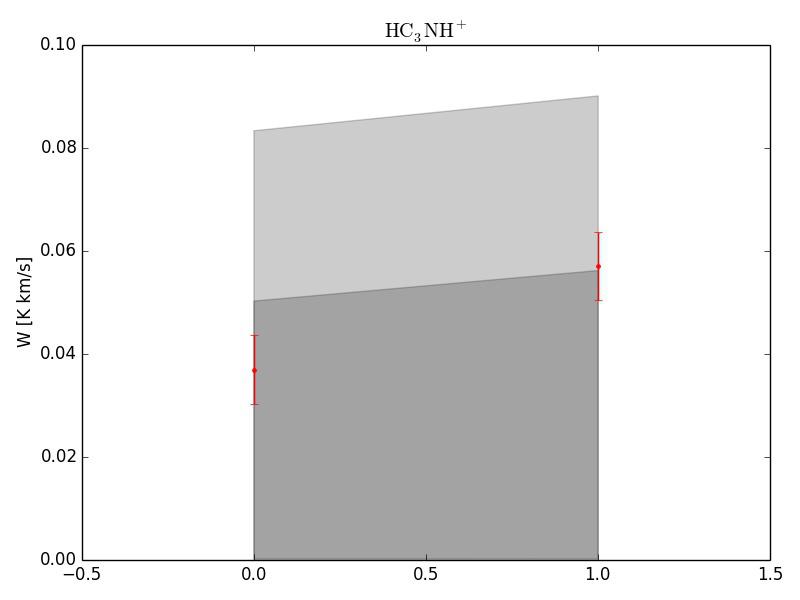}
\includegraphics[width=8cm]{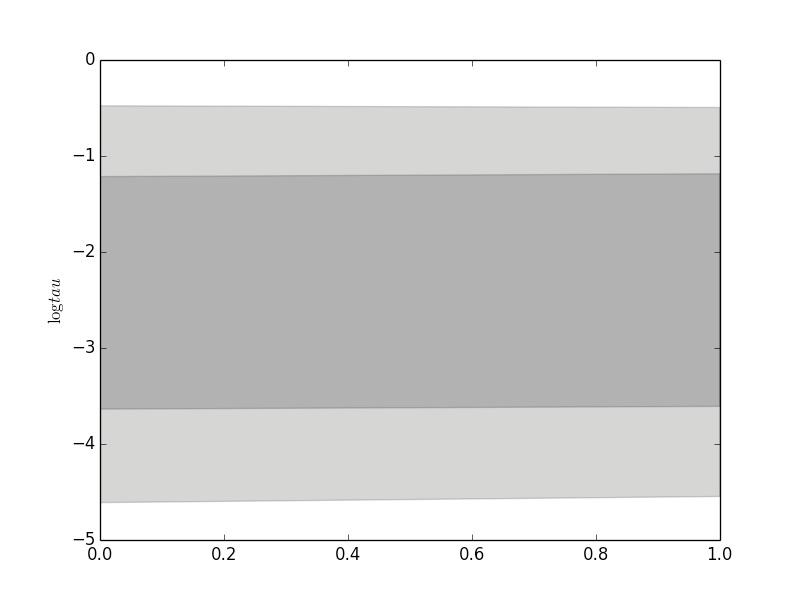}
\caption{\chem{HC_3NH^+}}
\label{HC3NHp}
\end{figure*}
\clearpage
\begin{figure*}
\includegraphics[width=16cm]{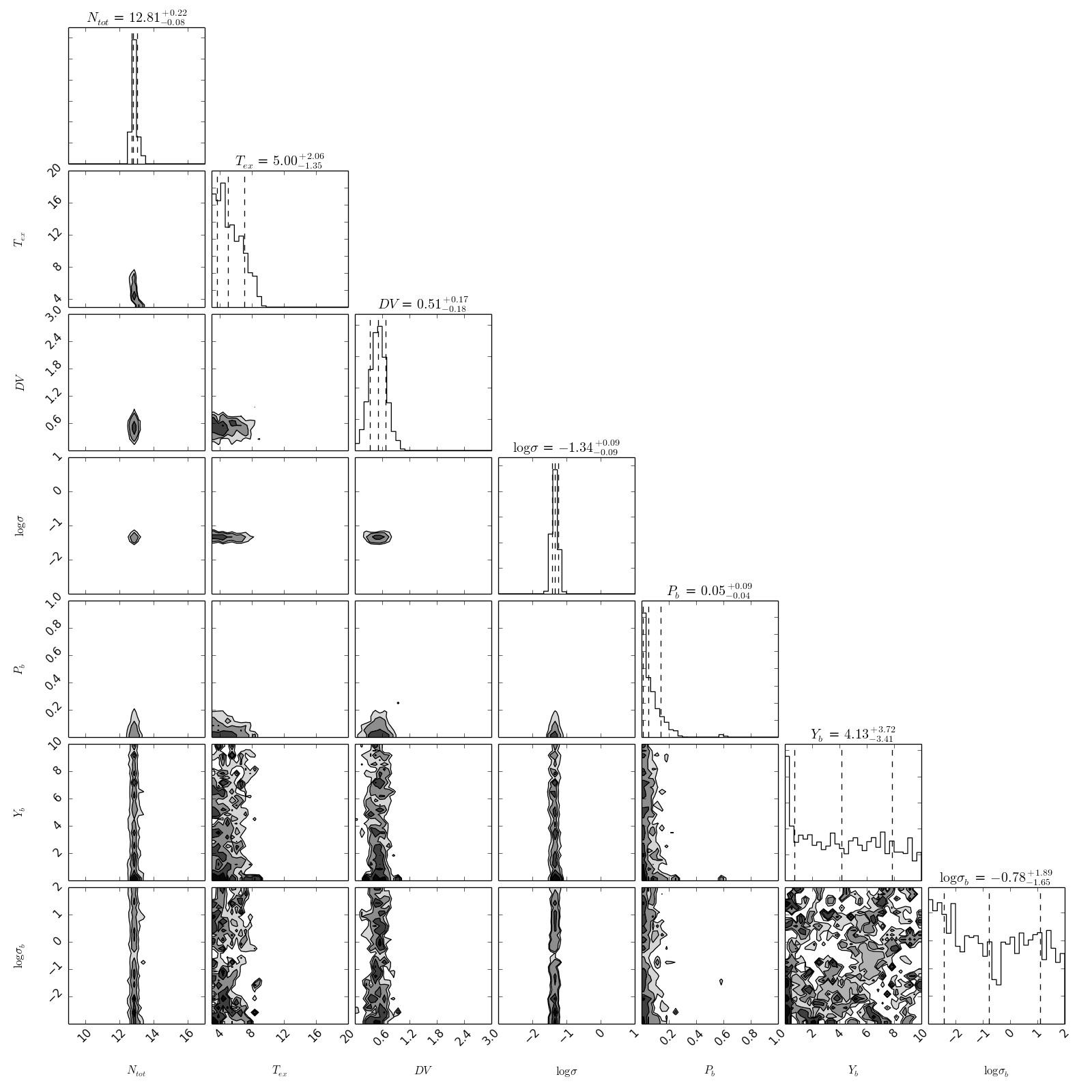}
\includegraphics[width=8cm]{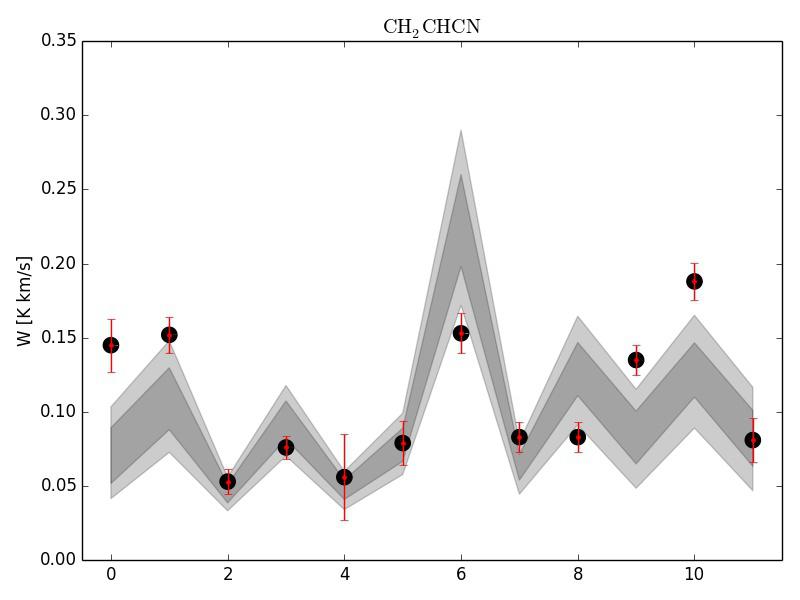}
\includegraphics[width=8cm]{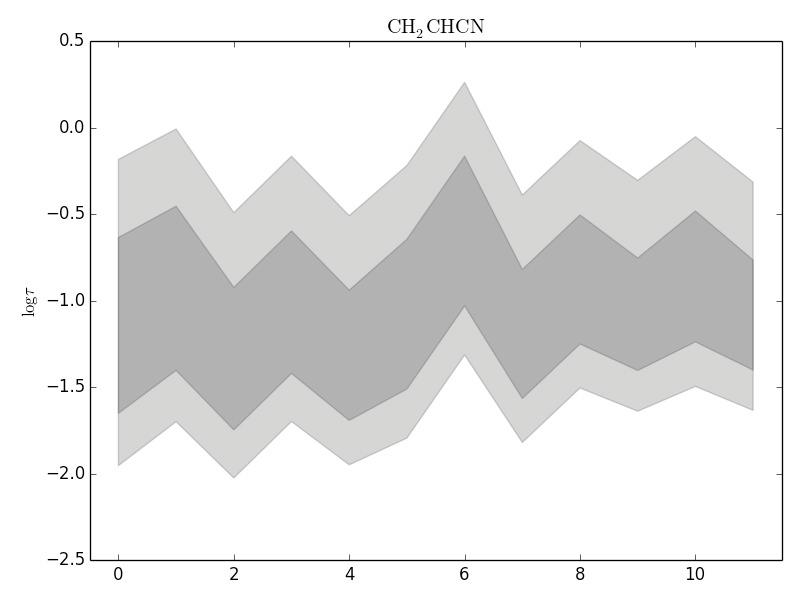}
\caption{\chem{CH_2CHCN}}
\label{CH2CHCN}
\end{figure*}
\clearpage
\begin{figure*}
\includegraphics[width=16cm]{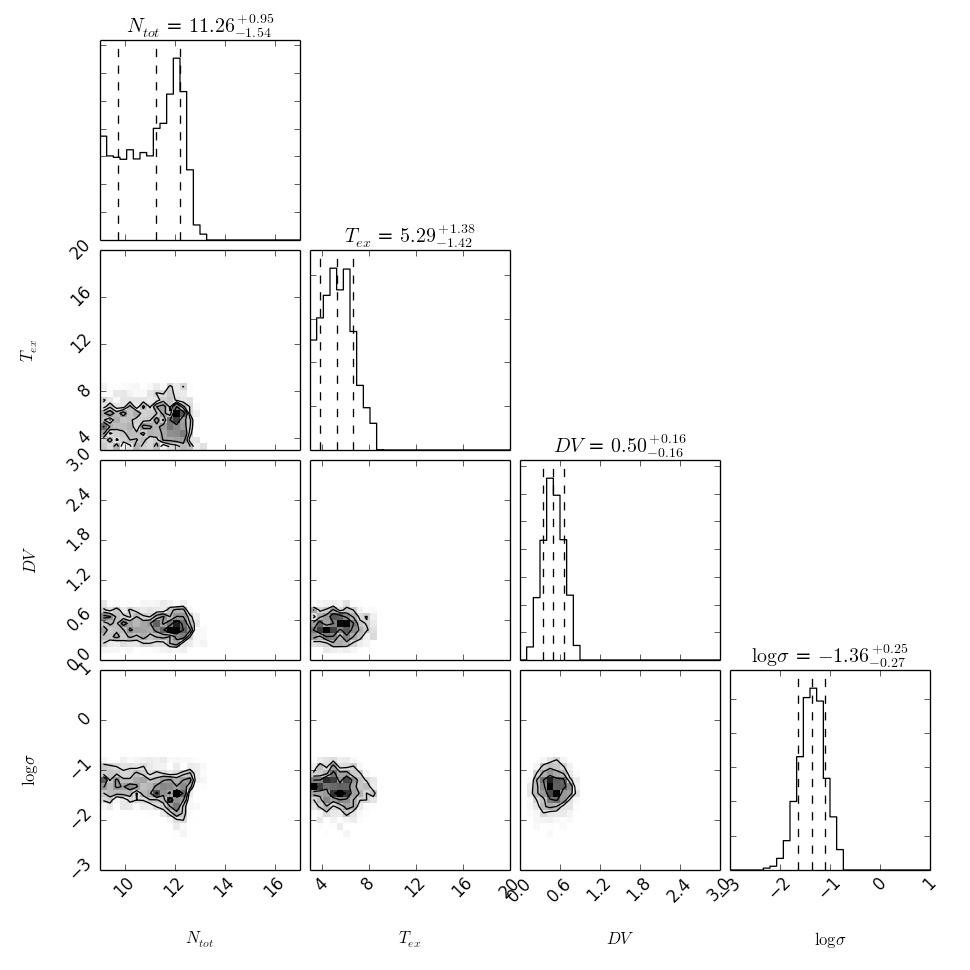}
\includegraphics[width=8cm]{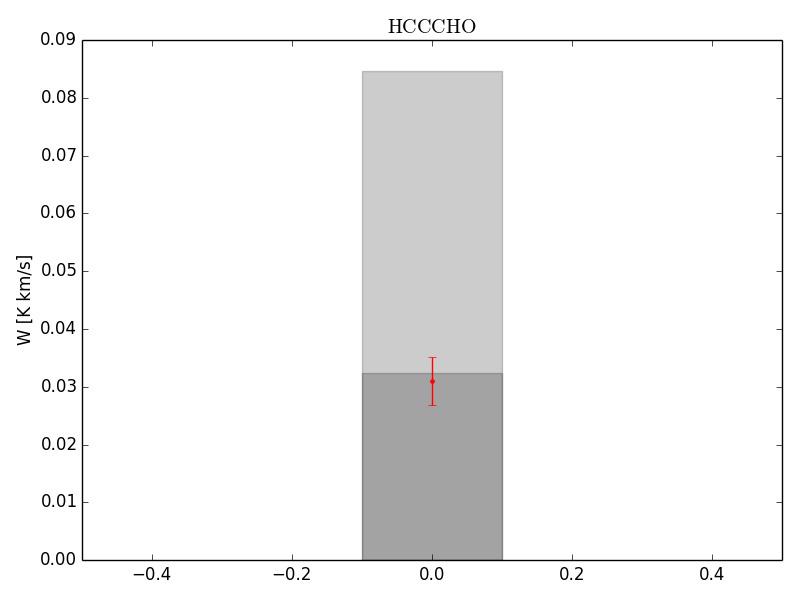}
\includegraphics[width=8cm]{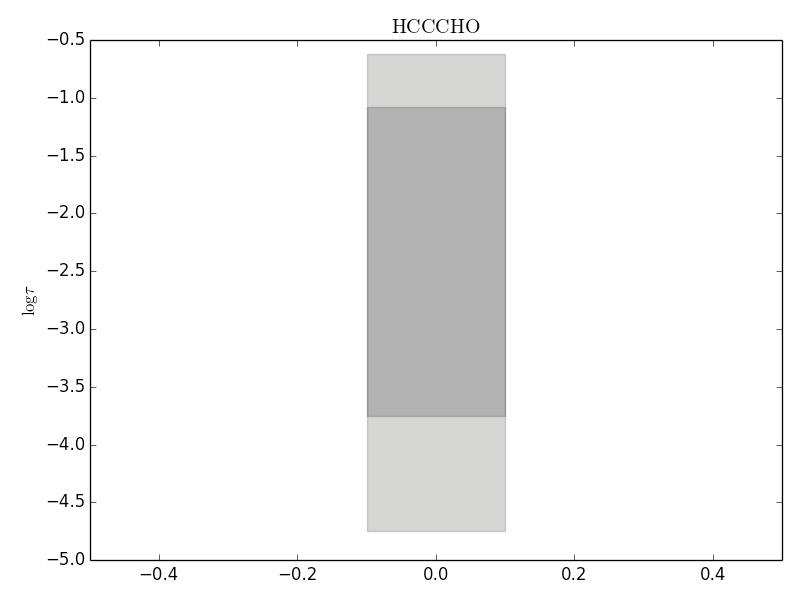}
\caption{\chem{HCCCHO}}
\label{HCCCHO}
\end{figure*}
\clearpage
\begin{figure*}
\includegraphics[width=16cm]{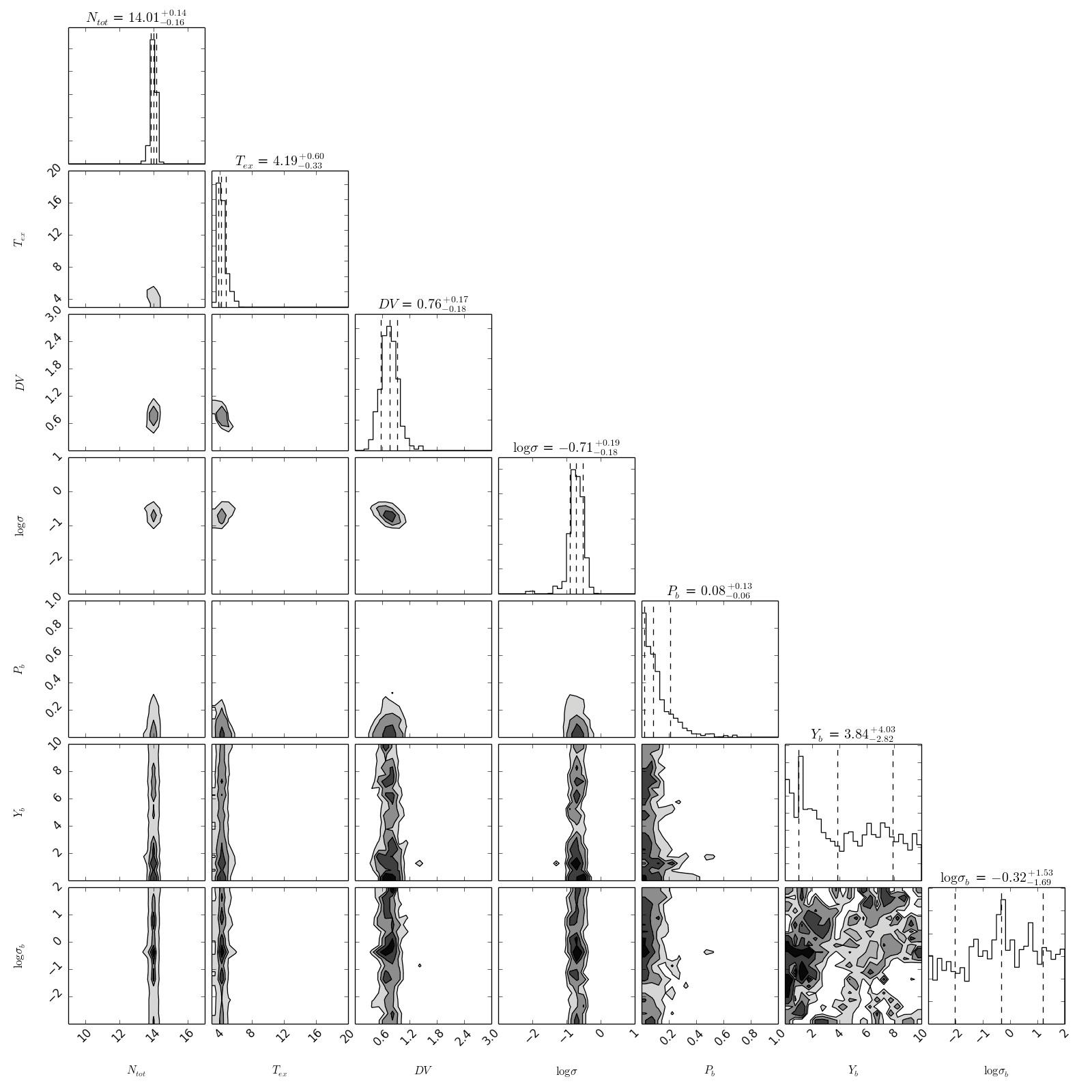}
\includegraphics[width=8cm]{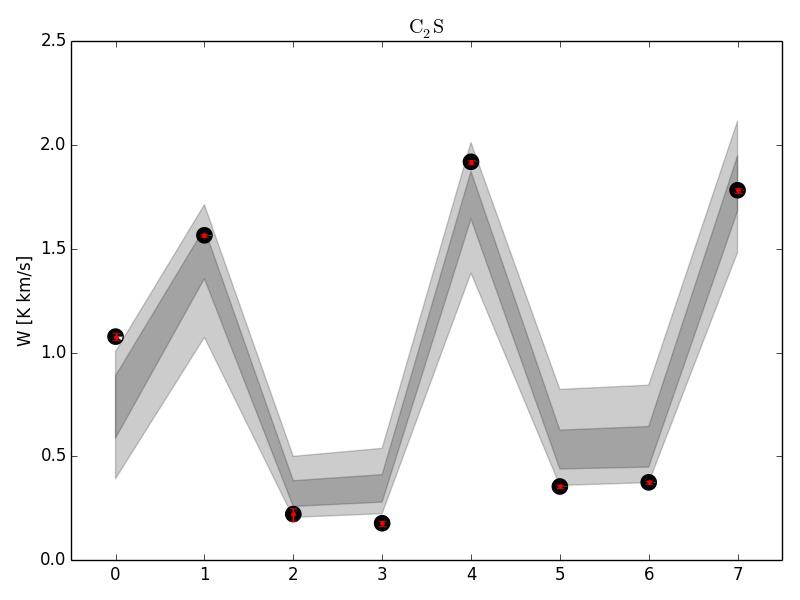}
\includegraphics[width=8cm]{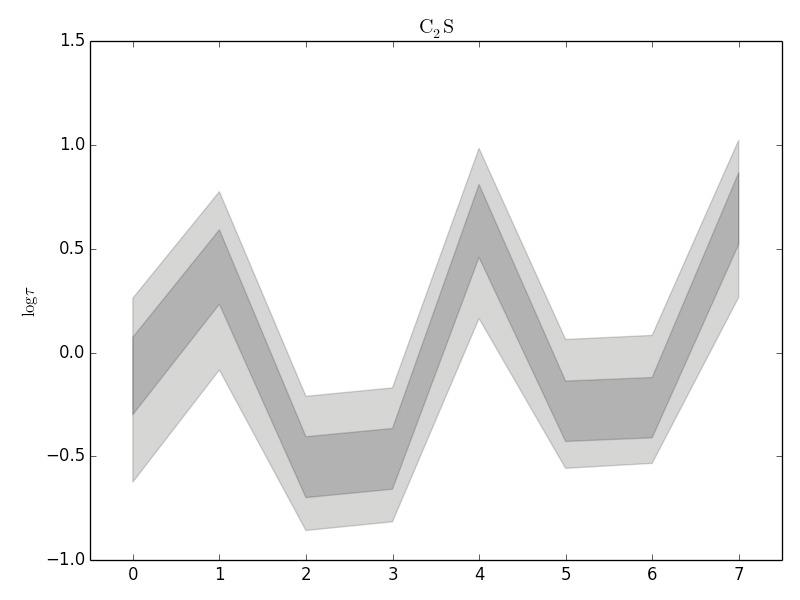}
\caption{\chem{C_2S}}
\label{C2S}
\end{figure*}
\clearpage
\begin{figure*}
\includegraphics[width=16cm]{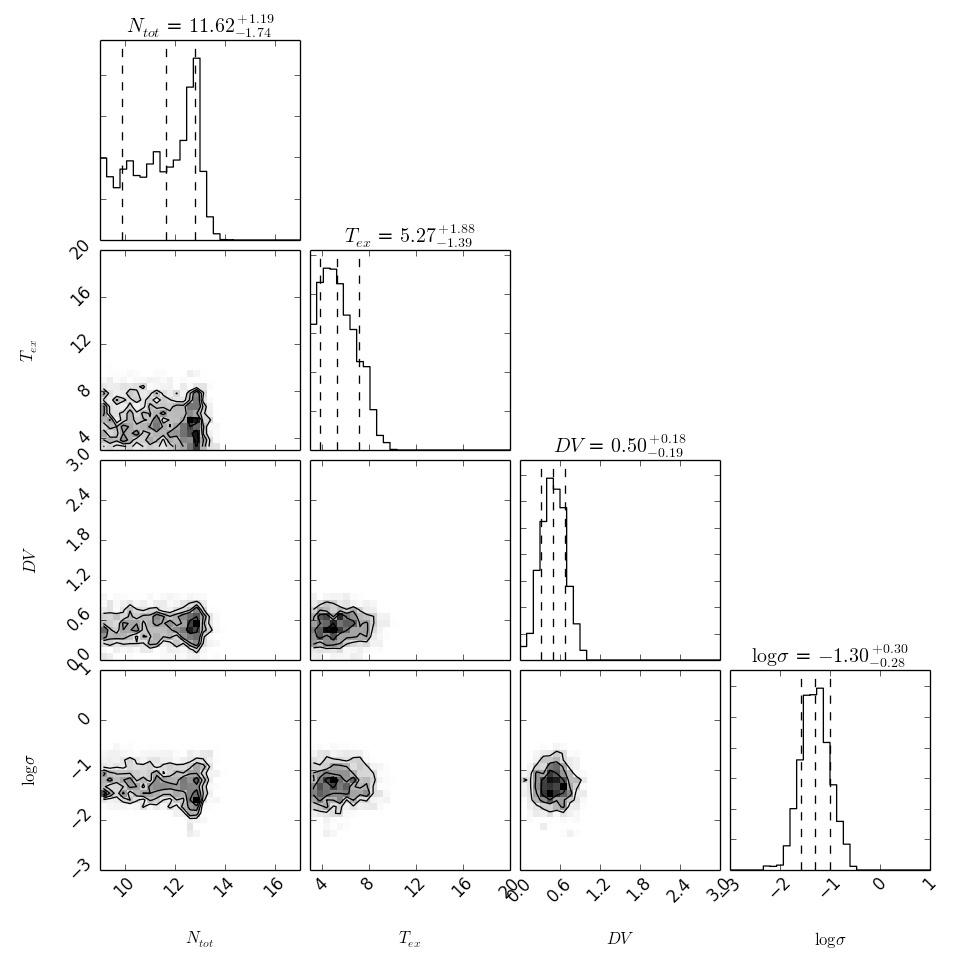}
\includegraphics[width=8cm]{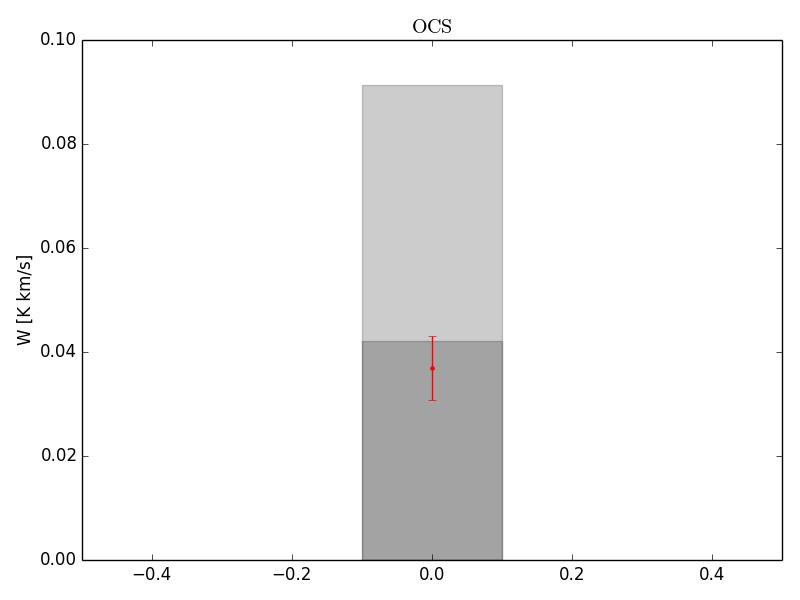}
\includegraphics[width=8cm]{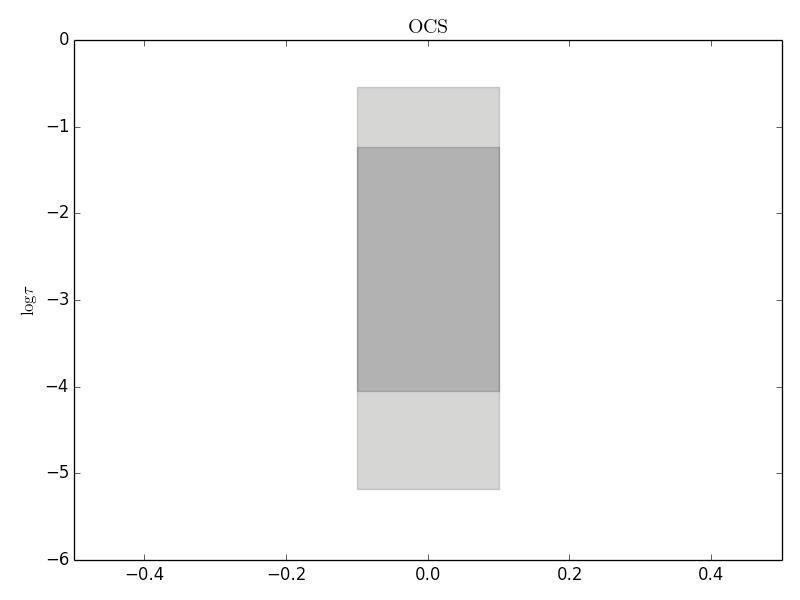}
\caption{\chem{OCS}}
\label{OCS}
\end{figure*}
\clearpage
\begin{figure*}
\includegraphics[width=16cm]{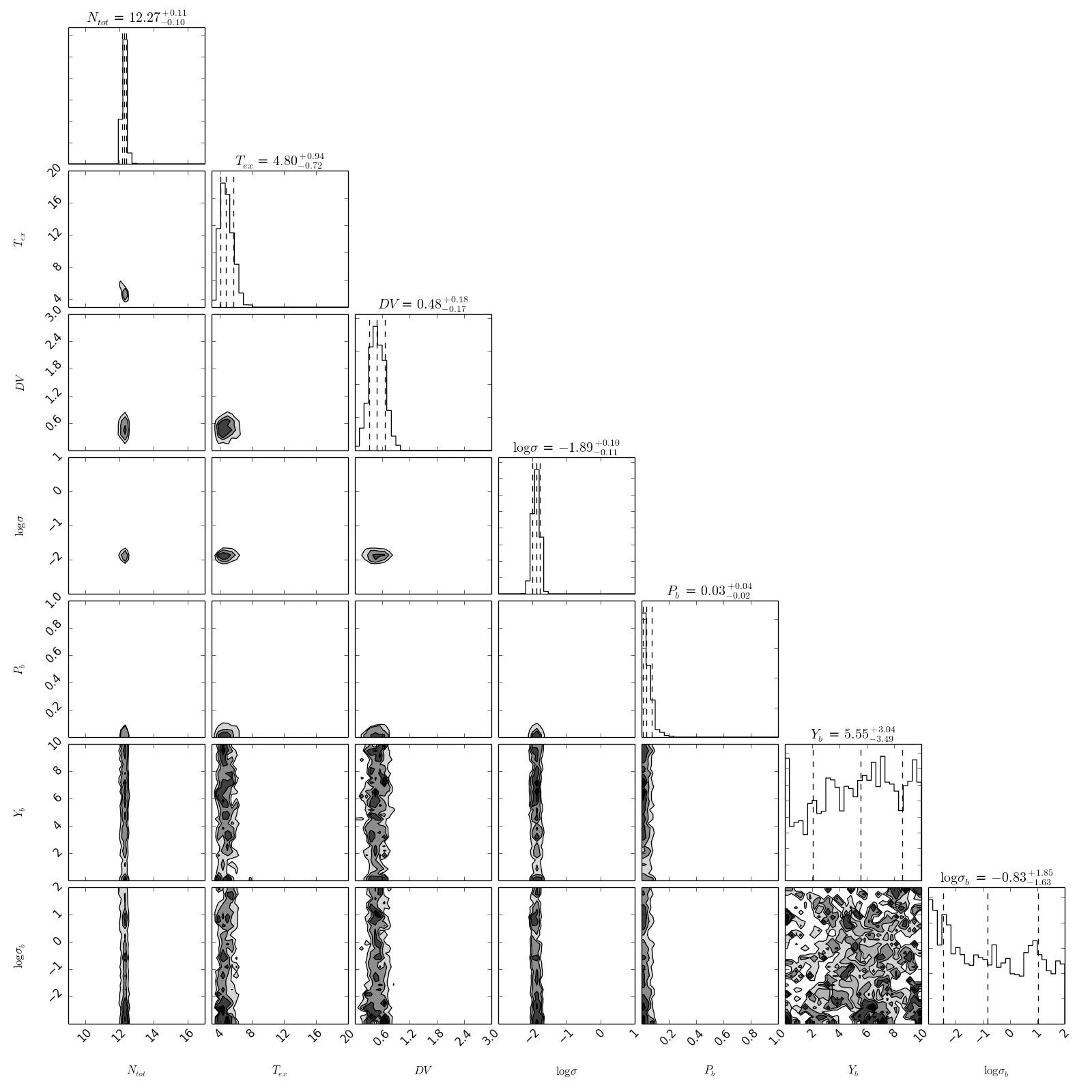}
\includegraphics[width=8cm]{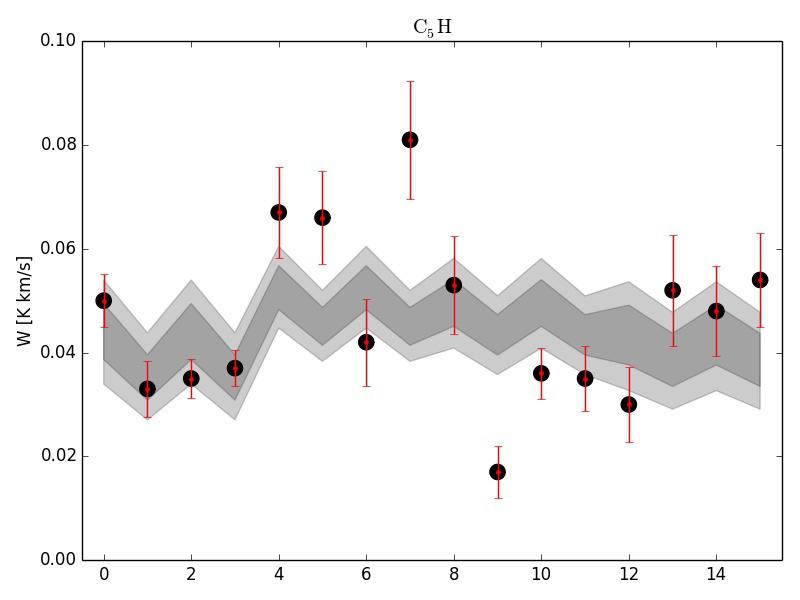}
\includegraphics[width=8cm]{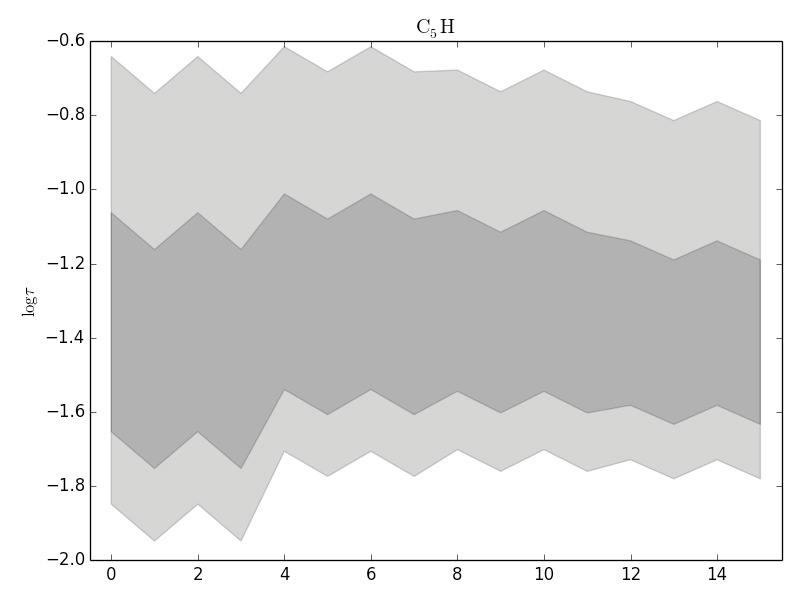}
\caption{\chem{C_5H}}
\label{C5H}
\end{figure*}
\clearpage
\begin{figure*}
\includegraphics[width=16cm]{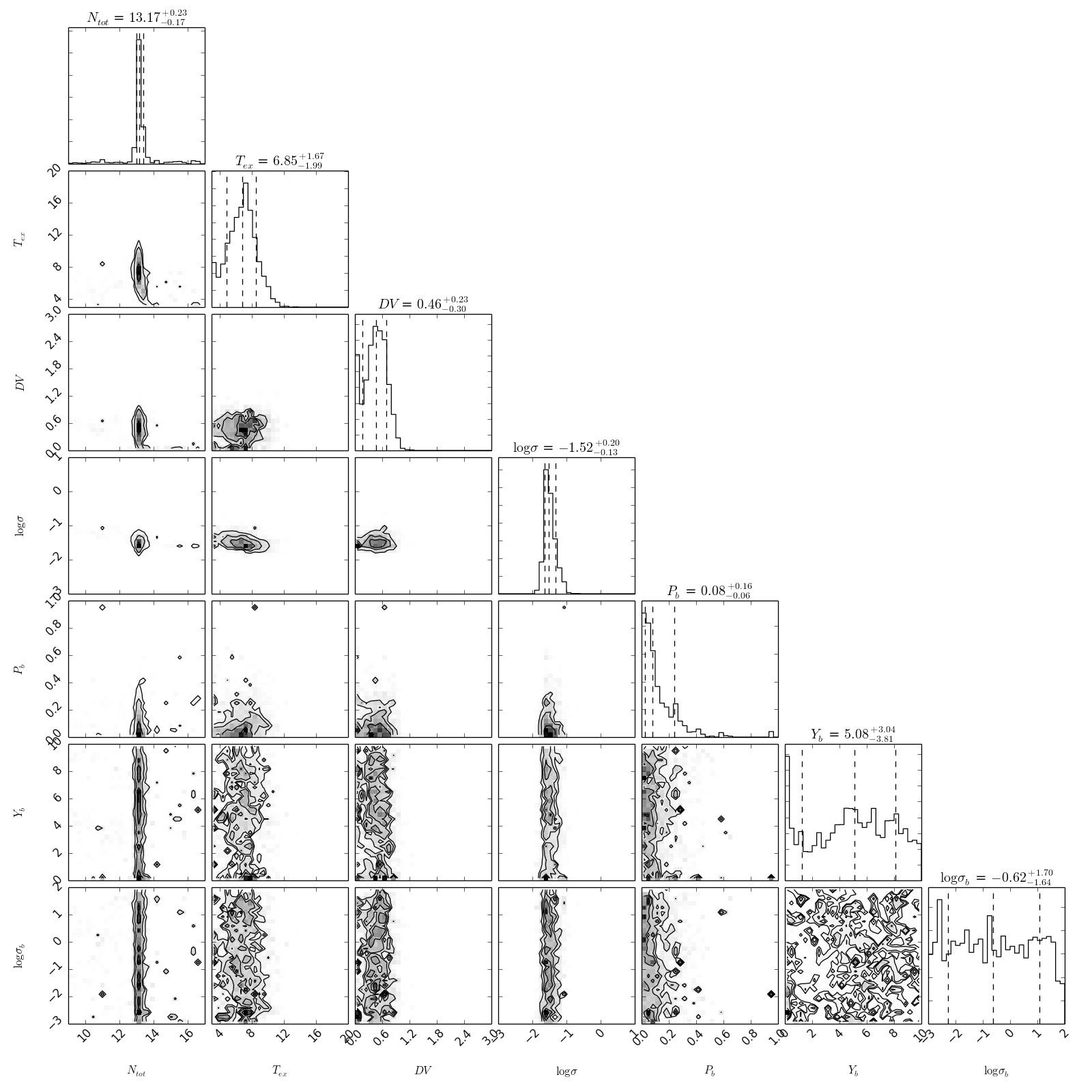}
\includegraphics[width=8cm]{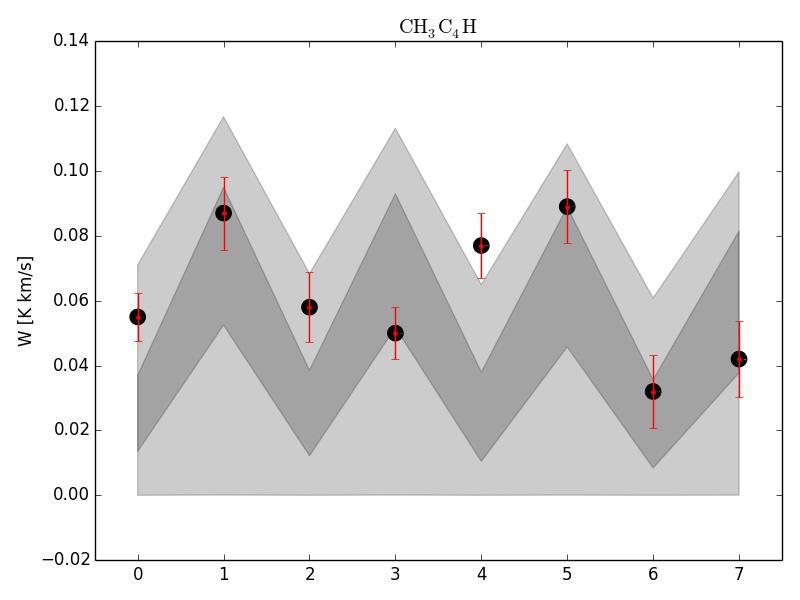}
\includegraphics[width=8cm]{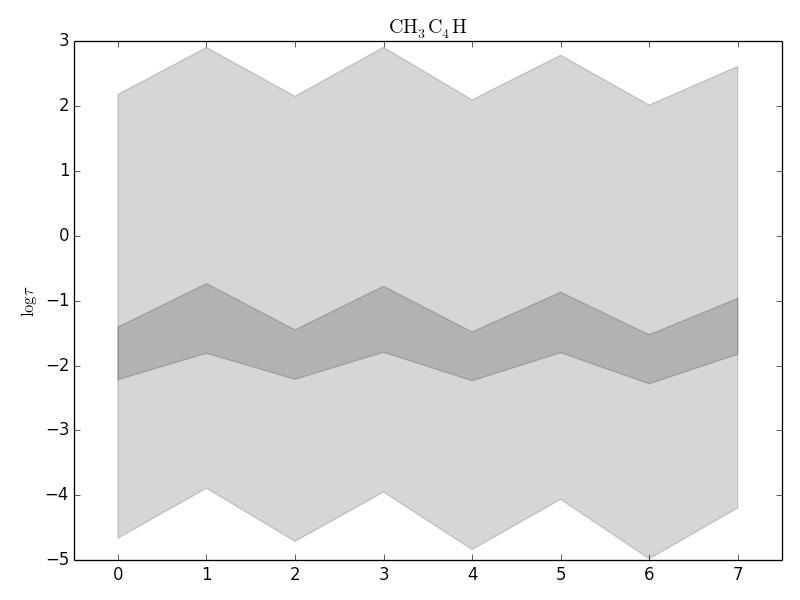}
\caption{\chem{CH_3C_4H}}
\label{CH3C4H}
\end{figure*}
\clearpage
\begin{figure*}
\includegraphics[width=16cm]{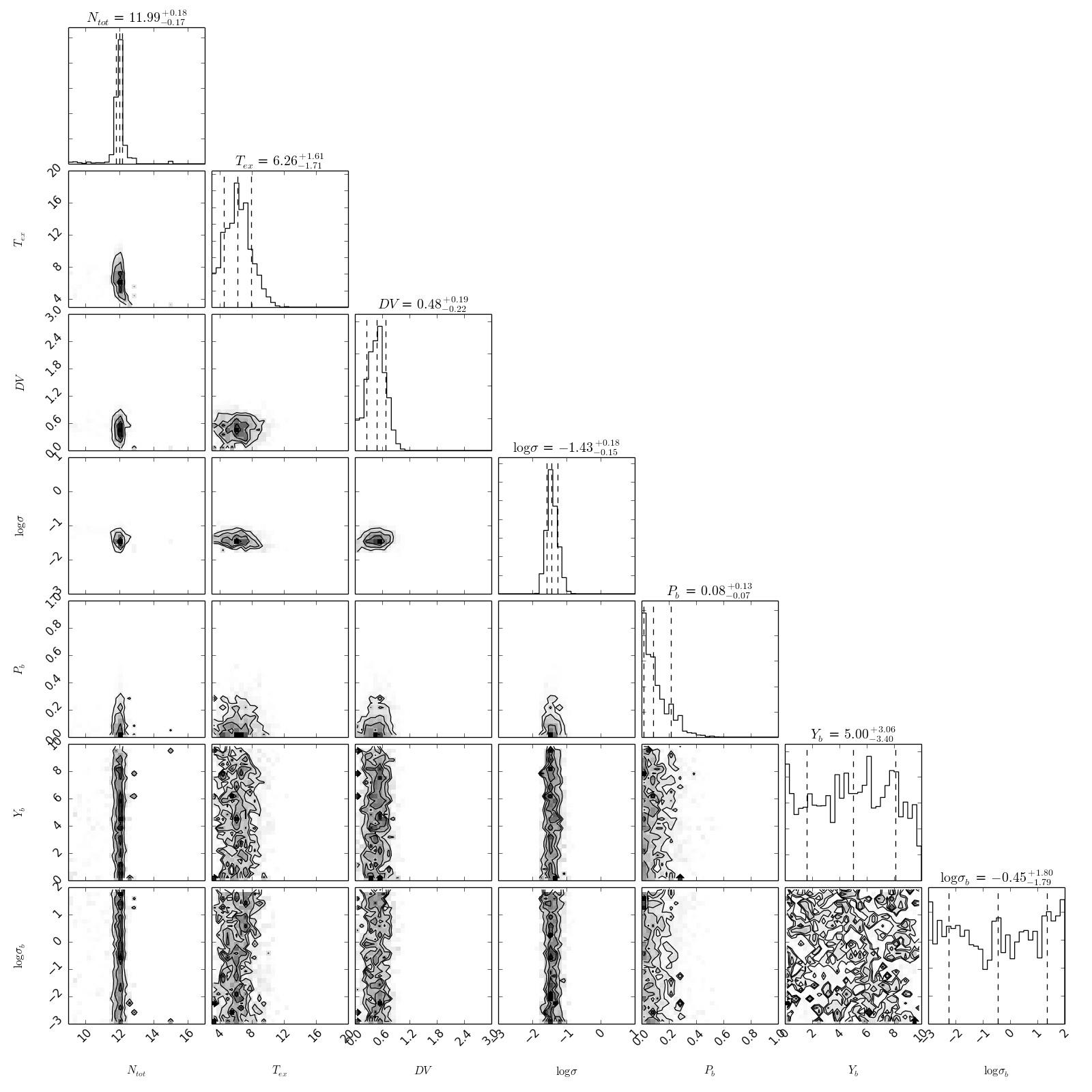}
\includegraphics[width=8cm]{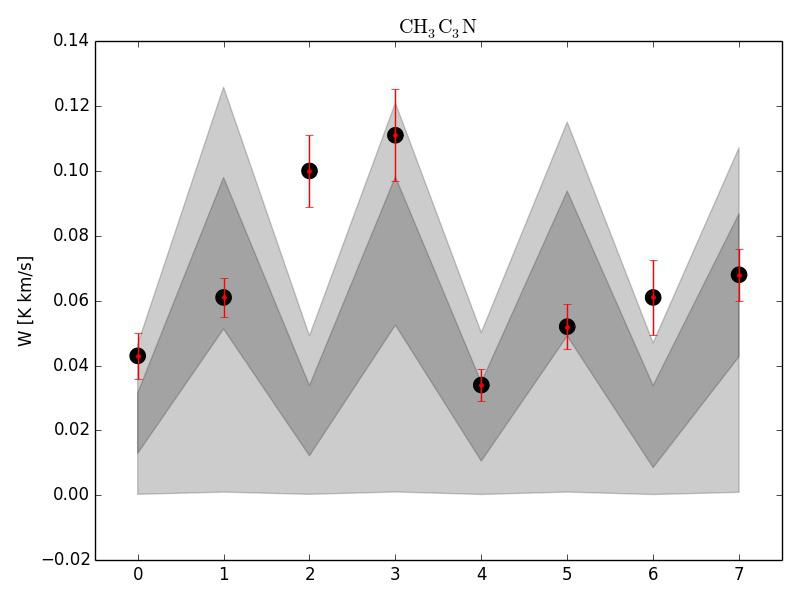}
\includegraphics[width=8cm]{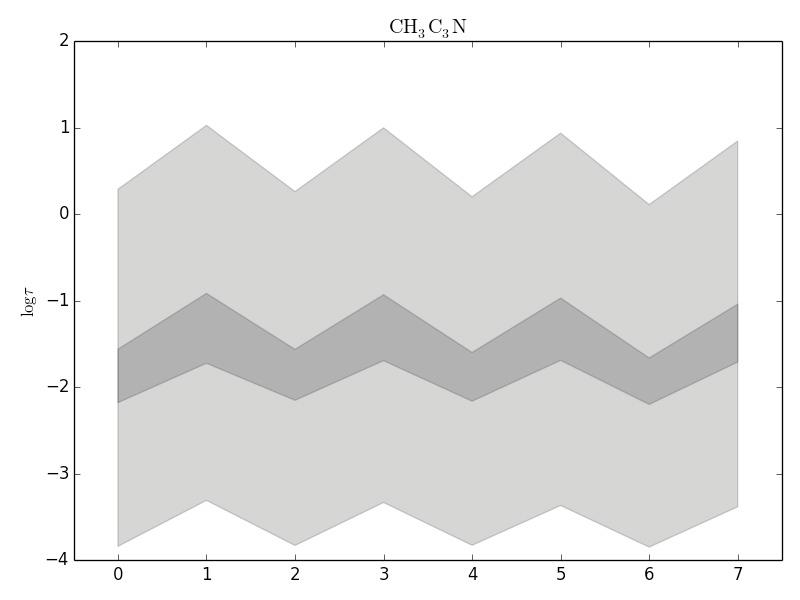}
\caption{\chem{CH_3C_3N}}
\label{CH3C3N}
\end{figure*}
\clearpage
\begin{figure*}
\includegraphics[width=16cm]{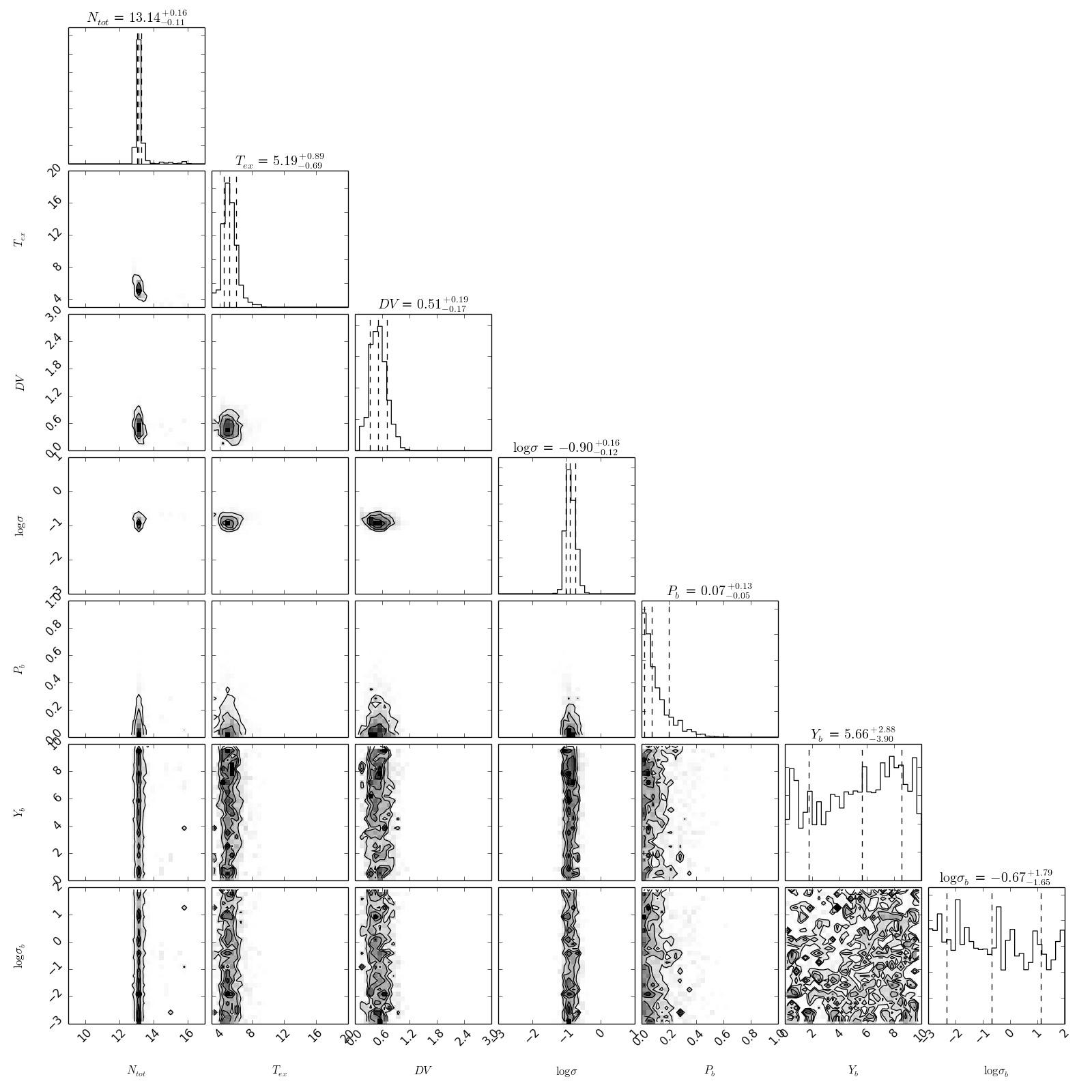}
\includegraphics[width=8cm]{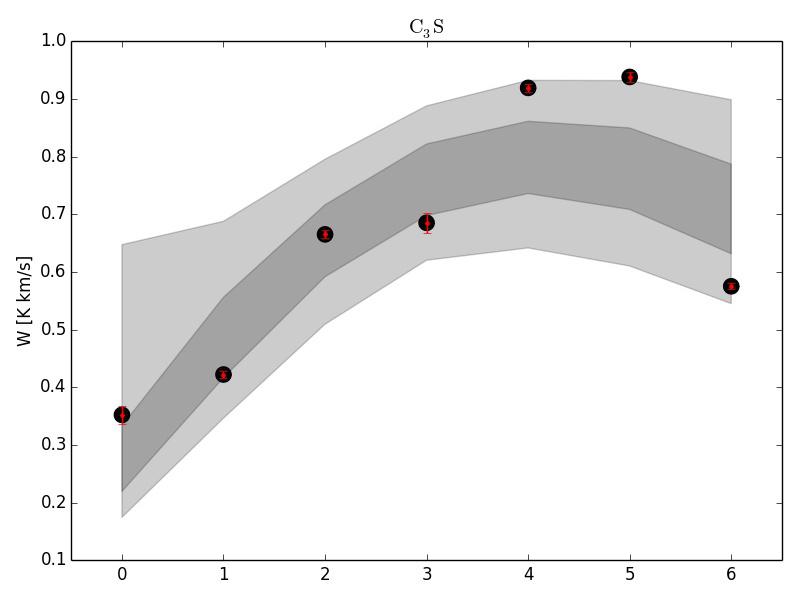}
\includegraphics[width=8cm]{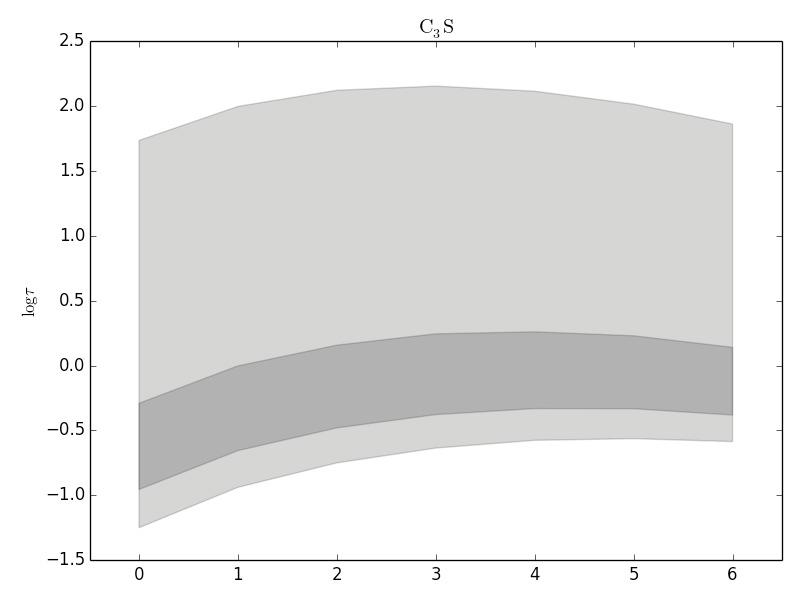}
\caption{\chem{C_3S}}
\label{C3S}
\end{figure*}
\clearpage
\begin{figure*}
\includegraphics[width=16cm]{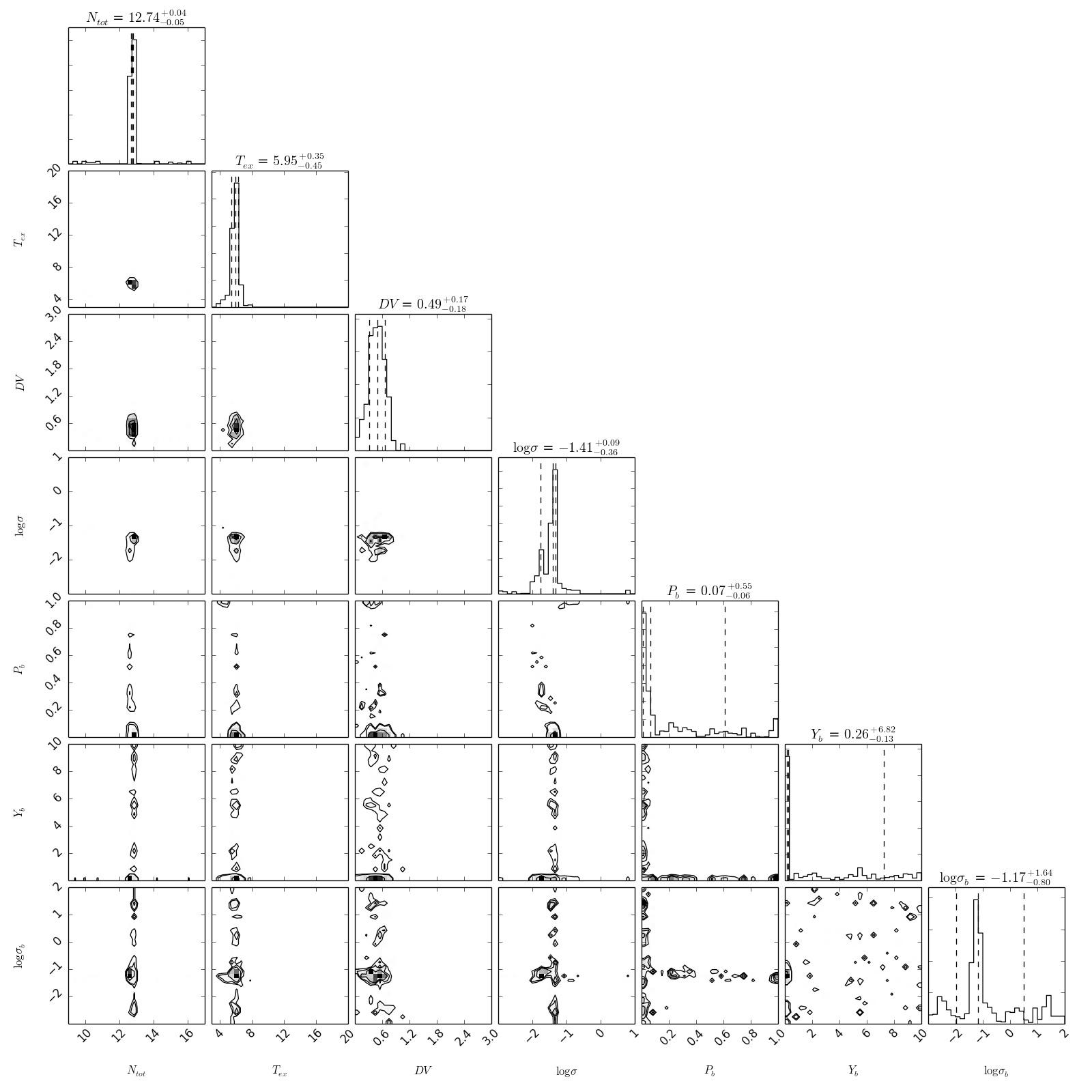}
\includegraphics[width=8cm]{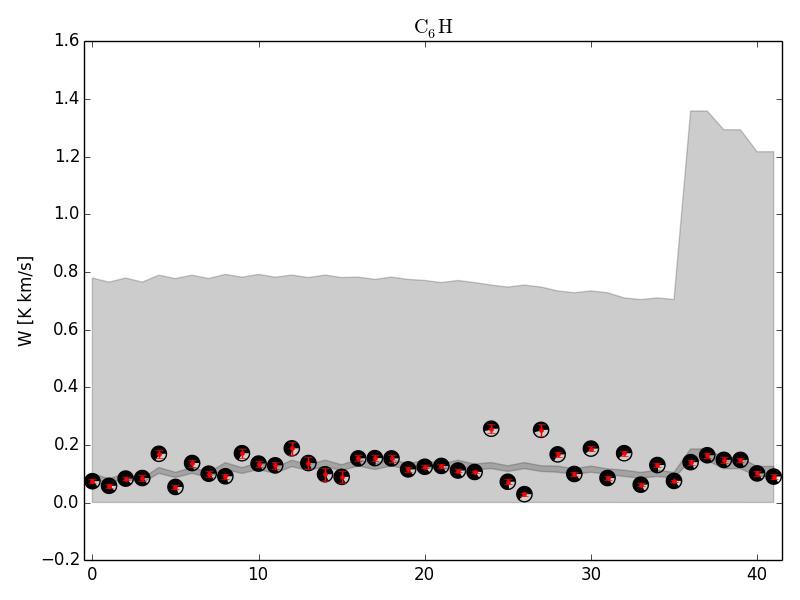}
\includegraphics[width=8cm]{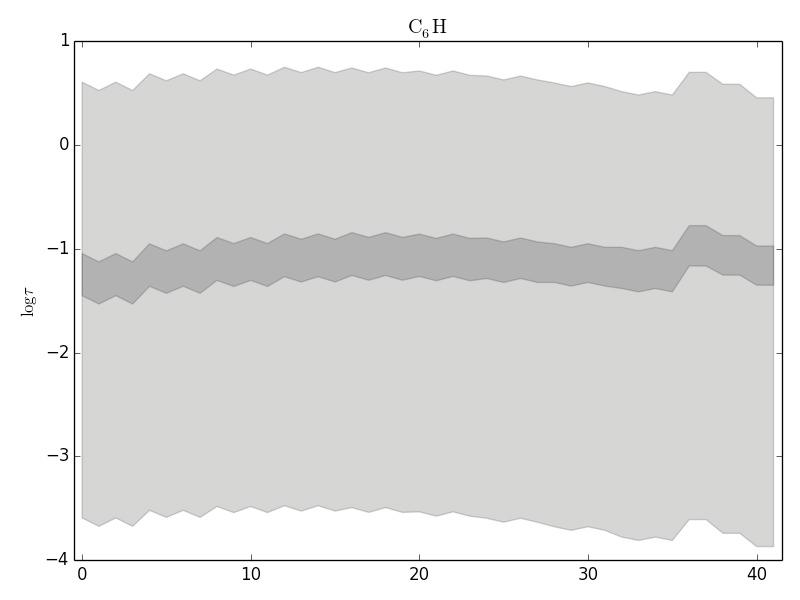}
\caption{\chem{C_6H}}
\label{C6H}
\end{figure*}
\clearpage
\begin{figure*}
\includegraphics[width=16cm]{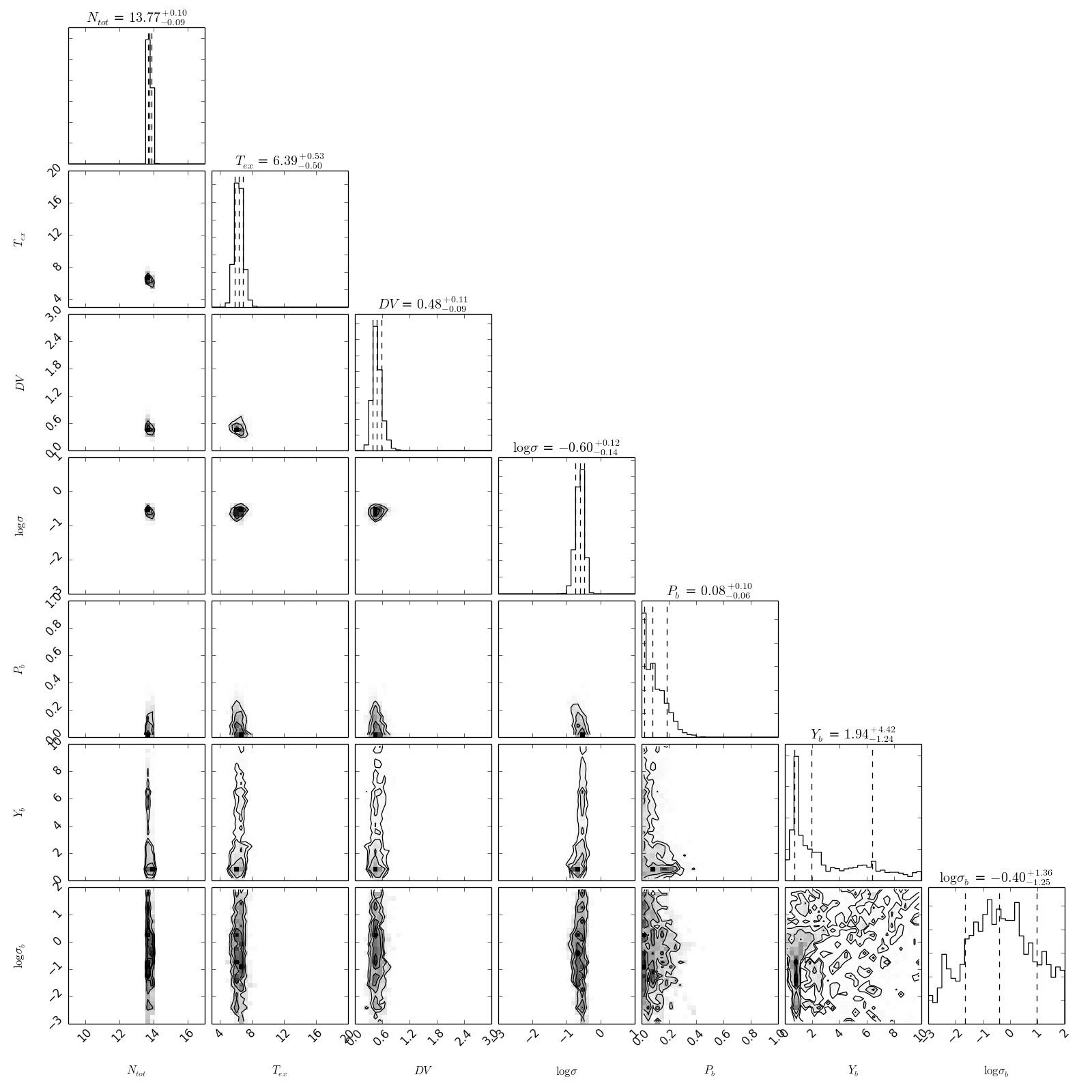}
\includegraphics[width=8cm]{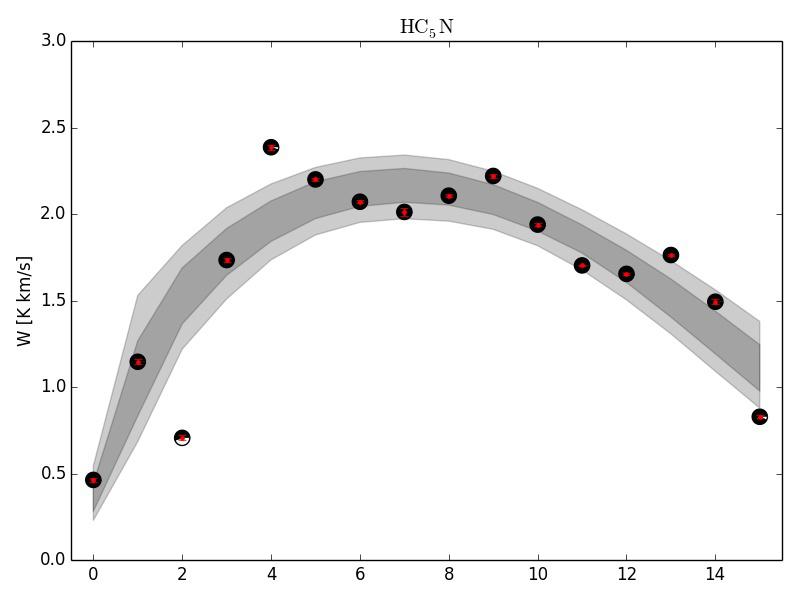}
\includegraphics[width=8cm]{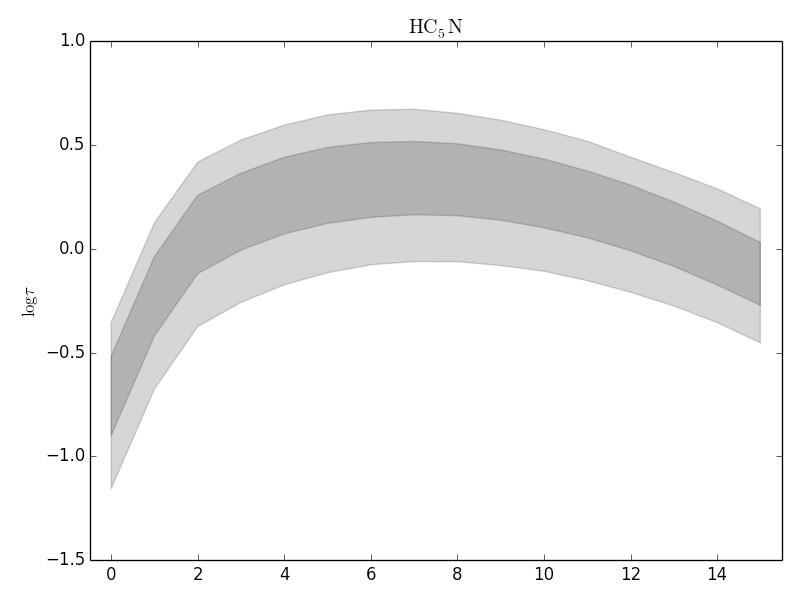}
\caption{\chem{HC_5N}}
\label{HC5N}
\end{figure*}
\clearpage
\begin{figure*}
\includegraphics[width=16cm]{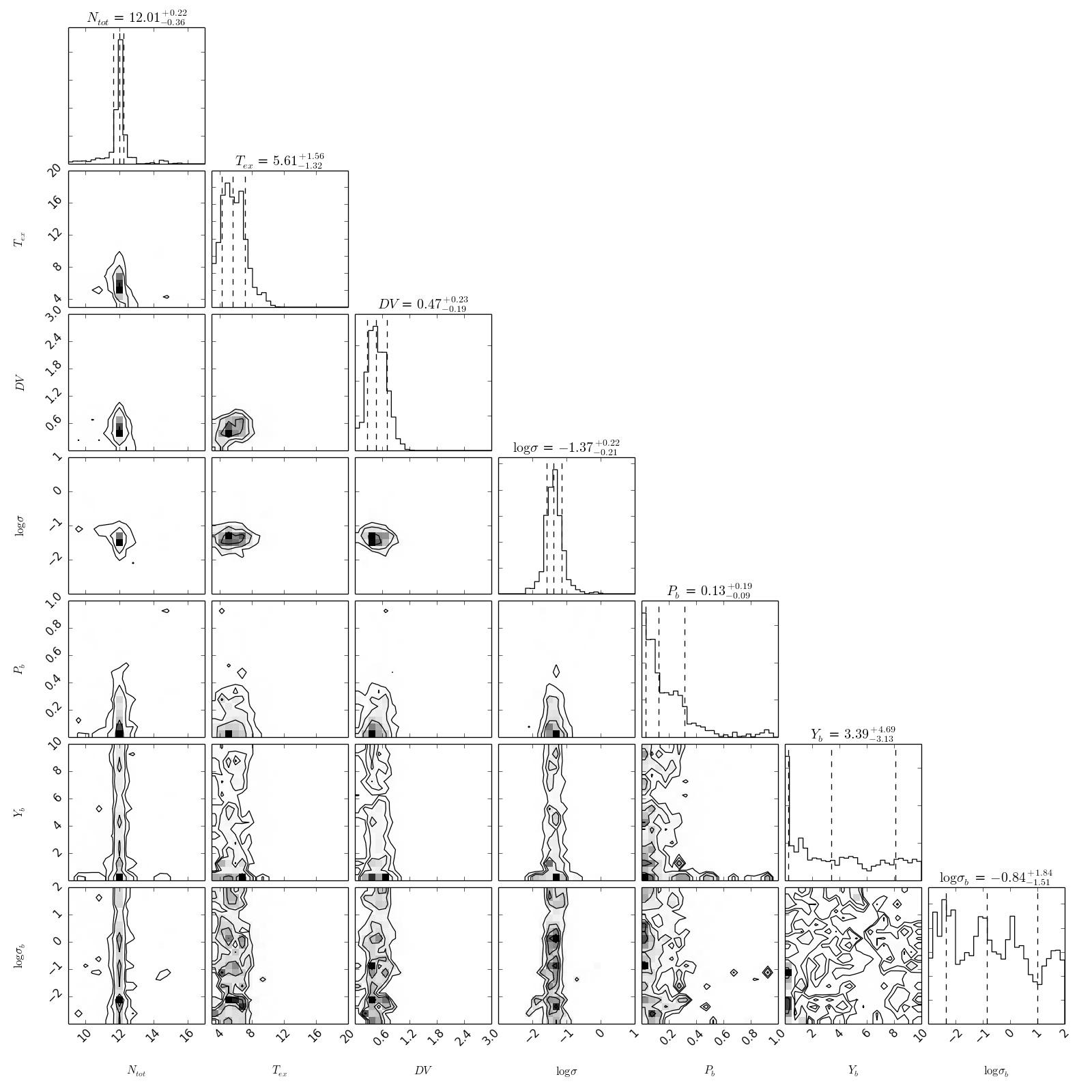}
\includegraphics[width=8cm]{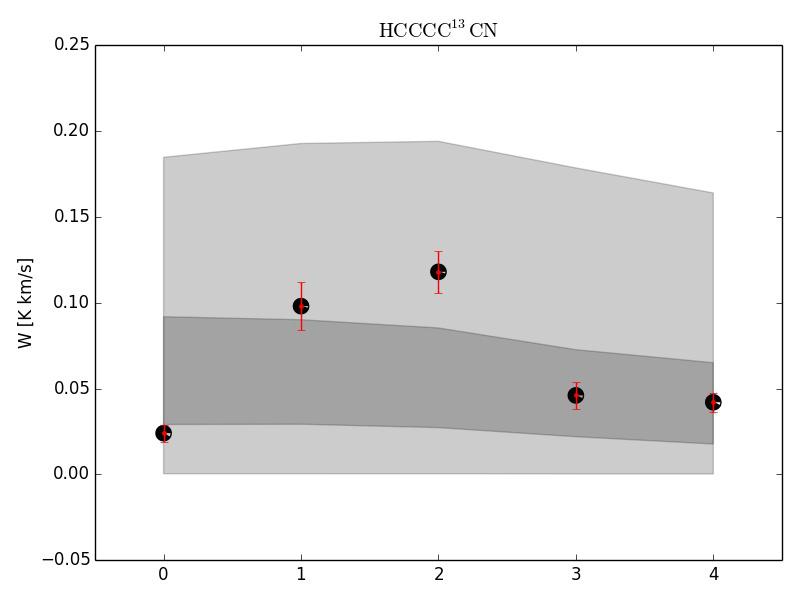}
\includegraphics[width=8cm]{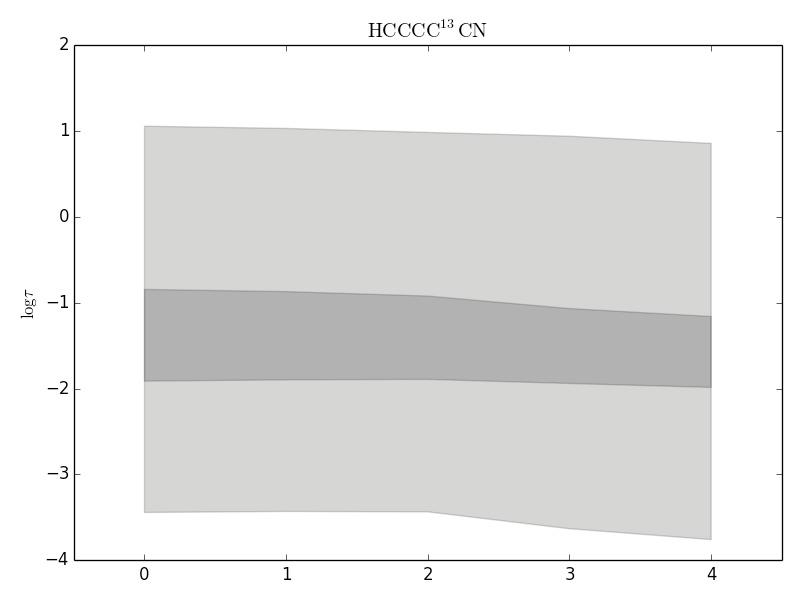}
\caption{\chem{HC_4^{13}CN}}
\label{HCCCC13CN}
\end{figure*}
\clearpage
\begin{figure*}
\includegraphics[width=16cm]{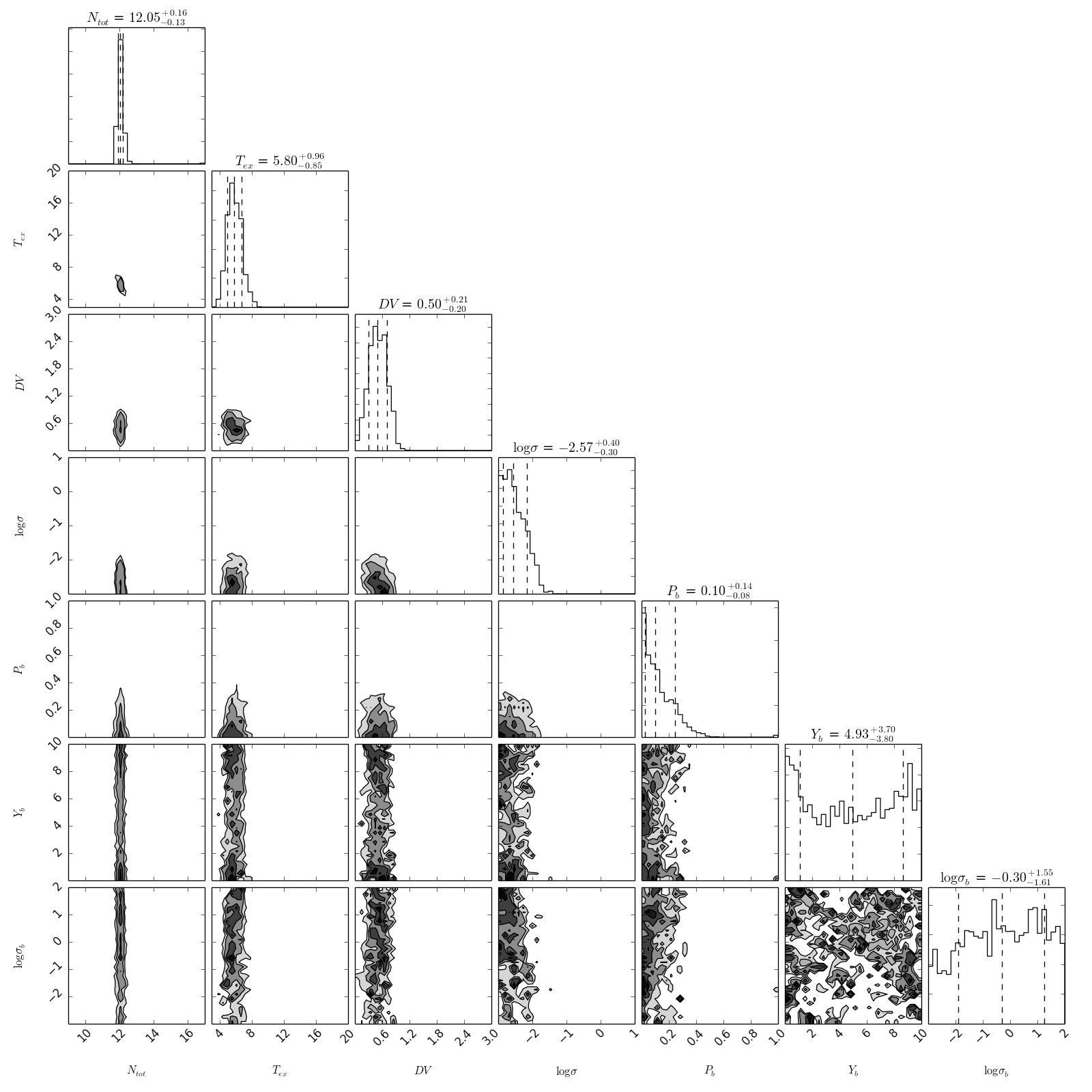}
\includegraphics[width=8cm]{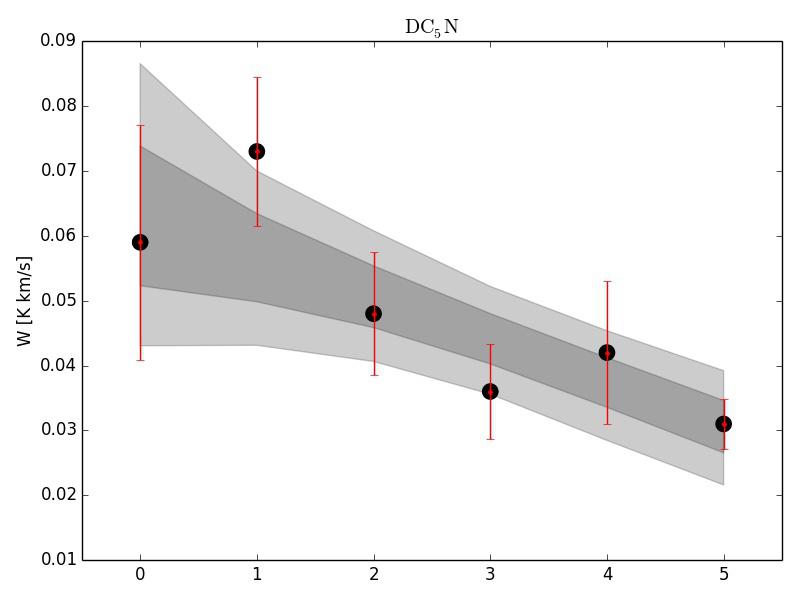}
\includegraphics[width=8cm]{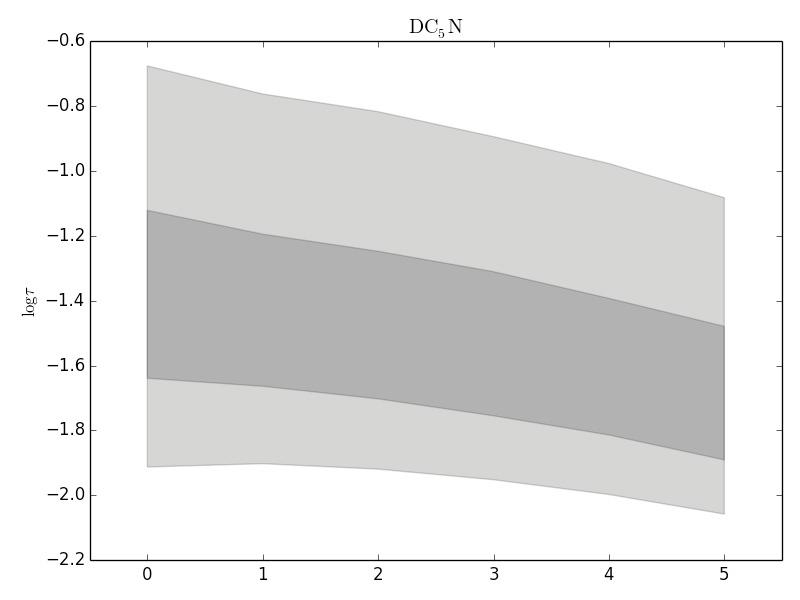}
\caption{\chem{DC_5N}}
\label{DC5N}
\end{figure*}
\clearpage
\begin{figure*}
\includegraphics[width=16cm]{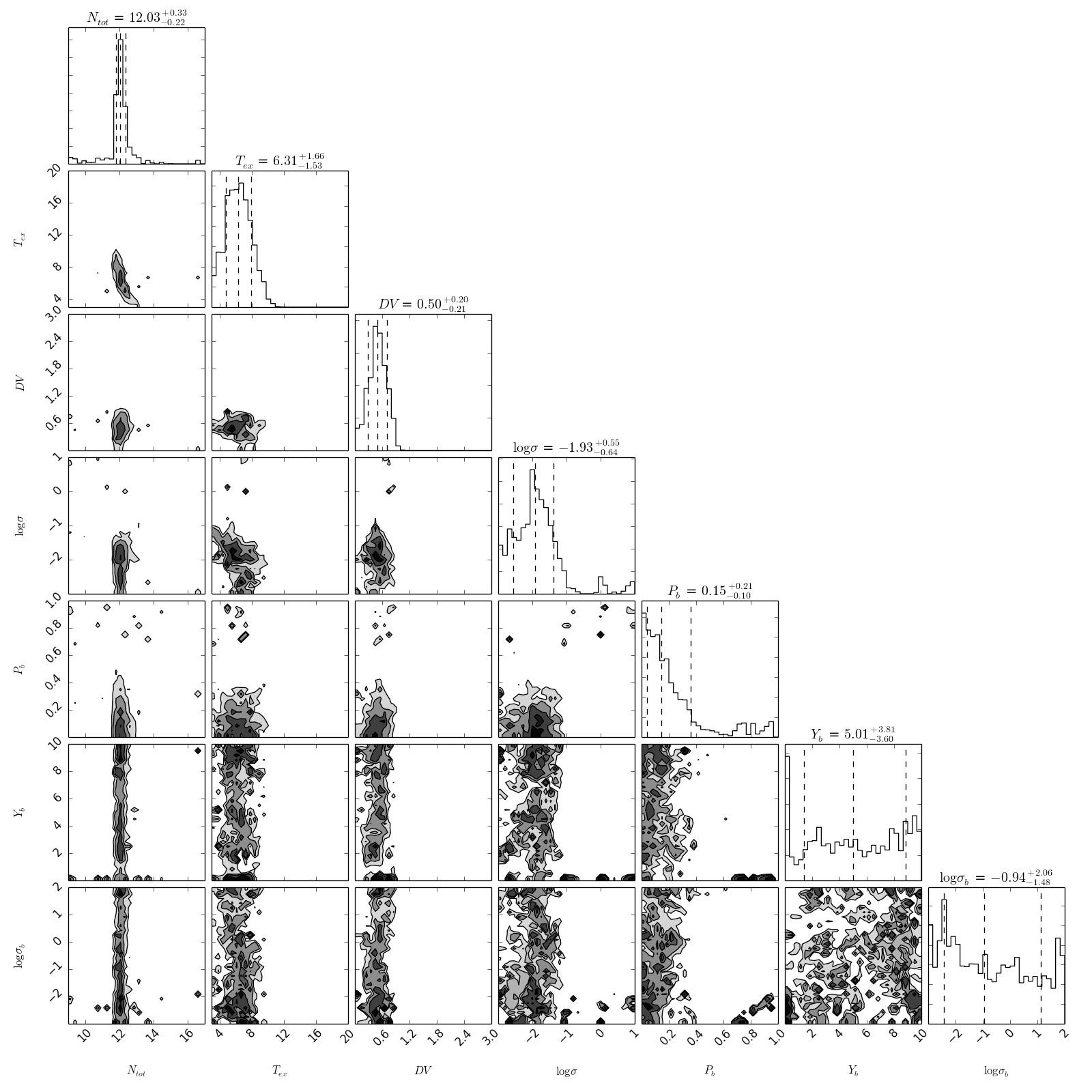}
\includegraphics[width=8cm]{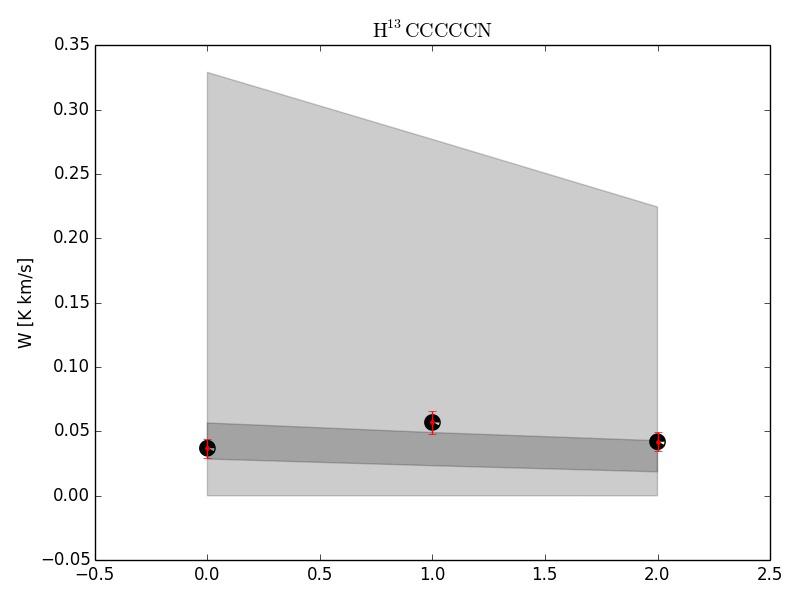}
\includegraphics[width=8cm]{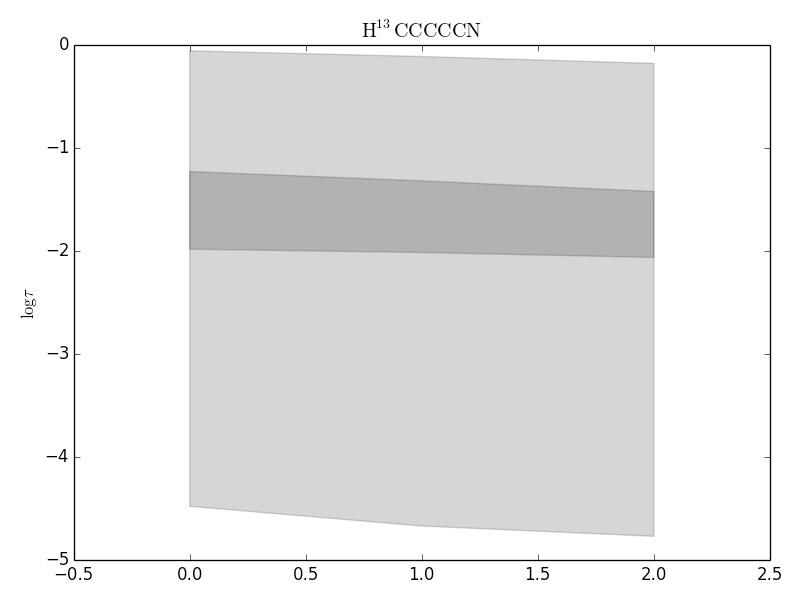}
\caption{\chem{H^{13}CC_4N}}
\label{H13CCCCCN}
\end{figure*}
\clearpage
\begin{figure*}
\includegraphics[width=16cm]{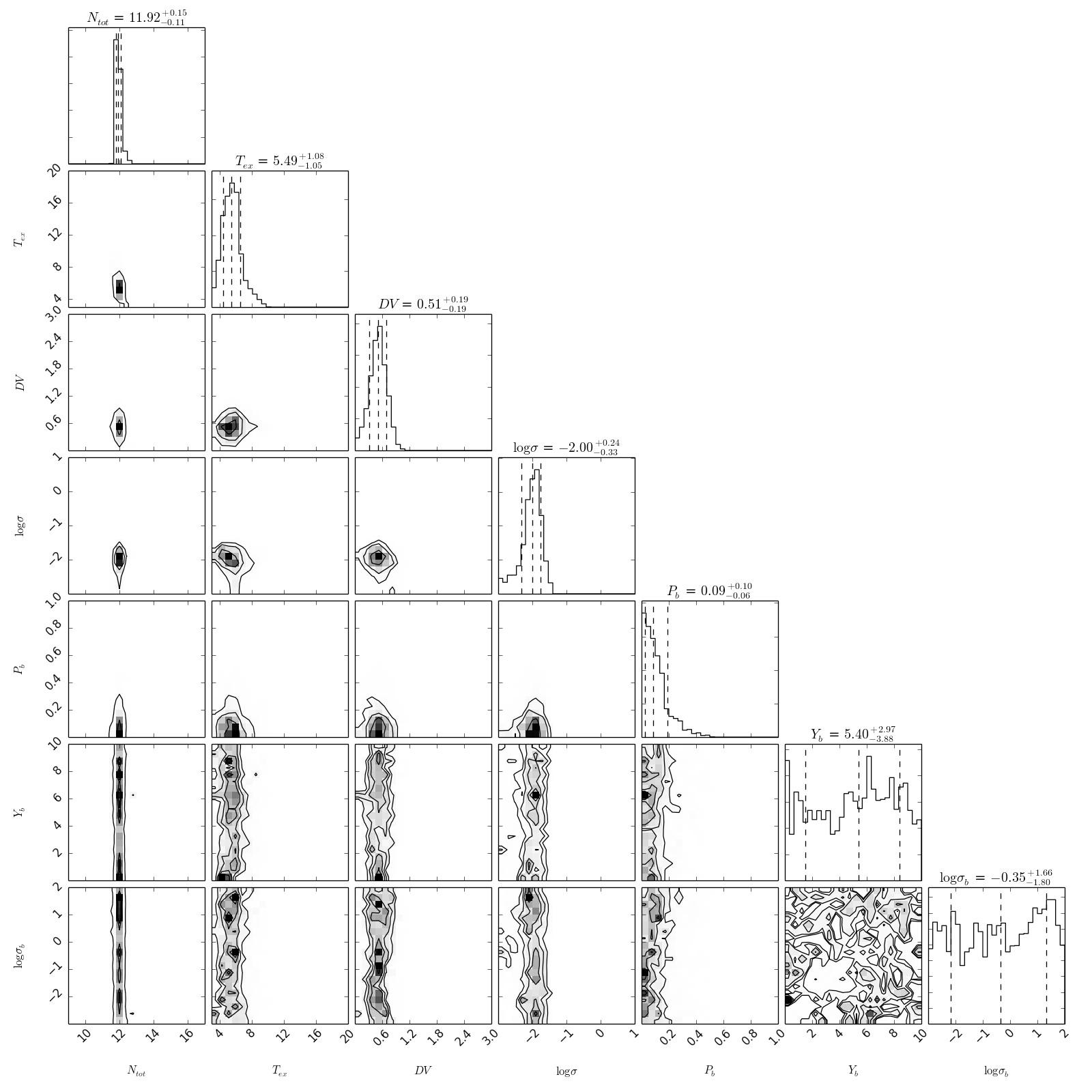}
\includegraphics[width=8cm]{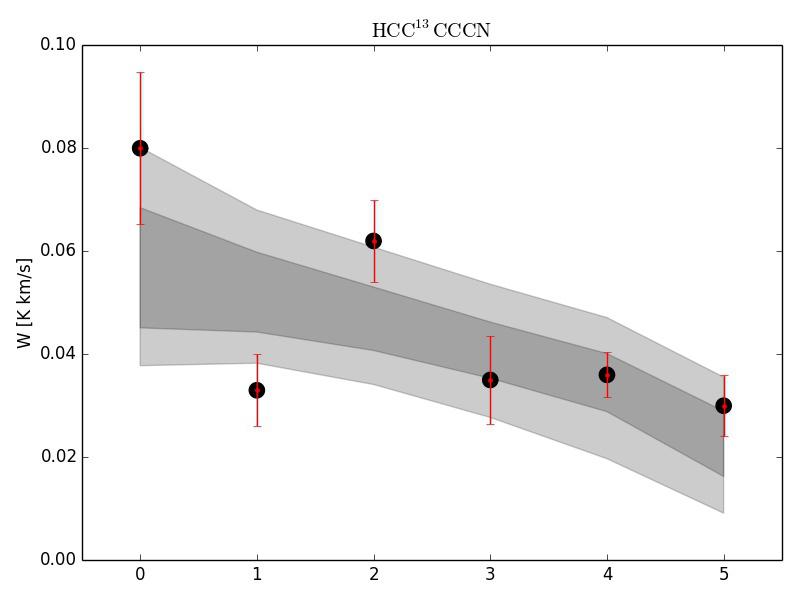}
\includegraphics[width=8cm]{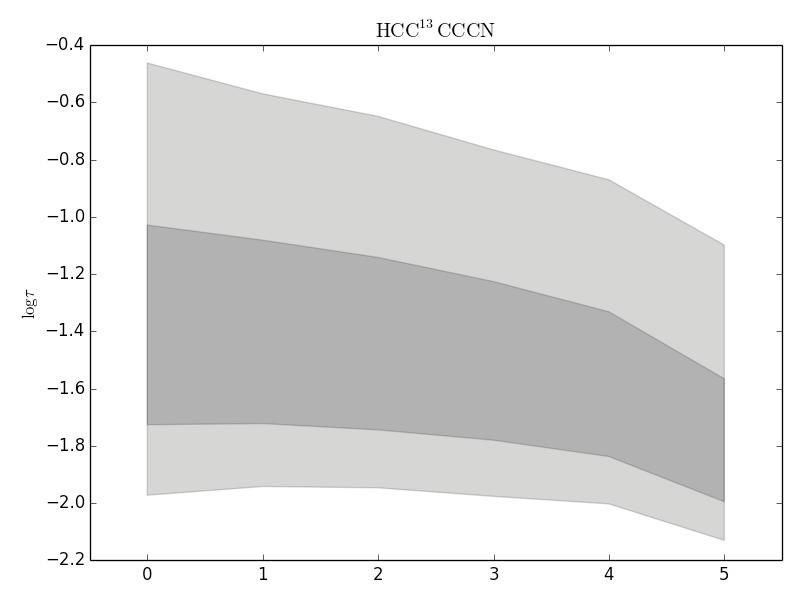}
\caption{\chem{HC_2^{13}CC_2N}}
\label{HCC13CCCN}
\end{figure*}
\clearpage
\begin{figure*}
\includegraphics[width=16cm]{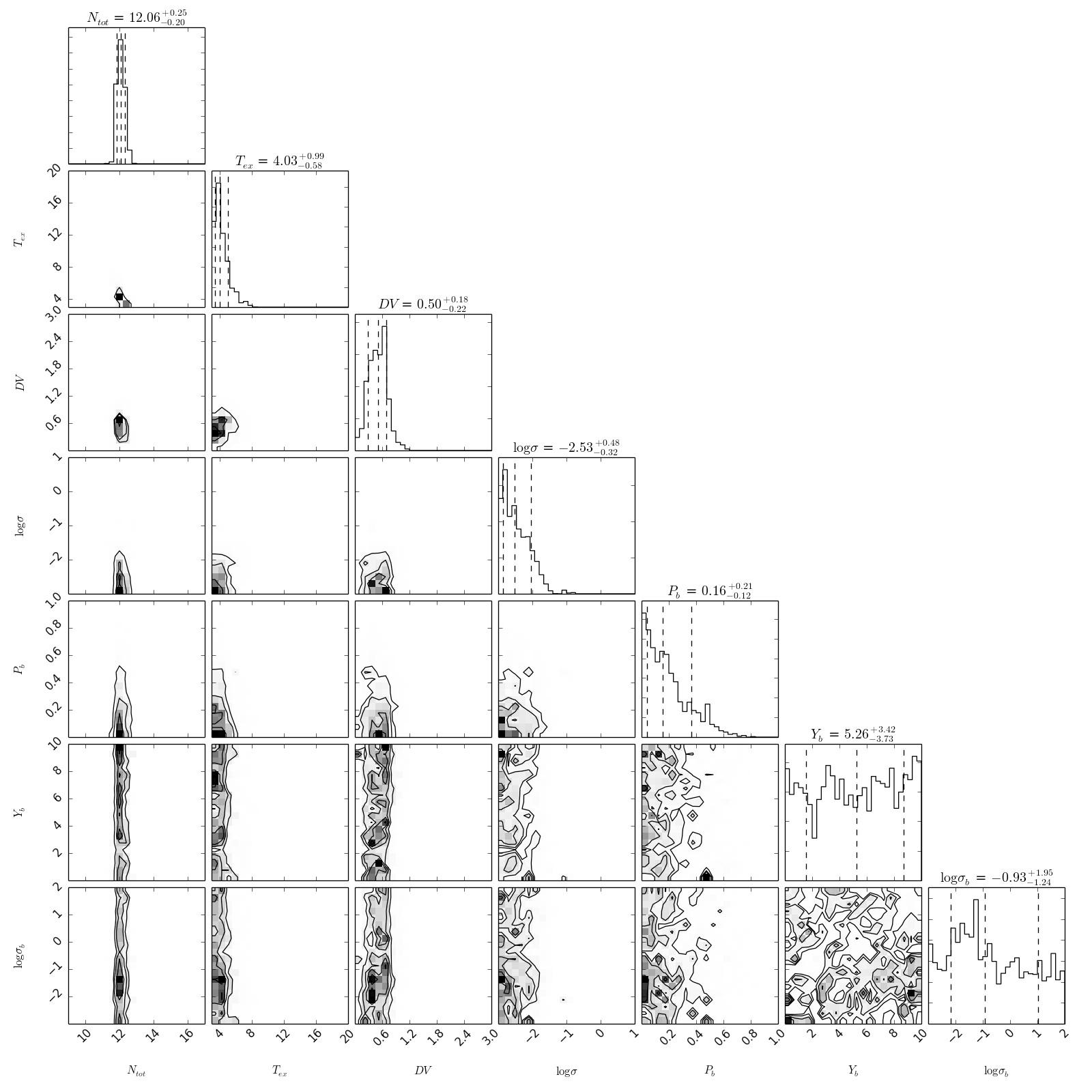}
\includegraphics[width=8cm]{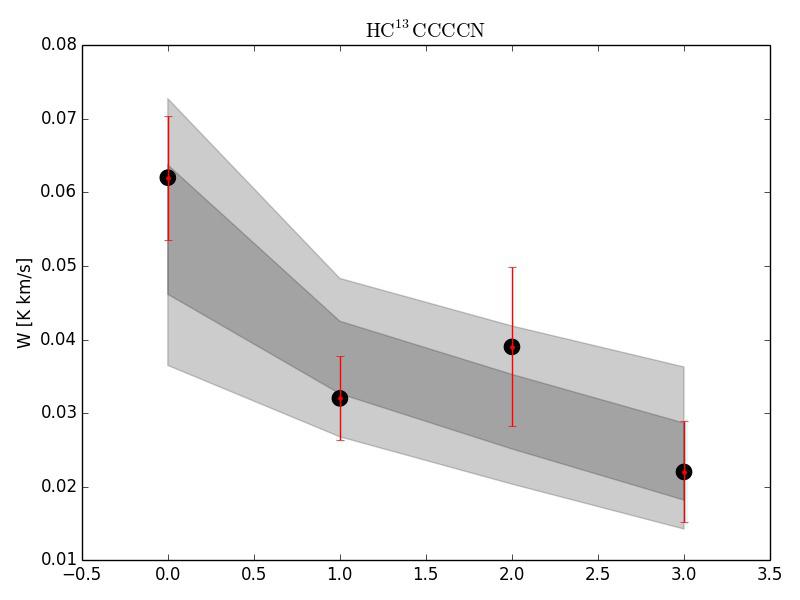}
\includegraphics[width=8cm]{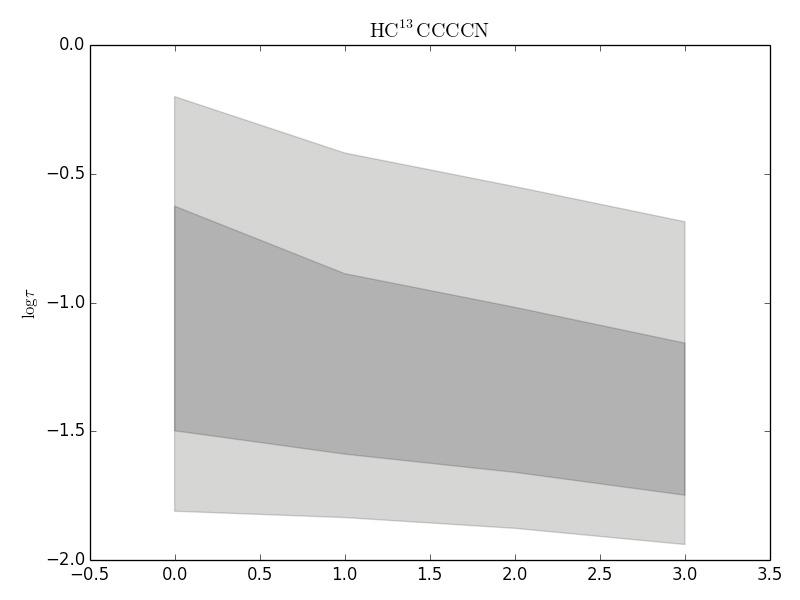}
\caption{\chem{HC^{13}CC_3N}}
\label{HC13CCCCN}
\end{figure*}
\clearpage
\begin{figure*}
\includegraphics[width=16cm]{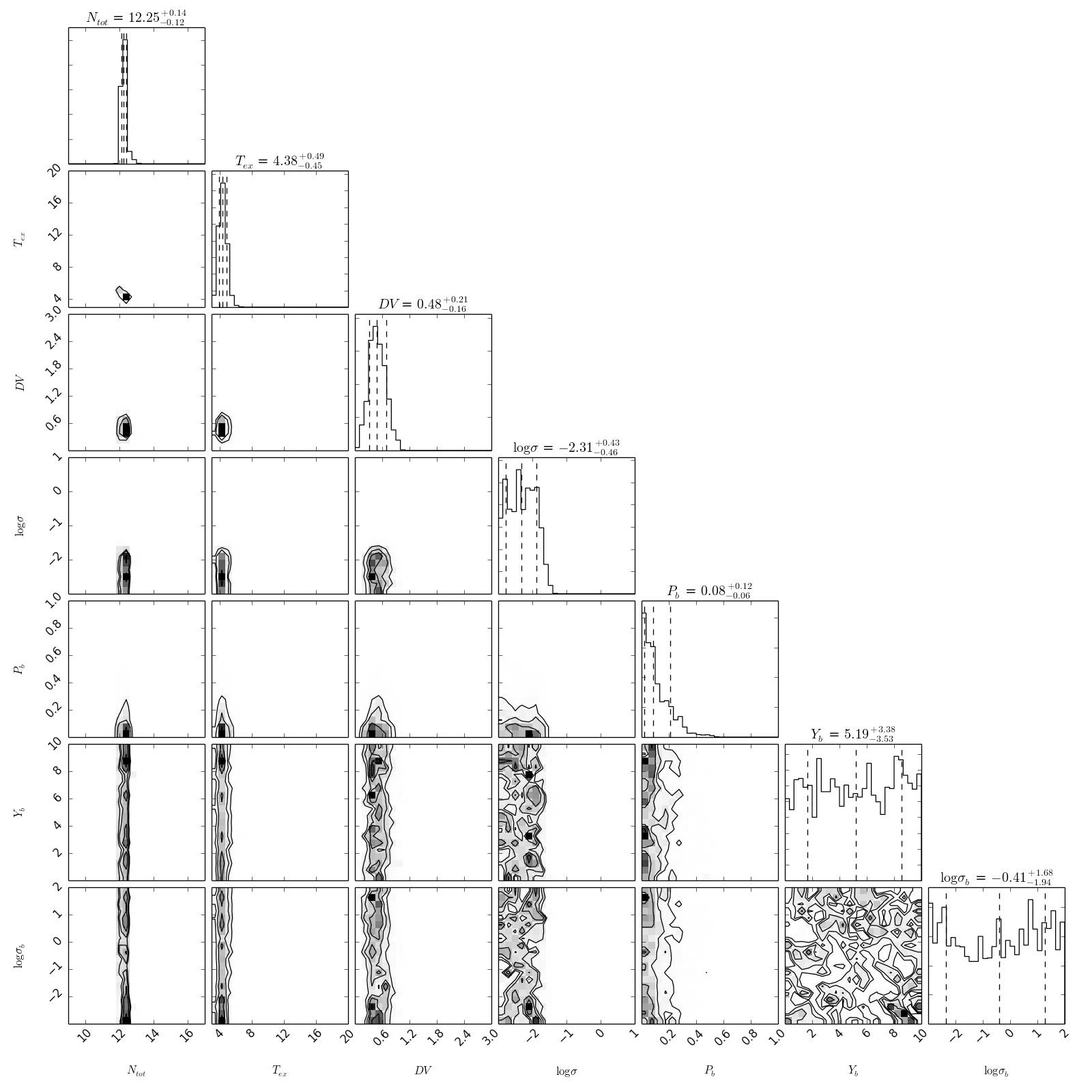}
\includegraphics[width=8cm]{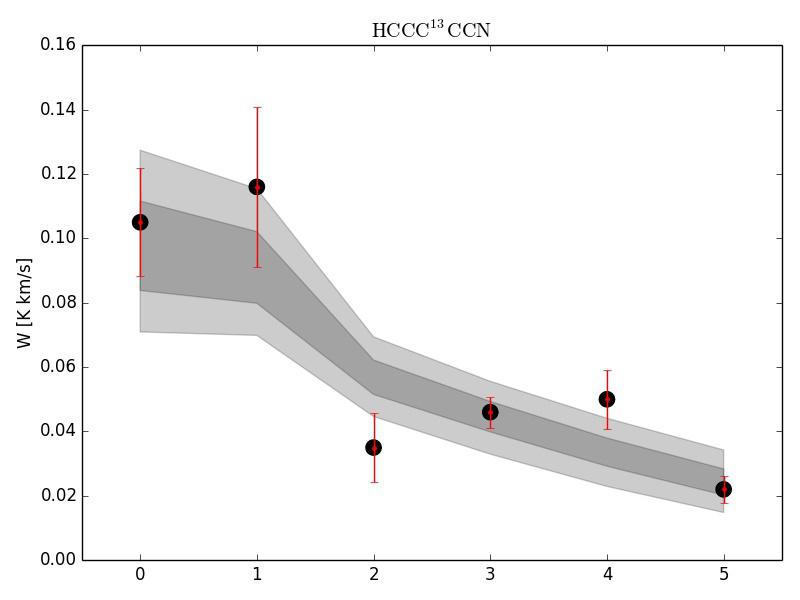}
\includegraphics[width=8cm]{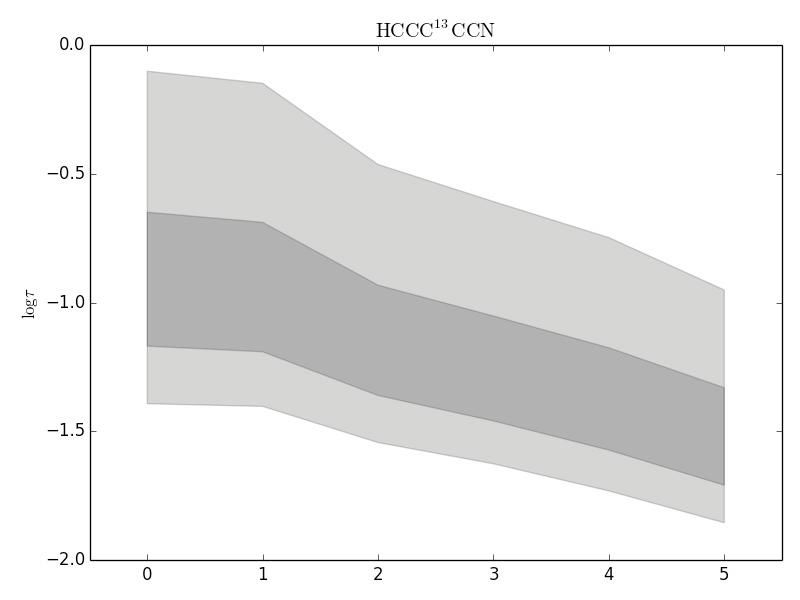}
\caption{\chem{HC_3^{13}CCN}}
\label{HCCC13CCN}
\end{figure*}
\clearpage
\begin{figure*}
\includegraphics[width=16cm]{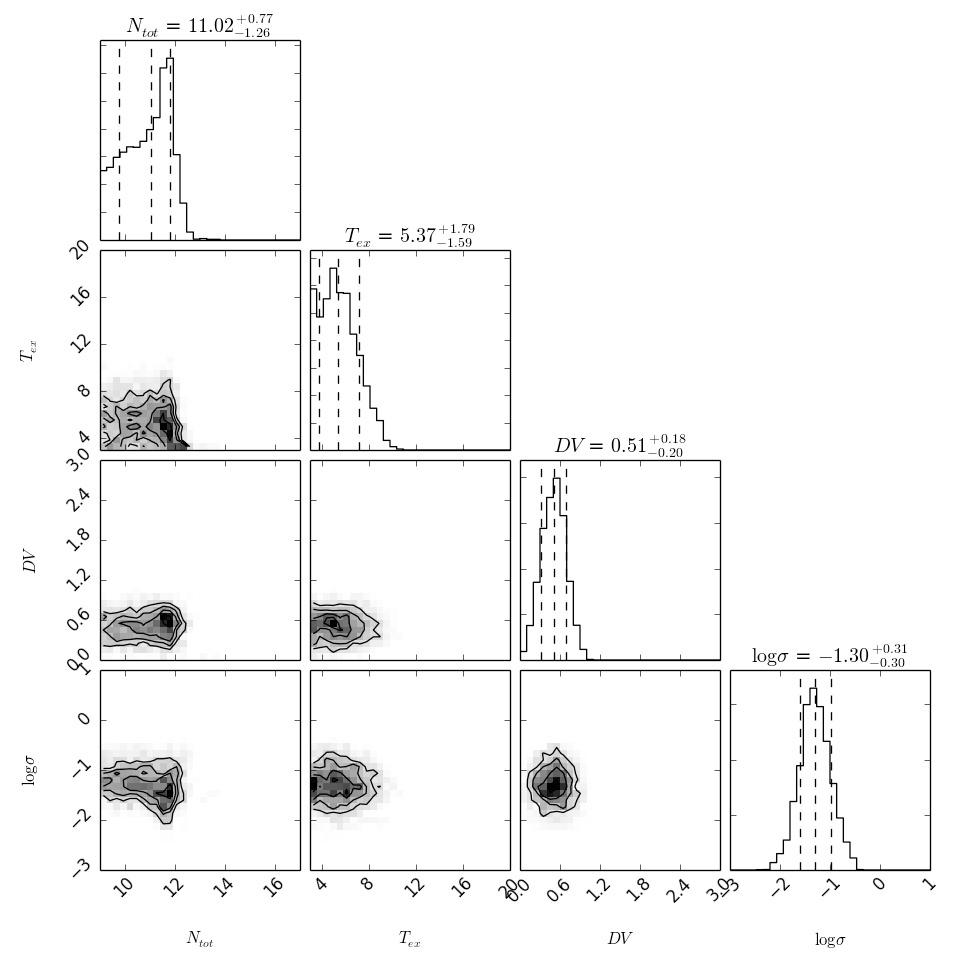}
\includegraphics[width=8cm]{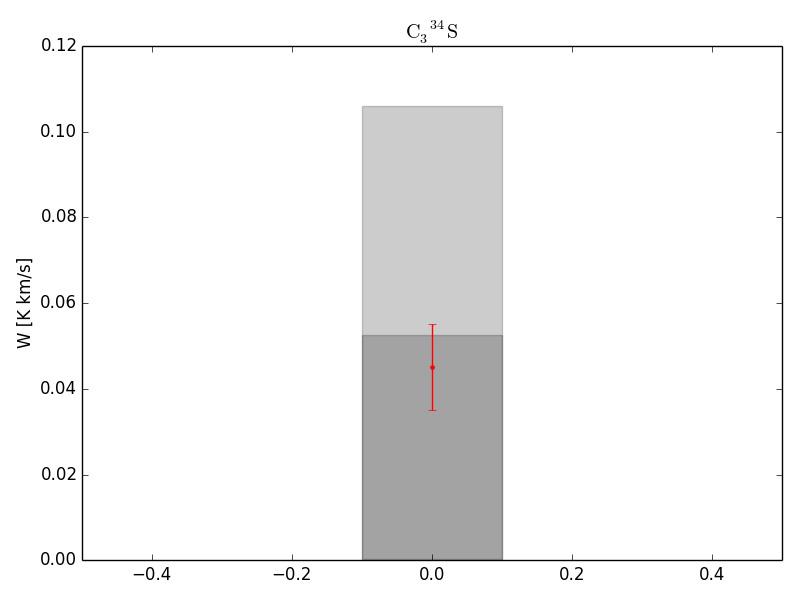}
\includegraphics[width=8cm]{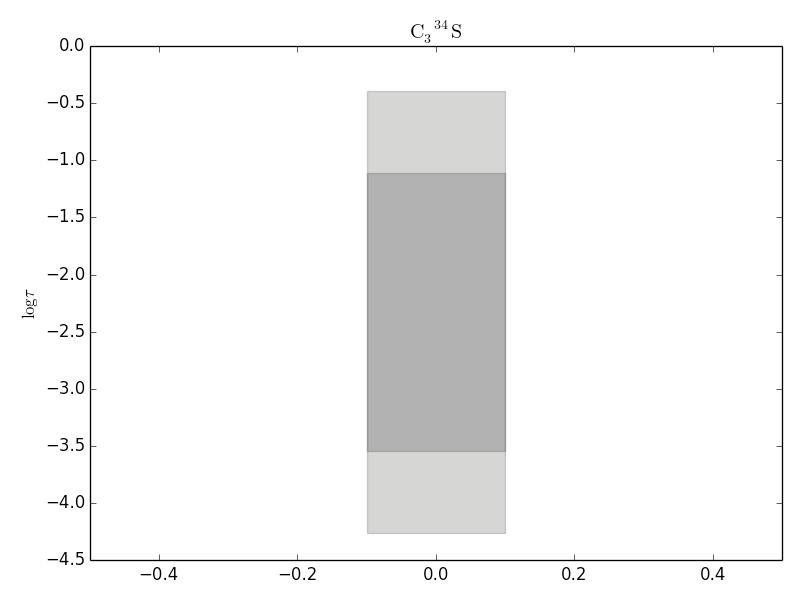}
\caption{\chem{C_3^{34}S}}
\label{C334S}
\end{figure*}
\clearpage
\begin{figure*}
\includegraphics[width=16cm]{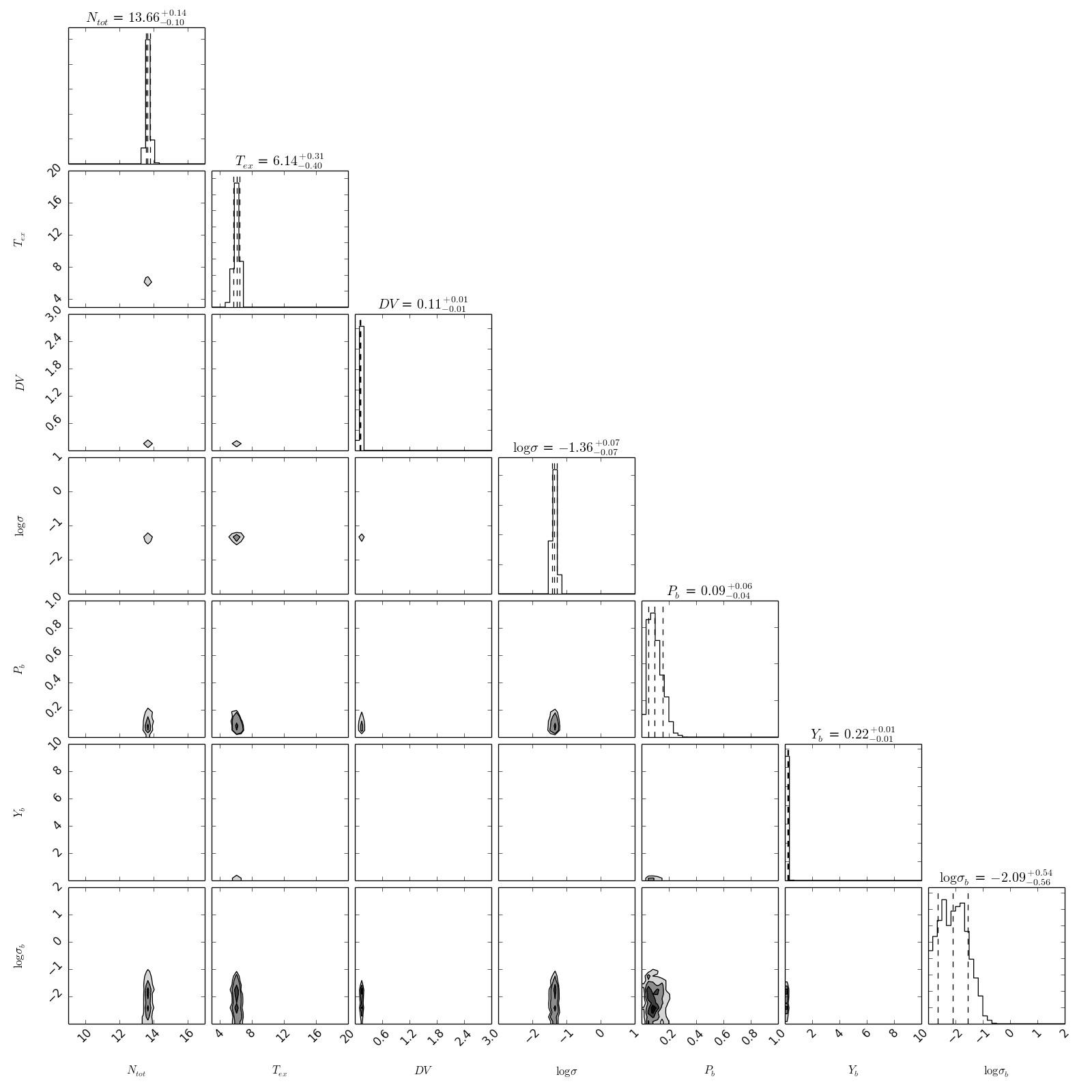}
\includegraphics[width=8cm]{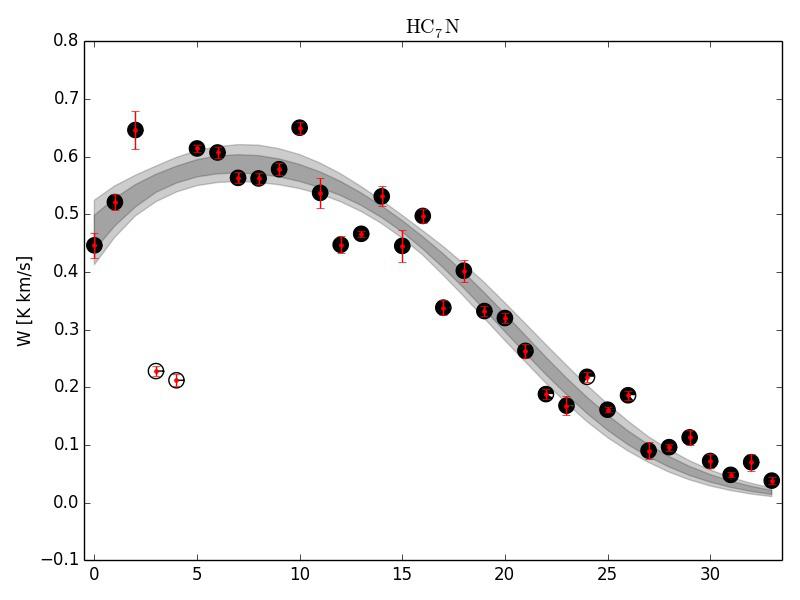}
\includegraphics[width=8cm]{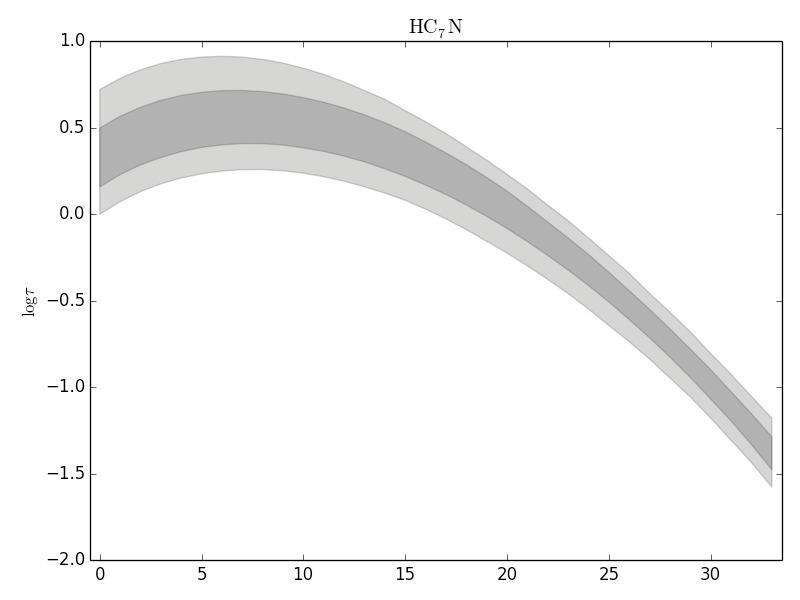}
\caption{\chem{HC_7N}}
\label{HC7N}
\end{figure*}
\clearpage
\begin{figure*}
\includegraphics[width=16cm]{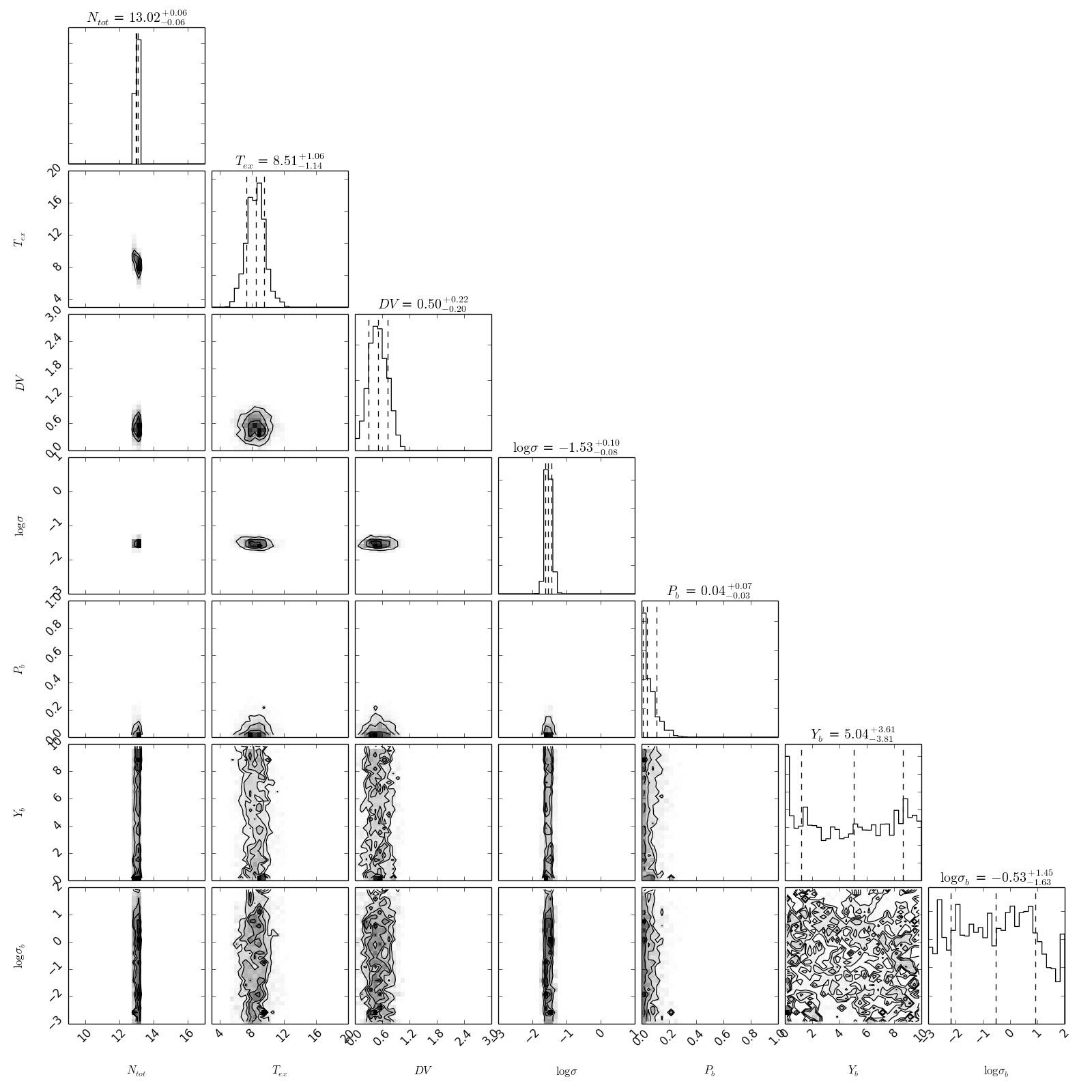}
\includegraphics[width=8cm]{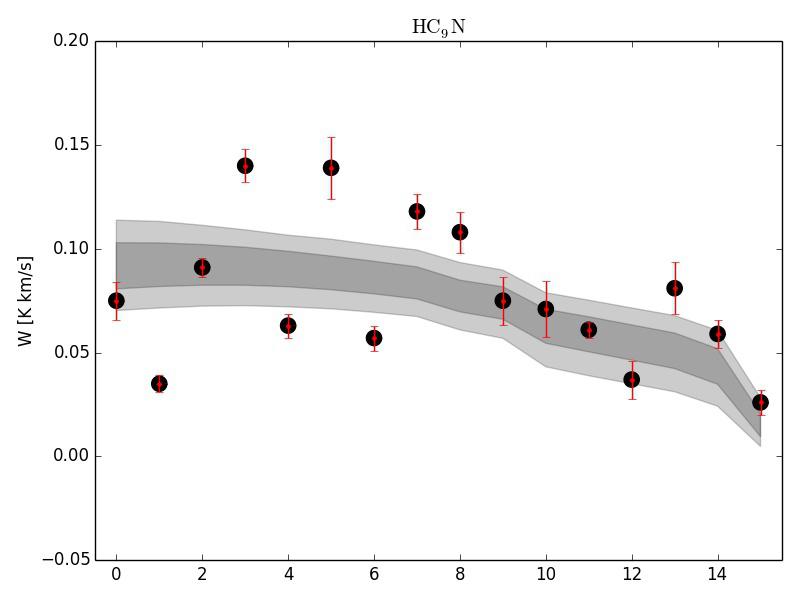}
\includegraphics[width=8cm]{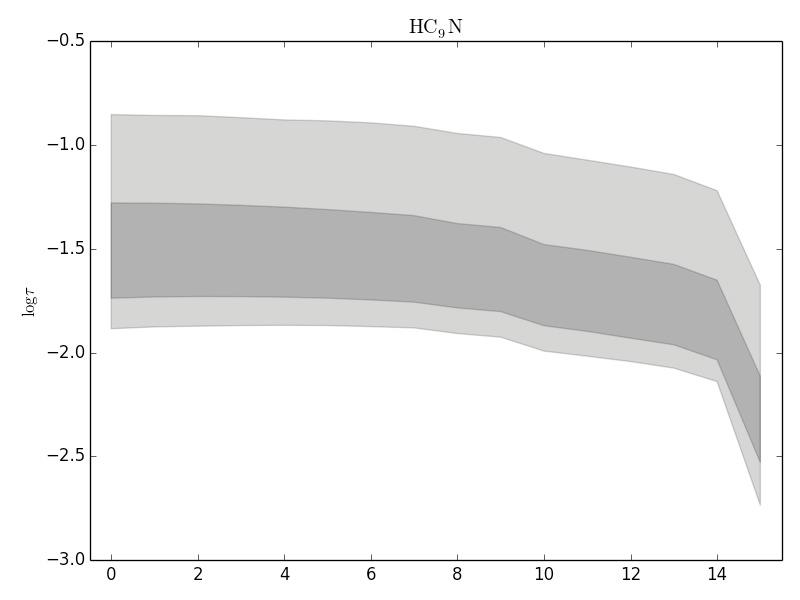}
\caption{\chem{HC_9N}}
\label{HC9N}
\end{figure*}
\clearpage

\bibliographystyle{apj} 
\bibliography{apj-jour,references} \clearpage

\end{document}